\documentclass[journal]{IEEEtran}
\ifCLASSINFOpdf
\else
 \usepackage[dvips]{graphicx}
\fi
\usepackage{multirow}
\usepackage{lineno}

\usepackage{hepunits}

\usepackage{cite}

\usepackage{authblk}

\begin{document}

\begin{titlepage}

\thispagestyle{empty}
\def\thefootnote{\fnsymbol{footnote}}       

\begin{center}
\mbox{ }

\end{center}
\begin{center}
\vskip 1.0cm
{\Huge\bf
MPX Detectors 
\vspace{2mm}

as LHC Luminosity Monitor
}
\vskip 1cm
{\LARGE\bf 
Andr\'e~Sopczak$^1$,
Babar Ali$^1$,
Nedaa Asbah$^2$,
Benedikt Bergmann$^1$,
Khaled Bekhouche$^3$,
Davide~Caforio$^1$,
Michael Campbell$^4$, \\
Erik Heijne$^1$,
Claude Leroy$^2$,
Anna Lipniacka$^5$,
Marzio Nessi$^4$,
Stanislav Posp\'i\v sil$^1$,
Frank Seifert$^1$, 
Jaroslav \v Solc$^1$, 
Paul Soueid$^2$,\\
Michal Suk$^1$,
Daniel~Ture\v cek$^1$,
Zden\v ek Vykydal$^1$ 
\bigskip
}

\Large 
$^1$Institute of Experimental and Applied Physics, 
Czech Technical University in Prague, Czech Republic\\
$^2$Group of Particle Physics, University of Montreal, Canada\\ 
$^3$Facult\'e des Sciences et de la Technologie, 
Universit\'e Mohammed Khidher de Biskra, Algeria\\
$^4$CERN, Switzerland\\
$^5$Department for Physics and Technology, Bergen University, Norway

\vskip 1.0cm
\centerline{\Large \bf Abstract}
\end{center}

\vskip 1.2cm
\hspace*{-0.5cm}
\begin{picture}(0.001,0.001)(0,0)
\put(,0){
\begin{minipage}{\textwidth}
\Large
\renewcommand{\baselinestretch} {1.2}
A network of 16 Medipix-2 (MPX) silicon pixel devices was installed in the ATLAS detector cavern at CERN.
It was designed to measure the composition and spectral characteristics of the radiation field in the ATLAS 
experiment and its surroundings. 
This study demonstrates
that the MPX network can also be used as a 
self-sufficient luminosity monitoring system. The MPX detectors 
collect
data independently 
of the ATLAS data-recording chain, and thus they provide independent 
measurements of the bunch-integrated 
ATLAS/LHC luminosity. 
In particular, the MPX detectors located close enough to the primary interaction point are used 
to perform van der Meer calibration scans with 
high precision. 
Results from the luminosity monitoring are presented for 2012 data taken at 
$\sqrt{\it s}=8$~TeV proton-proton collisions.
The characteristics of the LHC luminosity reduction rate are studied and the effects of beam-beam 
(burn-off) and beam-gas (single bunch) interactions are evaluated.
The systematic variations observed in the MPX luminosity measurements 
are below 0.3\% for one minute intervals.
\renewcommand{\baselinestretch} {1.}

\normalsize
\vspace{1.5cm}
\begin{center}
{\sl \large
Presented at the IEEE 2015 Nuclear Science Symposium, San Diego, USA 
\vspace{-6cm}
}
\end{center}
\end{minipage}
}
\end{picture}
\vfill

\end{titlepage}

\newpage
\setcounter{page}{1}

\title{\vspace*{-2mm} MPX Detectors \\ as LHC Luminosity Monitor}

\author{Andr\'e Sopczak, {\em Senior Member, IEEE},
Babar Ali,
Nedaa Asbah,
Benedikt Bergmann,
Khaled Bekhouche,
Davide~Caforio,
Michael Campbell,
Erik Heijne, {\em Fellow IEEE},
Claude Leroy, {\em Member IEEE},
Anna Lipniacka,
Marzio Nessi,
Stanislav Posp\'i\v sil, {\em Senior Member, IEEE},
Frank Seifert, 
Jaroslav \v Solc, 
Paul Soueid,
Michal Suk,
Daniel~Ture\v cek,
Zden\v ek Vykydal 
\thanks{A. Sopczak, B. Ali, B. Bergmann, D. Caforio, E. Heijne, S. Posp\'i\v sil, F. Seifert, J. \v Solc, M. Suk, D. Ture\v cek, and Z. Vykydal are with the 
Institute of Experimental and Applied Physics, 
Czech Technical University in Prague, Horska 3a, CZ-12800, 
Czech Republic (e-mail: andre.sopczak@cern.ch).}
\thanks{N. Asbah, C. Leroy, and P. Soueid are with the Group of Particle Physics, University of Montreal, 2900 boul. \'Edouard-Montpetit, Montr\'eal QC  H3T 1J4T, Canada.}
\thanks{K. Bekhouche is with the Facult\'e des Sciences et de la Technologie, Universit\'e Mohammed Khidher de Biskra, BP 145 RP, 07000 Biskra, Algeria.}
\thanks{M. Campbell and M. Nessi are with CERN, CH-1211 Geneva 23, Switzerland.}
  \thanks{A. Lipniacka is with the Department for Physics and Technology, University of Bergen, 
All\'eegaten 55,  N-5007 Bergen, Norway.}
}

\markboth{}
         {}

\maketitle

\begin{abstract}
A network of 16 Medipix-2 (MPX) silicon pixel devices was installed in the ATLAS detector cavern at CERN.
It was designed to measure the composition and spectral characteristics of the radiation field in the ATLAS 
experiment and its surroundings. 
This study demonstrates
that the MPX network can also be used as a 
self-sufficient luminosity monitoring system. The MPX detectors 
collect
data independently 
of the ATLAS data-recording chain, and thus they provide independent 
measurements of the bunch-integrated 
ATLAS/LHC luminosity. 
In particular, the MPX detectors located close enough to the primary interaction point are used 
to perform van der Meer calibration scans with 
high precision. 
Results from the luminosity monitoring are presented for 2012 data taken at 
$\mathbf{\sqrt{\it s}=8}$~TeV proton-proton collisions.
The characteristics of the LHC luminosity reduction rate are studied and the effects of beam-beam 
(burn-off) and beam-gas (single bunch) interactions are evaluated.
The systematic variations observed in the MPX luminosity measurements 
are below 0.3\% for one minute intervals.
\end{abstract}


\section{Introduction}
\label{sec:introduction}

A comprehensive analysis of data taken by a network of Medipix-2 (MPX) devices
as a self-sufficient luminosity monitoring system is presented. 
The MPX network was installed at different locations in the ATLAS 
detector~\cite{atlasCollaboration:2013} at CERN and in its cavern.
The MPX devices are based on the Medipix-2 hybrid silicon pixel detector which was
developed by the Medipix-2 Collaboration~\cite{medipixCollaboration:2013}. 
It consists of a $\approx 2$\,cm$^2$ silicon sensor matrix of $256\times256$ cells, 
bump-bonded to a readout chip. Each matrix element 
($55 \times 55\,\micron^2$ pixel, $300\,\micron$ thick) is
connected to its respective readout chain integrated on the readout chip. 
Pulse height discriminators determine the input energy window 
and 
provide noise suppression. 
A counter in each pixel records interacting quanta of radiation, photons, neutrons, electrons,
minimum ionizing particles, and ions with energy deposits falling within 
the preset energy window~\cite{mpx2ProjectProposal:2006,analysisRadiaField:2013}.

The ATLAS and CMS collaborations have 
elaborate systems
of luminosity measurements,
described in~\cite{improvedLumiDet:2013} and~\cite{cms:2013}, respectively. 
A comparative study of their results and the MPX luminosity monitoring
is beyond the scope of this article.
The methods and techniques described in this article were pioneered for 
high-precision luminosity determination
for the 2012 data-taking period at the LHC. 
These techniques and analysis methods are based on precision counting of 
particles passing the sensors. 
Similar requirements apply to analysis techniques used in other fields of research
(e.g. medical applications and space science) 
where high precision and long-term time-stability of measurements are needed.
MPX devices have already been successfully applied in these areas of research 
and the fundamental studies presented in this article can lead to further advances in their 
application.

The capability 
of the MPX devices for luminosity monitoring has been investigated 
before~\cite{sopczak:2013}. 
It is shown in this article that the MPX network is self-sufficient for luminosity monitoring.
In particular, van der Meer (vdM) scans~\cite{vdm} can be used for absolute luminosity calibration.
Detailed analysis of the MPX data allows the quantification of the long-term stability 
over one year of data-taking. It also provides short-term (minute by minute) precision. 
This information is crucial to evaluate the performance of the MPX network as a luminosity monitoring 
system.

The detection of charged particles in the MPX devices is based on 
the ionization energy deposited by particles passing through the silicon sensor. 
The signals are amplified and counted during an adjustable time window (frame) for 
each pixel. Neutral particles, however, need to be converted to charged particles before they can 
be detected. Therefore, a part of each silicon sensor is covered by a $^6$LiF converter. 

The MPX pixel detector can be operated in tracking 
or counting mode~\cite{mpx2ProjectProposal:2006,analysisRadiaField:2013}.
Every pixel records the number of hits within an adjustable time interval (acquisition time).

One of the important features of the MPX devices is the ability to record and identify clusters.
Clusters are defined as patterns of adjacent pixels with energy
deposits defined in~\cite[Sec.2.2]{analysisRadiaField:2013}. 
Different particles that traverse the device cause different cluster shapes. 
These shapes allow particle identification and the distinction between keV-MeV electrons, photons, 
energetic hadrons, alpha particles and ion-fragments. 
The energy deposited during the acquisition time can be estimated as well.
The data is stored frame-by-frame. 
After the data acquisition is closed, it takes about~6\,s 
to transmit the status of the full 65536 pixel matrix. 
The device is not sensitive during the readout process (dead-time).

The primary goal of the MPX network was to provide information on the radiation composition 
within the ATLAS cavern (including the thermal neutron component). The MPX detectors also allow the measurement
of beam-induced radioactivity during and after collisions. This real-time
measurement of the LHC-generated background radiation permits the validation of
background radiation simulation studies. Results from 2008-2011 MPX data-taking
have been released~\cite[Sec. 4]{analysisRadiaField:2013}. 

The use of the MPX network for relative luminosity measurements in proton-proton
collisions proposed in~\cite{mpx2ProjectProposal:2006} is studied here in detail.

\begin{figure}[htp]
\centering
\vspace*{2mm}
\includegraphics[width=\linewidth]{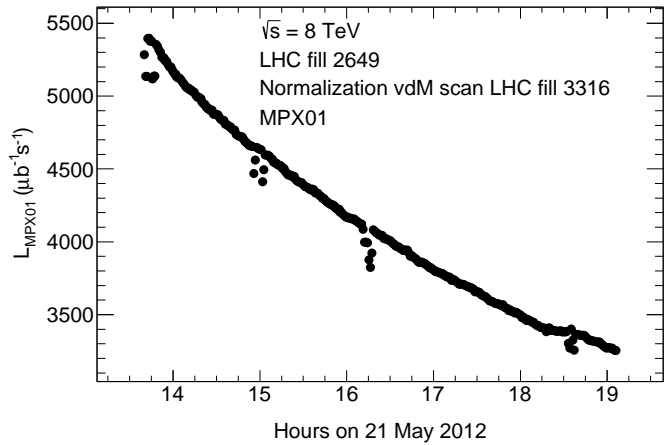}
\vspace*{-4mm}
\caption{Time history (CEST) of the MPX luminosity.
The small dips, visible as variations from the descending curve, 
correspond to times when the LHC operators performed small-amplitude 
beam-separation scans to optimize the luminosity,
The normalization between MPX01 hit rate and luminosity is based on 
van der Meer scans using LHC fill 3316, detailed in Sec.~\ref{sec:vdM_scans}.}
\label{fig:mpxMayRun}
\end{figure}

Thirteen out of the sixteen installed devices were used for the analysis.
Two detectors were noisy due to radiation damage. 
One detector was located too far away from the interaction point.
Table~\ref{tab:mpx_detectors} lists  the locations of the detectors and 
number of registered events (clusters)
per unit sensor area and per unit integrated luminosity.
Figure~\ref{fig:mpxMayRun} shows an example of the luminosity from hit counting 
measured with MPX01 for LHC fill 2649.

\begin{table}[bp]
\caption{MPX device locations with respect to the interaction point.
$Z$ is the longitudinal distance from the interaction point and
$R$ is the distance from the beam axis. 
Only devices with low cluster rates are used for the heavy blob (thermal
neutron) counting analysis as indicated.
Ordering in the table is given with decreasing particle flux.
$^*$rejected during analysis (Sec.~\ref{sec:lumi_th_neutron_counting}).
}
\label{tab:mpx_detectors}
\vspace*{-4mm}
\begin{center}
\renewcommand{\arraystretch}{1.2}
             \begin{tabular}{crrccc}
                        \hline\hline
                Device  & $Z$~  & $R$~  & Measured MPX clusters per             & Used for\\
                        & (m) & (m) & unit sensor area and per unit         & th.\,neutron\\
                        &     &     & luminosity $(\rm cm^{-2}/ nb^{-1})$   & analysis\\
                        \hline
                        MPX01 & 3.42    & 0.77  & 55000   & No\\
                        MPX13 & $-3.42$ & 2.44  & 380     & No\\
                        MPX02 & 3.42    & 2.50  & 230     & No\\
                        MPX04 & 7.12    & 1.30  & 110     & No\\
                        MPX05 & 7.20    & 2.36  & 47      & No\\
                        MPX03 & 2.94    & 3.57  & 31      & No\\
                        MPX06 & 7.20    & 3.36  & 20      & Yes$^*$\\   
                        MPX09 & 15.39   & 1.56  & 5.8     & Yes\\
                        MPX12 & 7.23    & 6.25  & 3.9     & Yes\\
                        MPX08 & 4.02    & 4.40  & 1.2     & Yes\\
                        MPX10 & 22.88   & 5.19  & 1.0     & Yes\\
                        MPX07 & 0.35    & 4.59  & 0.45    & Yes\\
                        MPX11 & 4.86    & 16.69 & 0.30    & Yes\\
                        \hline\hline
                \end{tabular}
\end{center}
\end{table}

The paper is structured as follows.
Section~\ref{sec:lumi_hit_counting} describes the luminosity measurements from hit counting, and 
Section~\ref{sec:lumi_th_neutron_counting} describes the luminosity measurements from heavy blob 
(thermal neutron) counting.
The relation between hits and clusters used to evaluate the statistical precision is 
discussed in Section~\ref{sec:statistics}.
Details of the analysis of MPX data taken during LHC vdM scans for an absolute luminosity calibration
are given in Section~\ref{sec:vdM_scans}.
The short-term MPX precision evaluated from a detailed study of LHC luminosity curves is presented in 
Section~\ref{sec:short-term}. Conclusions are given in Section~\ref{sec:conclusions}.

\section{MPX Luminosity from Hit Counting}
\label{sec:lumi_hit_counting}

The six MPX devices with highest cluster rate, specified in Table~\ref{tab:mpx_detectors}
are used (MPX01-05,13).
Each device measures the luminosity independently and is cross-checked with the other devices.
This is an intrinsic advantage of the MPX network.

Each MPX device has its own acquisition time window independent of the other devices.
In order to compare luminosity measurements from different devices, a common time window
is introduced, called Luminosity Block (LB)
which is typically one minute long.

A small number of noisy pixels could have a significant effect on the 
luminosity measurement.
Three independent methods differing in 
procedure and criteria 
for noisy pixel removal have been tested. 
The results of their respective luminosity measurements are compared as 
potential sources of systematic uncertainty.
\begin{itemize}
\item{Method 1: A pixel is defined as noisy if its count rate is more than five
standard deviations from the average.
If a pixel is found to be noisy in a 24 hour reference period 
(chosen as the day in 2012 that had the largest number of noisy pixels) 
it is removed from the entire 2012 data-taking period.
A linear interpolation is made between the rates (number of hits per second) 
in different frames. 
The hit rate at the middle of an LB is obtained from this interpolation.
}
\item{Method 2: Noisy pixel removal  is done  frame by frame,
i.e. a different set of noisy pixels is removed in each frame.
Noisy pixels are those with a counting rate that differs from the mean by more
than a luminosity-dependent threshold.
The MPX luminosity from frames falling within an LB is used without an
interpolation. 
A correction is made for the relative duration of the MPX frames and of the LB.
}
\item{Method 3: Noisy pixel removal  is done  frame by frame.
The counts of 15 frames (the frame under investigation and 7 frames before and after) 
are summed and a pixel is removed if the sum of these counts is above a threshold.
An interpolation of the frame hit rate at the time of each LB is 
done as in method~1.}
\end{itemize}
All three methods show a significant increase in the number of noisy pixels with time, 
when applied to MPX01 data.
This might indicate possible radiation damage in the readout chip.
In method~1 (method~2), the number of noisy pixels in MPX01 increases from less 
than 10 (300) in April 2012 to about 300 (1800) at the end of November 2012.
The other devices have a smaller number of noisy pixels since they are exposed to a 
much lower particle flux (Table~\ref{tab:mpx_detectors}).

The luminosities measured with the three methods were compared in
short (frame-by-frame) and long (7 months) time periods. 
Depending on the MPX device considered, the frame-by-frame agreement 
varies from a few percent to less than 0.1\% (for MPX01). 
The largest variation is between method 2 and the other two methods.
We have determined that most of this variation is attributed to the
conversion between frames and LBs in method~2. 
In the following, method 1 is used and thus
the same noisy pixels are removed 
for the whole 2012 data-taking period.

During the analysis of the MPX data, time-shifts between the three readout 
PCs were noticed. Therefore, they were synchronized off-line 
by analyzing the rising and falling luminosity curves when an LHC 
fill starts and ends. 
A time accuracy better than the LB length was therefore achieved.

In the process of performing the luminosity determination with MPX devices, the activation of the 
ATLAS detector material was investigated and found to have a negligible effect. 

First, the hit rates per frame $N_{\rm hits}/t_{\rm acquisition}$ 
are converted into hit rates per LB 
for each MPX device separately. The procedure is described below. 
Frames within the time window of the LB are selected.
The hit rates of these frames are averaged. Thus, one hit rate is stored per LB. 
Given that the acquisition times vary between 5 and 120\,s the number of frames 
used per LB varies for the six MPX devices. If there is no hit rate for 
a given LB, the previous LB hit rate is used.
In the MPX luminosity analysis only those LBs are used for which all six
MPX devices (MPX01-05 and MPX13) were operational.    

The hit rate for the MPX01 device is normalized to units of luminosity 
by multiplying with the factor
$n_{\rm f}= 1.5628\cdot 10^{-3}~{\rm \micro b^{-1}/ hit}$,
derived in Sec.~\ref{sec:vdM_scans}. 
Then, the other devices are normalized to MPX01
based on the average hit rate for the June to November 2012 running period.
Table~\ref{tab:MPX_SUM_used} summarizes the normalization factors.

The average luminosity (MPXav) for all other devices 
(excluding the one under consideration) is calculated LB-by-LB.
Using this normalization factor, each MPX device obtains an equal weight, 
although the MPX devices have largely varying particle fluxes (hit rates).

\begin{figure*}[htp]
\vspace*{-0.7cm}
\centering
\includegraphics[width=0.49\linewidth]{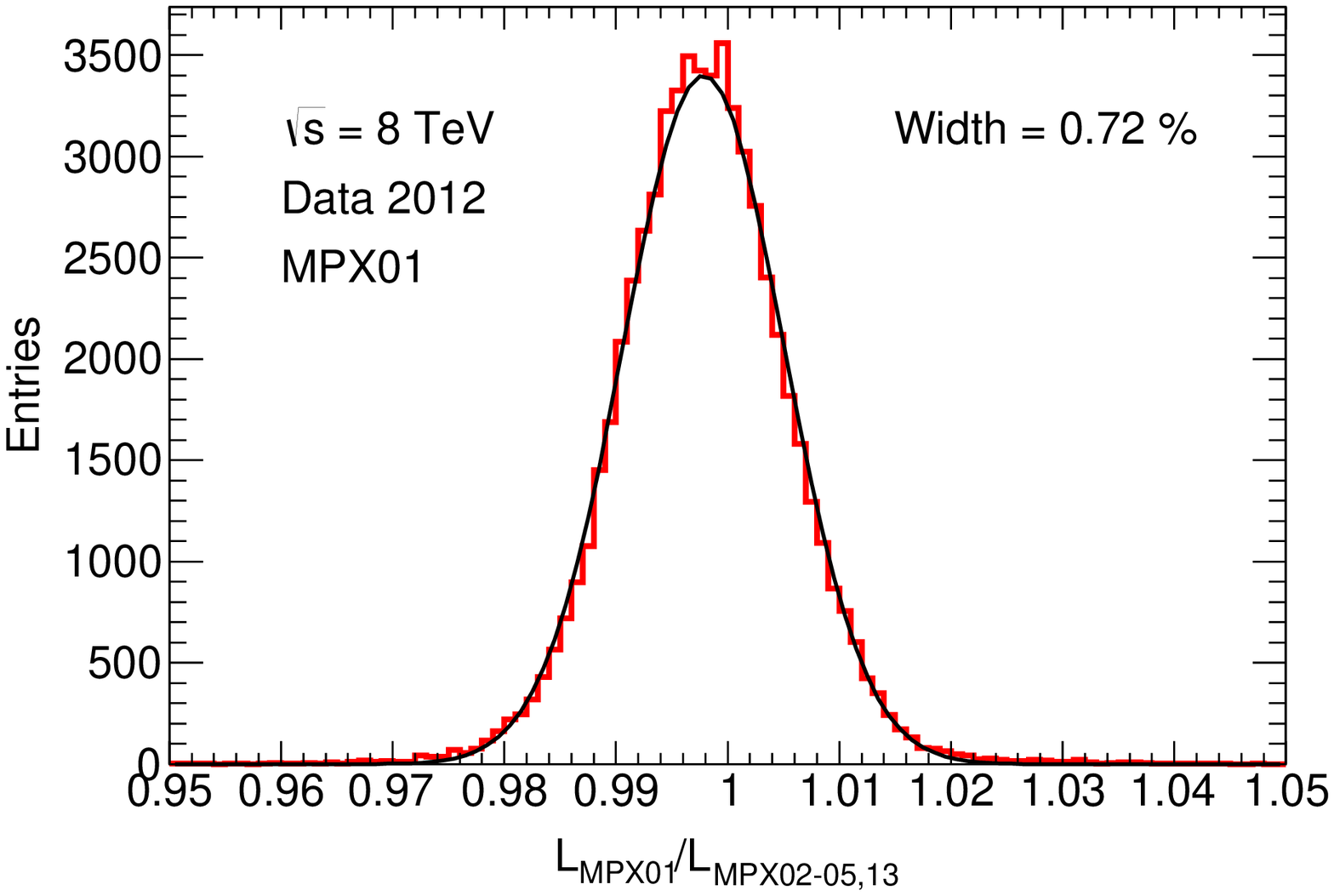}
\includegraphics[width=0.49\linewidth]{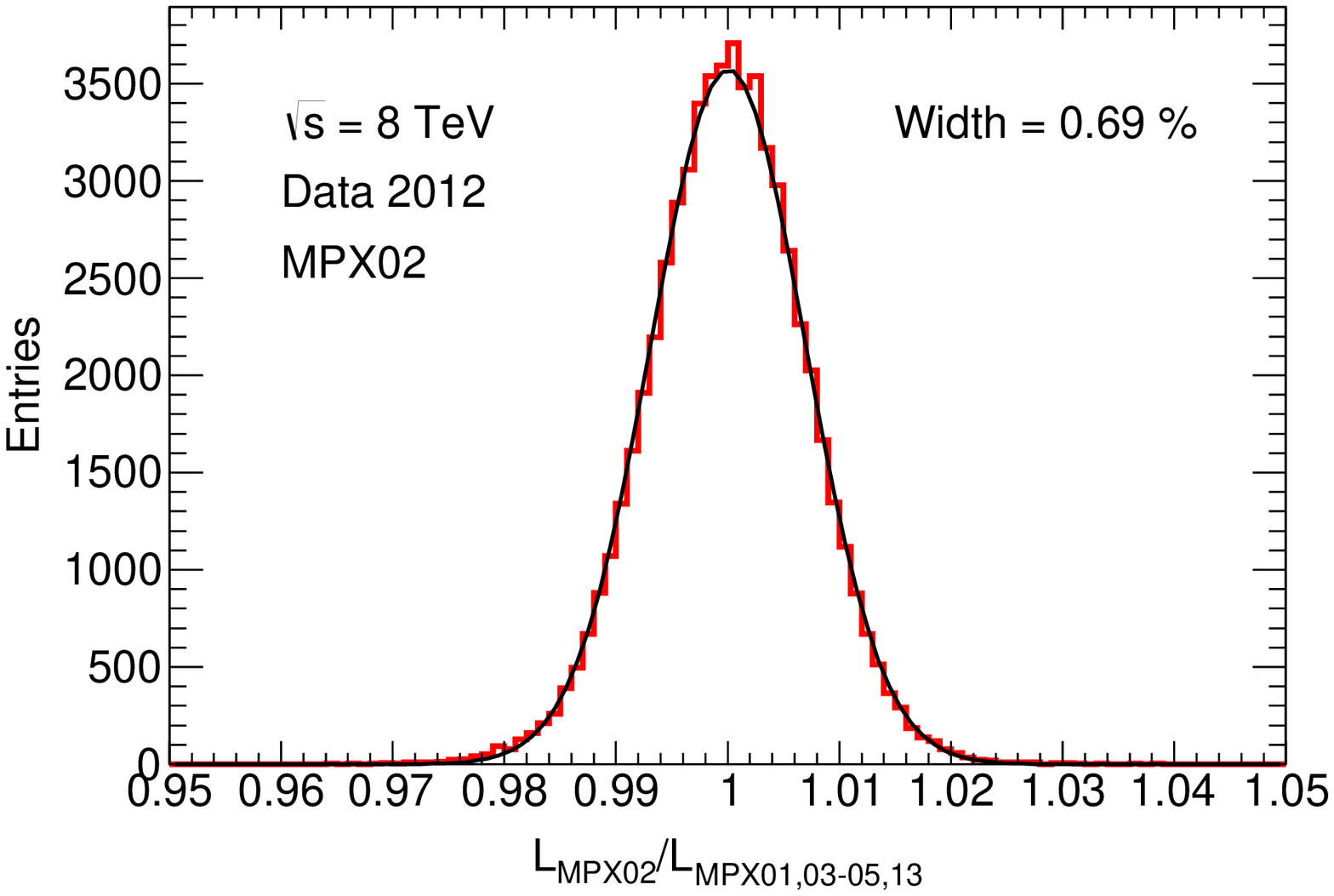}
\includegraphics[width=0.49\linewidth]{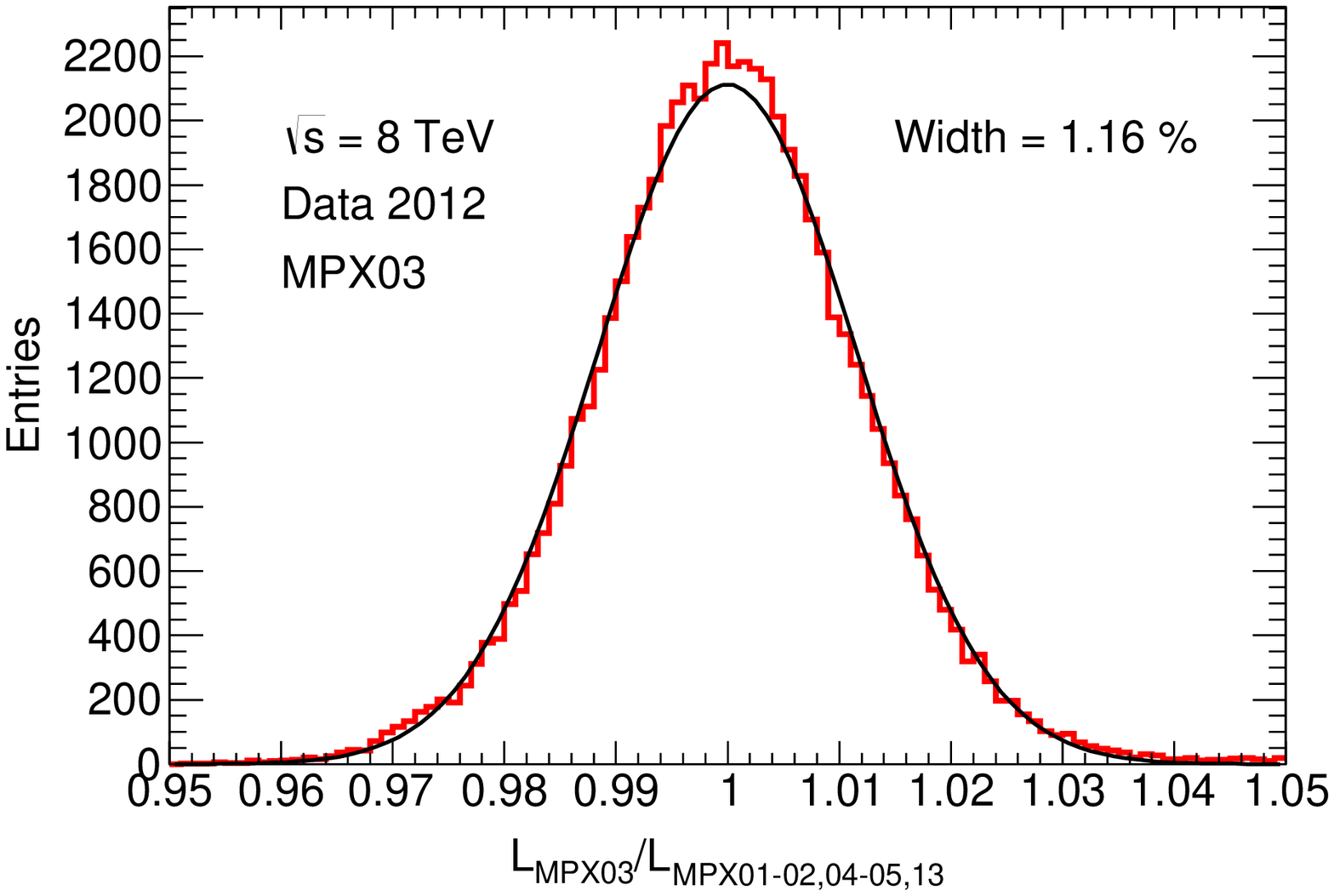}
\includegraphics[width=0.49\linewidth]{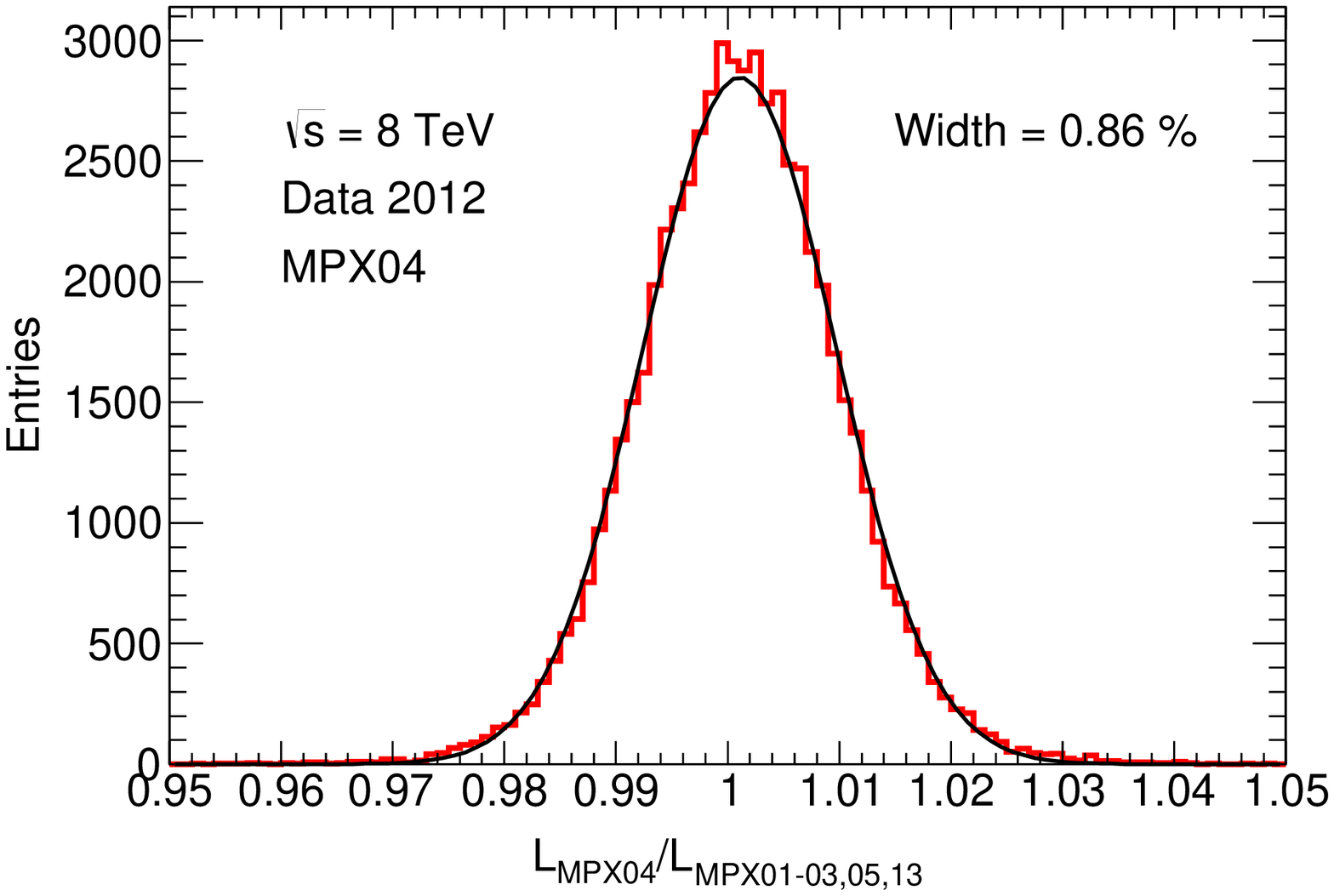}
\includegraphics[width=0.49\linewidth]{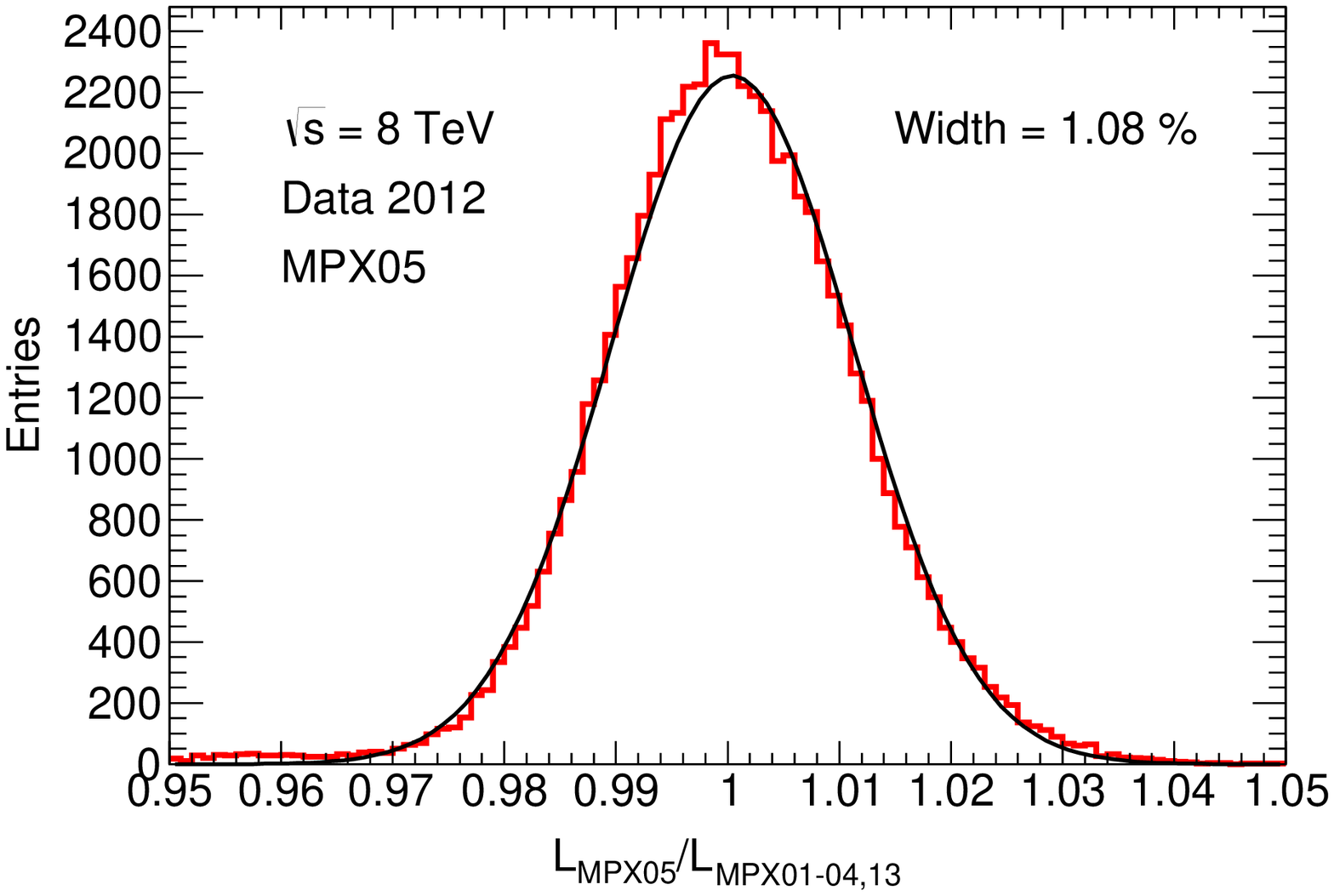}
\includegraphics[width=0.49\linewidth]{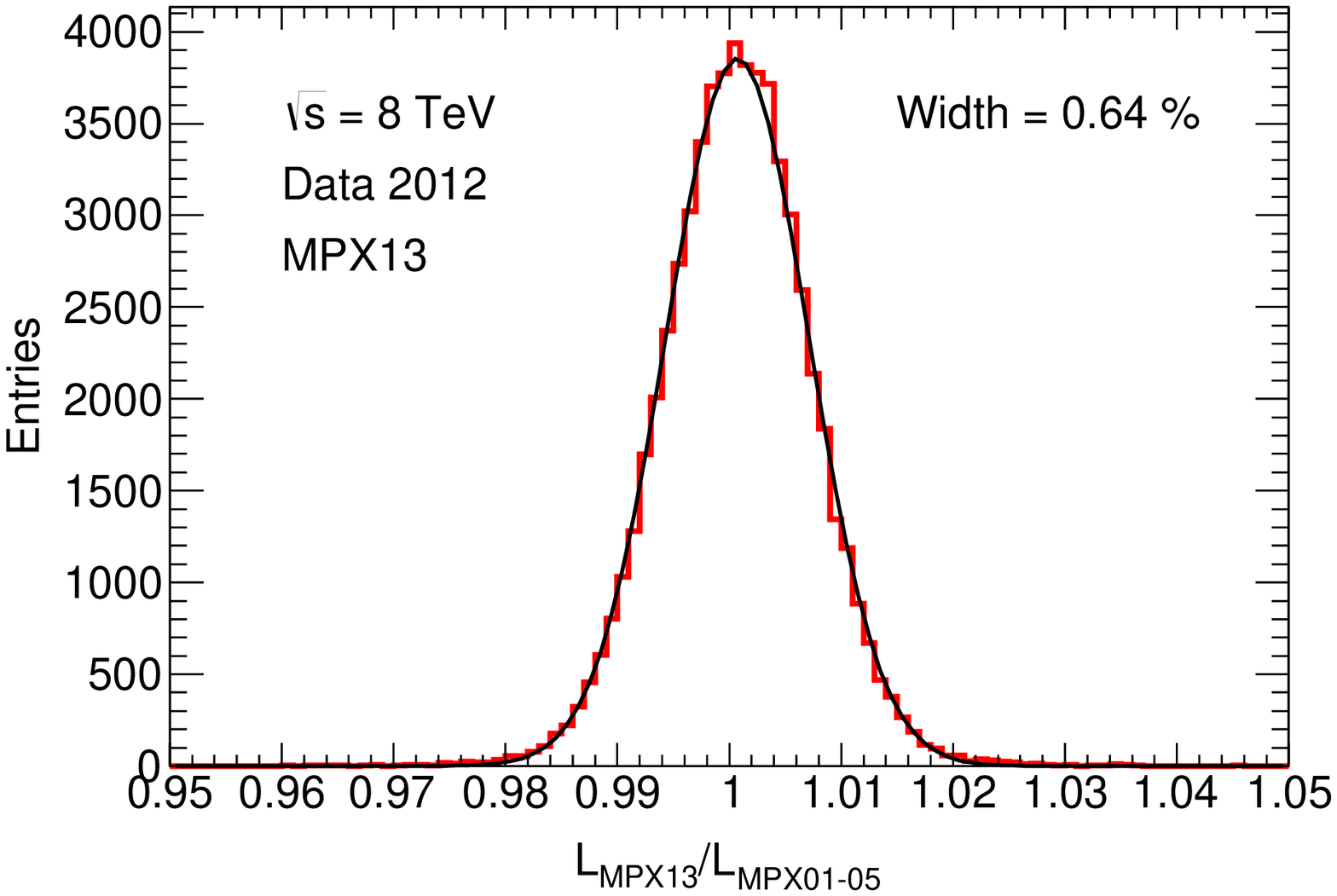}
\vspace*{-2mm}
\caption{Luminosity ratio MPX/MPXav for the six
         MPX devices with highest particle flux (MPX01-05 and MPX13). 
         The distributions are approximated by Gaussian fits.
         The width of the Gaussian is an estimate of the LB-by-LB 
         uncertainties.
         LHC fills from June to November 2012.
}
\label{fig:internal_hit}
\end{figure*}

The ratio MPX/MPXav is calculated LB-by-LB for the six MPX devices. 
Figure~\ref{fig:internal_hit} shows the luminosity ratio
per LB for the data-taking period April-November 2012 for all devices, except for MPX01 using the 
data from June to November. 
A single Gaussian fit is applied using the statistical uncertainty $\rm \sqrt{entries}$ in each bin.
Table~\ref{tab:MPX_SUM_used} summarizes the Gaussian fit values. 
The width of these fits vary between 0.6\% and 1.2\% depending on the 
MPX device.

\begin{figure*}[hbp]
\vspace{-0.7cm}
\centering
\includegraphics[width=0.49\linewidth]{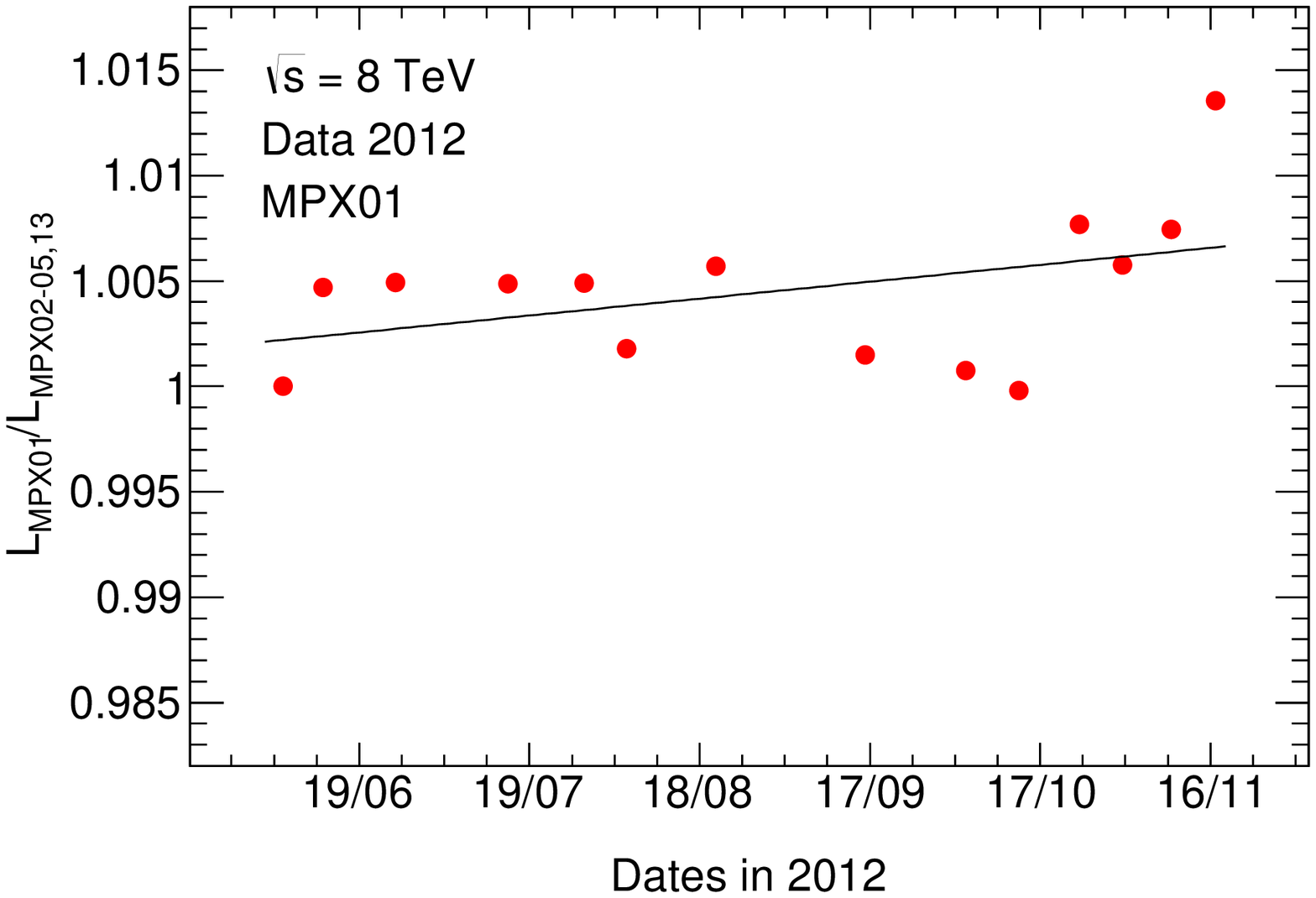}
\includegraphics[width=0.49\linewidth]{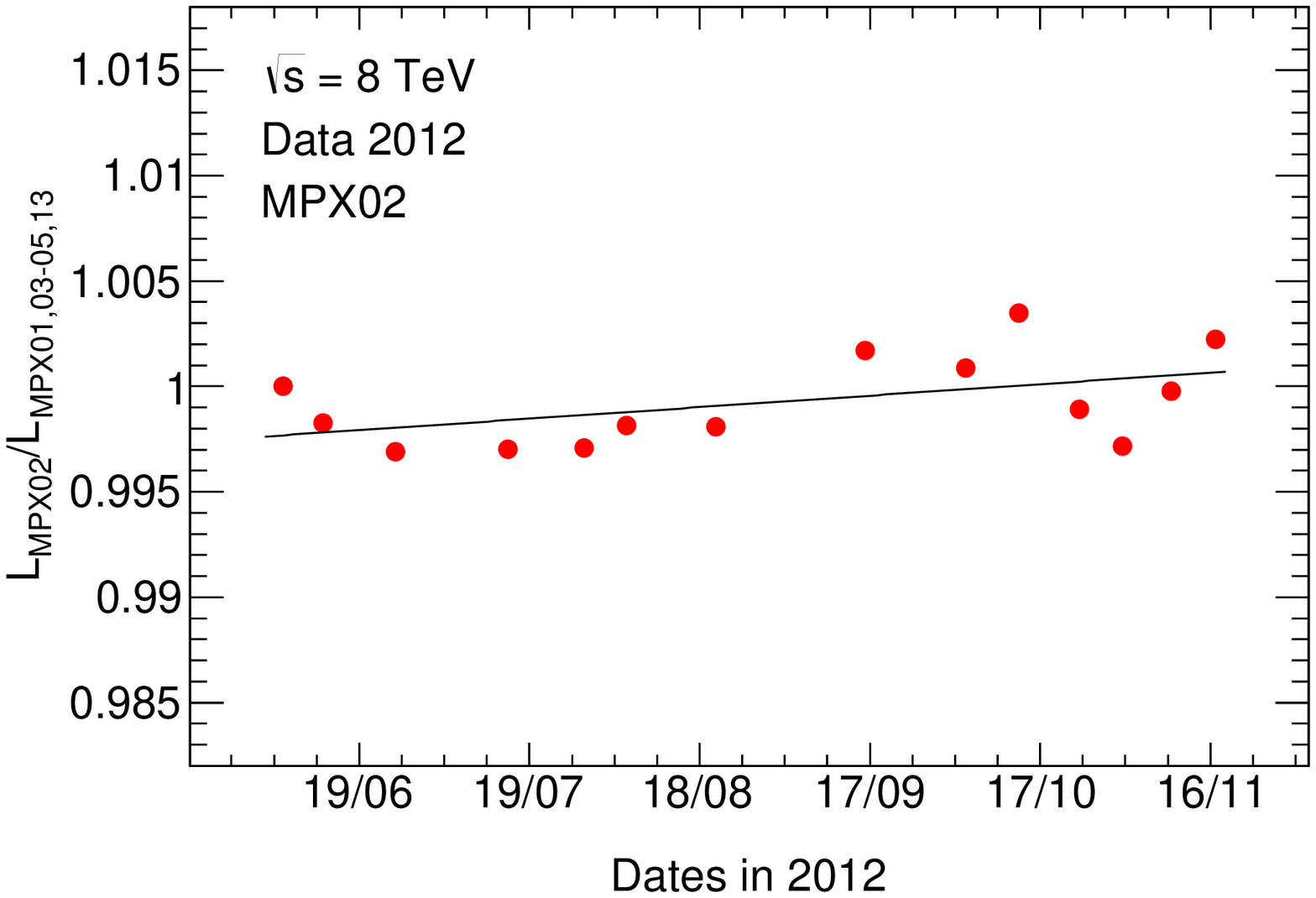}
\includegraphics[width=0.49\linewidth]{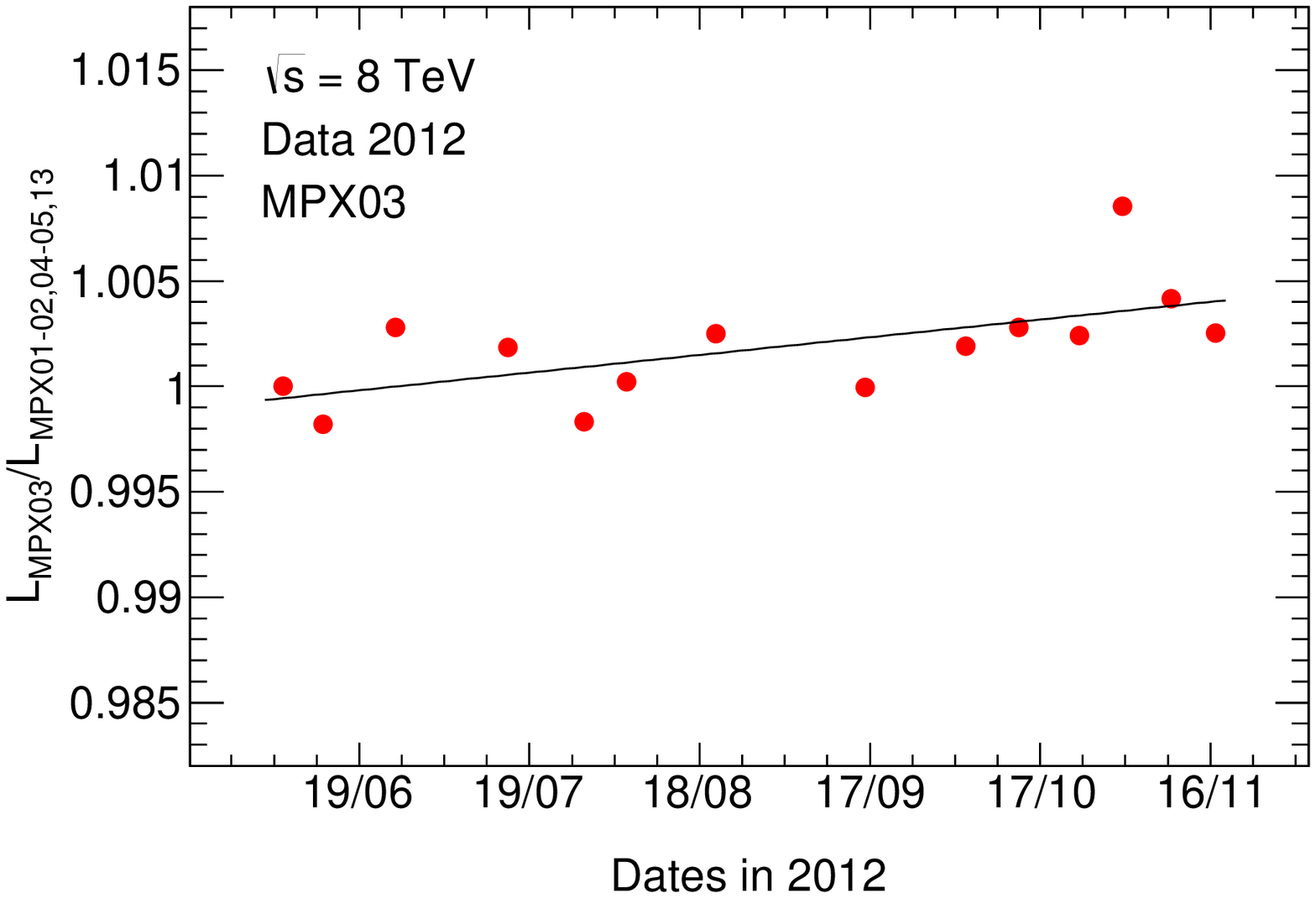}
\includegraphics[width=0.49\linewidth]{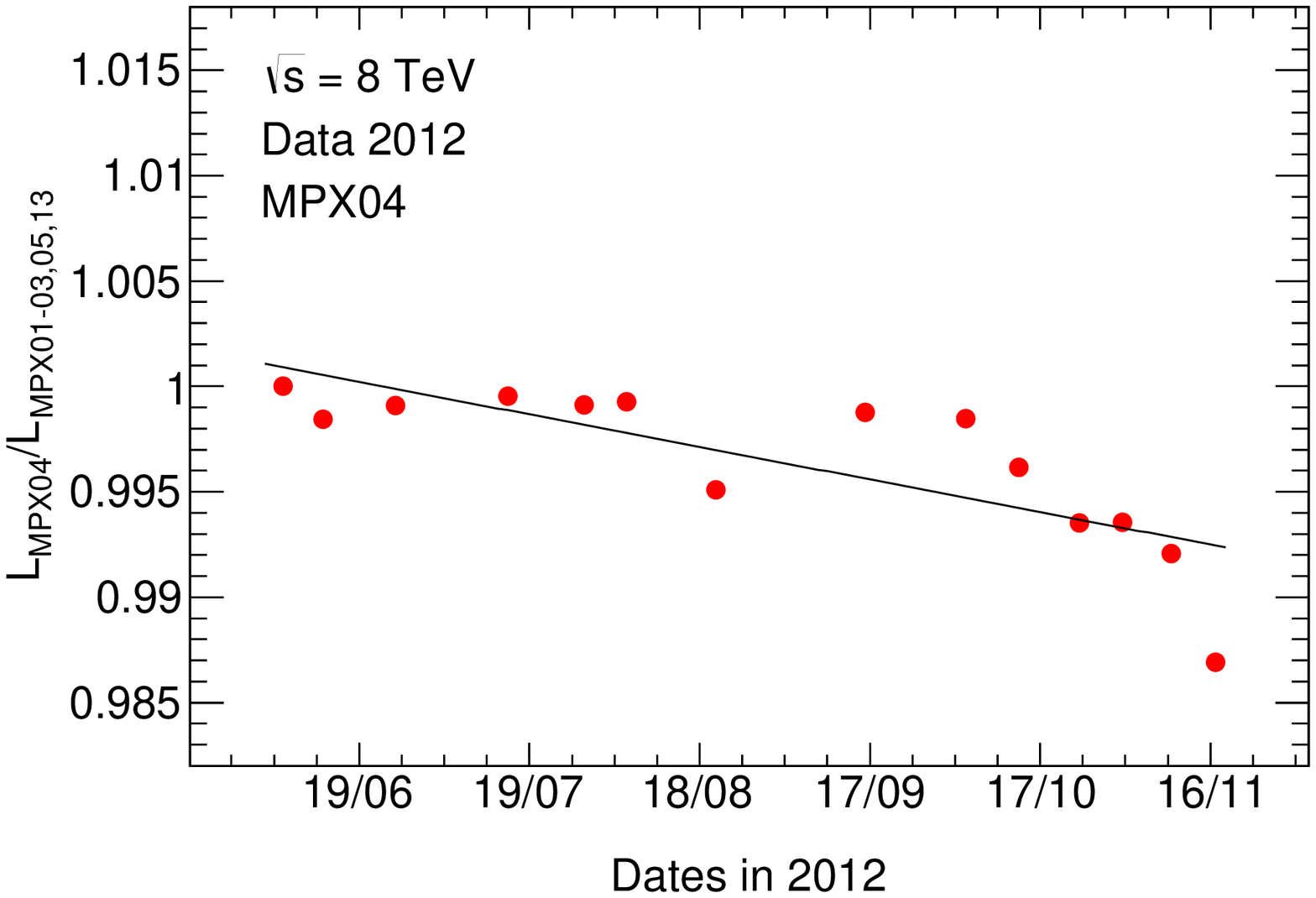}
\includegraphics[width=0.49\linewidth]{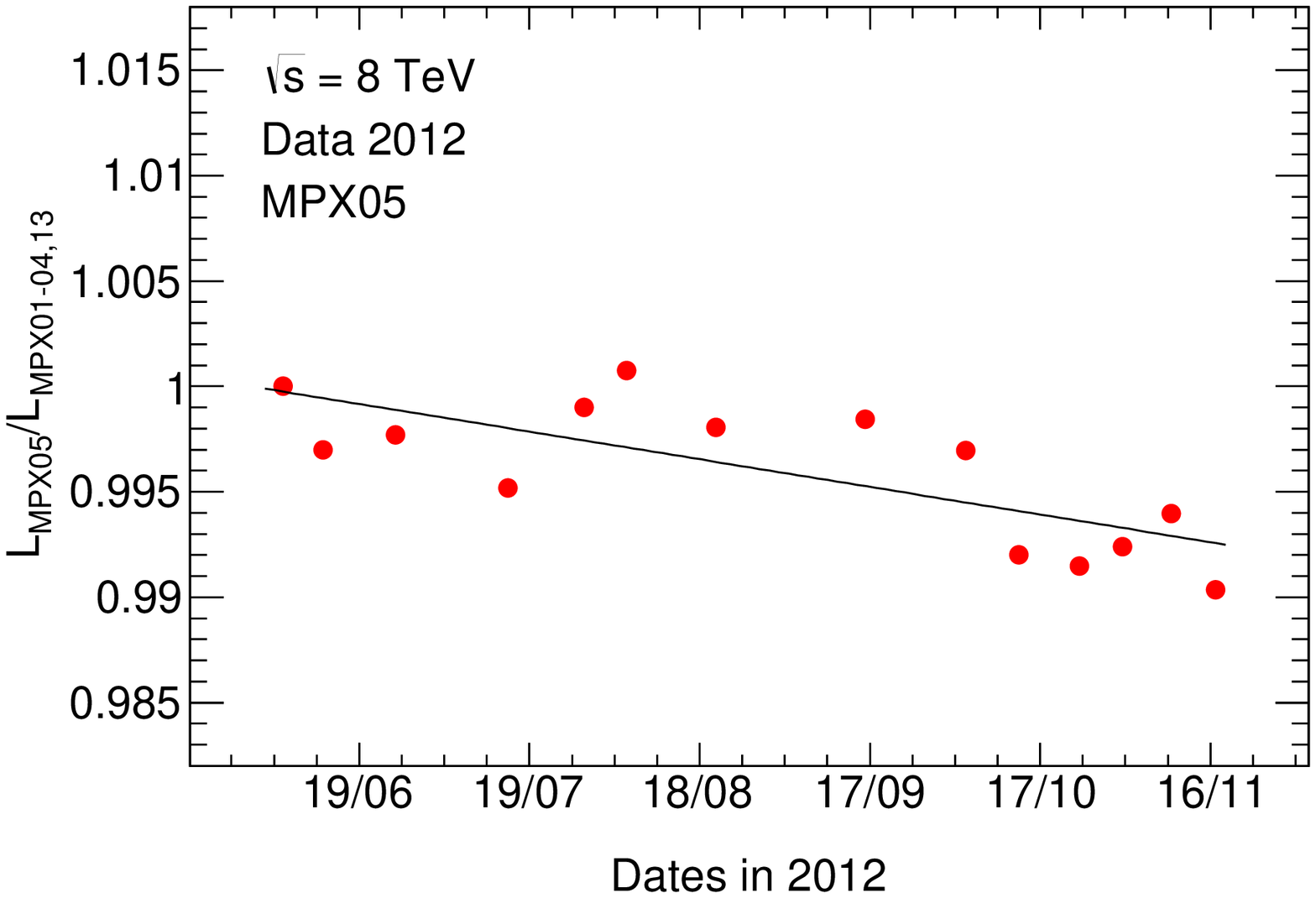}
\includegraphics[width=0.49\linewidth]{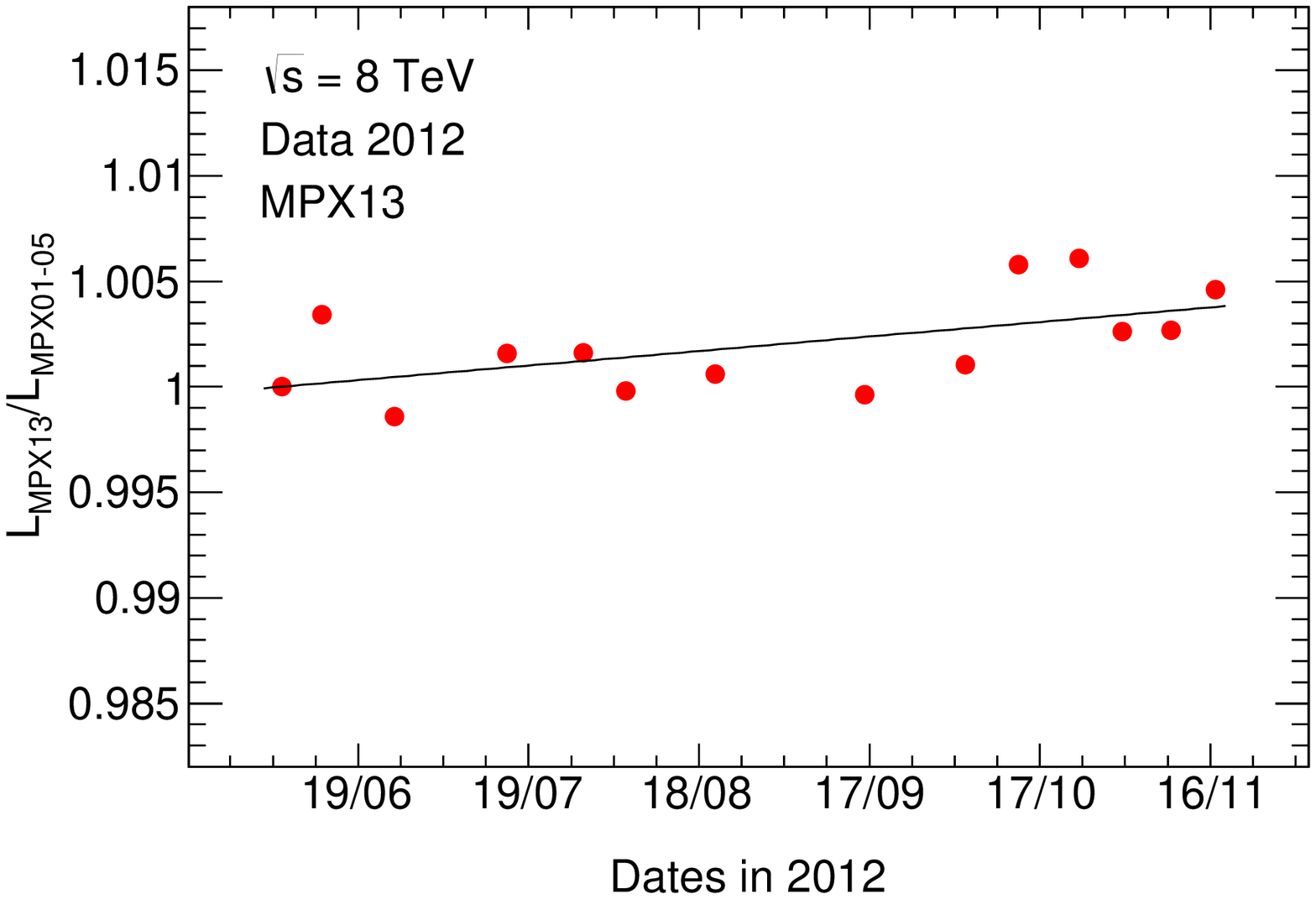}
\vspace*{-2mm}
\caption{Time history of the luminosity ratio MPX/MPXav for the six
         MPX devices with highest particle flux (MPX01-05 and MPX13)
         used for hit counting.
         A linear fit is applied to determine the slopes.
         The 2012 data
         is divided 
         into 14 time periods with an equal number of LBs.
         The data is scaled such that the value of the first bin is unity.
         The size of the statistical error bar is below the size of the data point.
         It is identical for every data point of the same device.
         The size of the error bar which would give $\chi^2/{\rm ndf}=1$ is 
         given in Table~\ref{tab:MPX_SUM_used14} for each MPX device.
         LHC fills from June to November 2012.
}
\label{fig:internal_hit14}
\end{figure*}

In addition, the long-term time stability of the six MPX devices is studied. 
For this study the LBs for which all the MPX devices were operational are
grouped into 14 time periods, such that each time period contains the same number
of LBs. 
The luminosity ratio of an individual MPX device
to the average of all other MPX devices (MPXav) is calculated for each time period 
and given in Fig.~\ref{fig:internal_hit14}.
A normalization is applied
such that the ratio is unity in time period 1 for each MPX device.   
  
A linear fit is applied to the MPX/MPXav luminosity ratio versus 
time for the June to November 2012 data-taking period. 
The slope of the linear fit is taken as a measure of time stability. 
The obtained slope values and their uncertainties are summarized in 
Table~\ref{tab:MPX_SUM_used14}. 
The variance of these slope measurements is 0.69 [\% per 200 days]$^2$.
The resulting standard deviation of 0.83 [\% per 200 days] is used as an estimation of the 
systematic uncertainty.

In summary, for the six high statistics MPX devices the width of the 
fluctuations LB-by-LB is between (0.6-1.2)\%,
and the time-stability from June to November 2012 is better than 1\%.
This gives us an indication of the time stability of MPX luminosity monitoring.

\section{MPX Luminosity from Heavy Blob (Thermal Neutron) Counting}
\label{sec:lumi_th_neutron_counting}

Thermal neutrons are detected by MPX devices via $\mathrm{{}^{6}Li(n,\alpha){}^{3}H}$ reactions 
in a $\mathrm{{}^{6}LiF}$ converter layer with thickness of 2-3~mg/cm$^2$ 
on average~\cite[Sec. 2.3]{analysisRadiaField:2013}. 
In MPX tracking mode, tritons and alpha particles are registered by Si-sensors as so-called heavy 
blobs, HB (large round-shaped pixel clusters). 
The typical detection efficiency for thermal 
neutrons is 1\%, determined from individual calibrations of the 
MPX devices in a thermal neutron field~\cite[Sec. 2.3]{analysisRadiaField:2013}.
Hence, the HB count rate is used as a measure of instantaneous 
luminosity since neutrons are generated in the LHC collisions. 

The MPX06 to MPX12 devices are used for the HB (thermal neutron) counting
since the pixel matrix occupancy for these devices is sufficiently small for pattern recognition.
A dedicated study was performed to determine the misidentification of heavy blobs which are 
lost due to the overlap with other clusters~\cite[Sec. 2.2]{analysisRadiaField:2013}. 
The resulting correction factors, specific to each MPX device, depend on 
the number of clusters per frame (i.e., on the LHC collision rate, on the device location and on the acquisition time).
The precision of these correction factors was estimated to be below 1\%
with the exception of MPX06 with the largest pixel occupancy.
Therefore, MPX06 was not used for the precision study and combination with the other devices.

An analytic model for the cluster overlap probability has been developed and agrees
with the experimental results for simple cluster shapes.

The distribution of heavy blobs per frame recorded within the MPX12 region covered  
by a $\mathrm{{}^{6}LiF}$ converter is well described by a Poisson 
distribution, demonstrated in~\cite{sopczak:2013}.

\begin{table*}[bp]
\vspace*{4mm}
\caption{\label{tab:MPX_SUM_used}
Normalization factor $1/n_{\rm f}$, 
width of Gaussian fits of luminosity ratio MPX/MPXav
for the six MPX devices (MPX01-05 and MPX13), shown in Fig.~\ref{fig:internal_hit}.
The $\chi^2/{\rm ndf}$ values indicate that in addition to the statistical uncertainties, 
some systematic uncertainties are present. 
For a realistic width uncertainty determination of an individual device, the errors are 
scaled to obtain $\chi^2/{\rm ndf}=1$. 
}
\begin{center}
\renewcommand{\arraystretch}{1.3}
\begin{tabular}{cccccc}
\hline\hline
MPX  & $1/n_{\rm f}$ ($\rm hit / \micro b^{-1}$) 
&$R_{\rm width}$ (\%)& $\sigma R_{\rm width}$ (\%)& $\chi^2/{\rm ndf}$  & $\sigma R_{\rm width'}$\\ \hline
01   &639.88 & 0.7221      & 0.0024              & 596/97                     & 0.0058              \\
02   &4.0393 & 0.6912      & 0.0021              & 238/97                     & 0.0034              \\
03   &0.5079 & 1.1599      & 0.0038              & 398/97                     & 0.0076              \\
04   &1.3106 & 0.8638      & 0.0028              & 461/97                     & 0.0061              \\
05   &0.5534 & 1.0839      & 0.0035              & 662/97                     & 0.0091              \\
13   &5.8902 & 0.6394      & 0.0020              & 360/97                     & 0.0038              \\
\hline\hline
\end{tabular}
\end{center}
\vspace*{-1cm}
\end{table*}

\begin{table*}[hbp]
\caption{\label{tab:MPX_SUM_used14}
         Slope of time history of the luminosity ratio MPX/MPXav for the six
         MPX devices with highest particle flux (MPX01-05 and MPX13)
         used for hit counting.
The slope values and the uncertainties are given per second, and in percent for 200 days. 
For a realistic slope uncertainty determination of an individual device, the errors are 
scaled to obtain $\chi^2/{\rm ndf}=1$. 
$\sigma R$ is the size of the error bars which are the same for each
data point (Fig.~\ref{fig:internal_hit14}).}
\begin{center}
\renewcommand{\arraystretch}{1.3}
\begin{tabular}{cccccc}
\hline\hline
MPX  & Slope       & $\sigma$Slope    & Slope     & $\sigma$Slope & $\sigma R$\\
     & ($10^{-10}\,{\rm s}^{-1}$) & ($10^{-10}\,{\rm s}^{-1}$) & (\%/200d) & (\%/200d)  &  \\ \hline
01   &   3.09    &1.94  &   0.534   &  0.336   & 0.00349 \\
02   &   2.10    &1.08  &   0.364   &  0.187   & 0.00194 \\
03   &   3.24    &1.19  &   0.559   &  0.205   & 0.00214 \\
04   & $-5.95$   &1.36  & $-1.029$  &  0.235   & 0.00245 \\
05   & $-5.06$   &1.30  & $-0.874$  &  0.225   & 0.00234 \\
13   &   2.66    &1.10  &   0.459   &  0.191   & 0.00198 \\
\hline\hline
\end{tabular}
\end{center}
\vspace{-1cm}
\end{table*}


For luminosity determination, {\it overlap corrected number of heavy blobs} (cHB) 
is used for each MPX device.
The number of cHB per frame is converted into LBs, similar to the hit analysis. 
Frames which lie within the time window of the LB are selected.
The numbers of cHB of these frames are averaged for all operational MPX devices.
Only those LBs for which all MPX devices (MPX07-12) 
were operational are used.    

These LBs are grouped into 14 time periods, such that each time 
period has an equal number of LBs, in the same way as it was done for the hit analysis. 
For each time period the number of cHB are summed for each MPX device.
The summed cHB are converted into luminosity 
by using a normalization factor such that the luminosity ratio MPX/MPXav 
of time period 1 is unity.
In order to calculate the luminosity ratio MPX/MPXav for each time period, the weighted 
luminosity average of all the devices (excluding the device under consideration) 
used in MPXav is determined. 
For each device and each time period the statistical uncertainty is 
$1/\sqrt{N_{\rm HB}}$, where $N_{\rm HB}$ is the summed number of heavy blobs.

\begin{figure*}[hbp]
\vspace*{-0.7cm}
\centering
\includegraphics[width=0.49\linewidth]{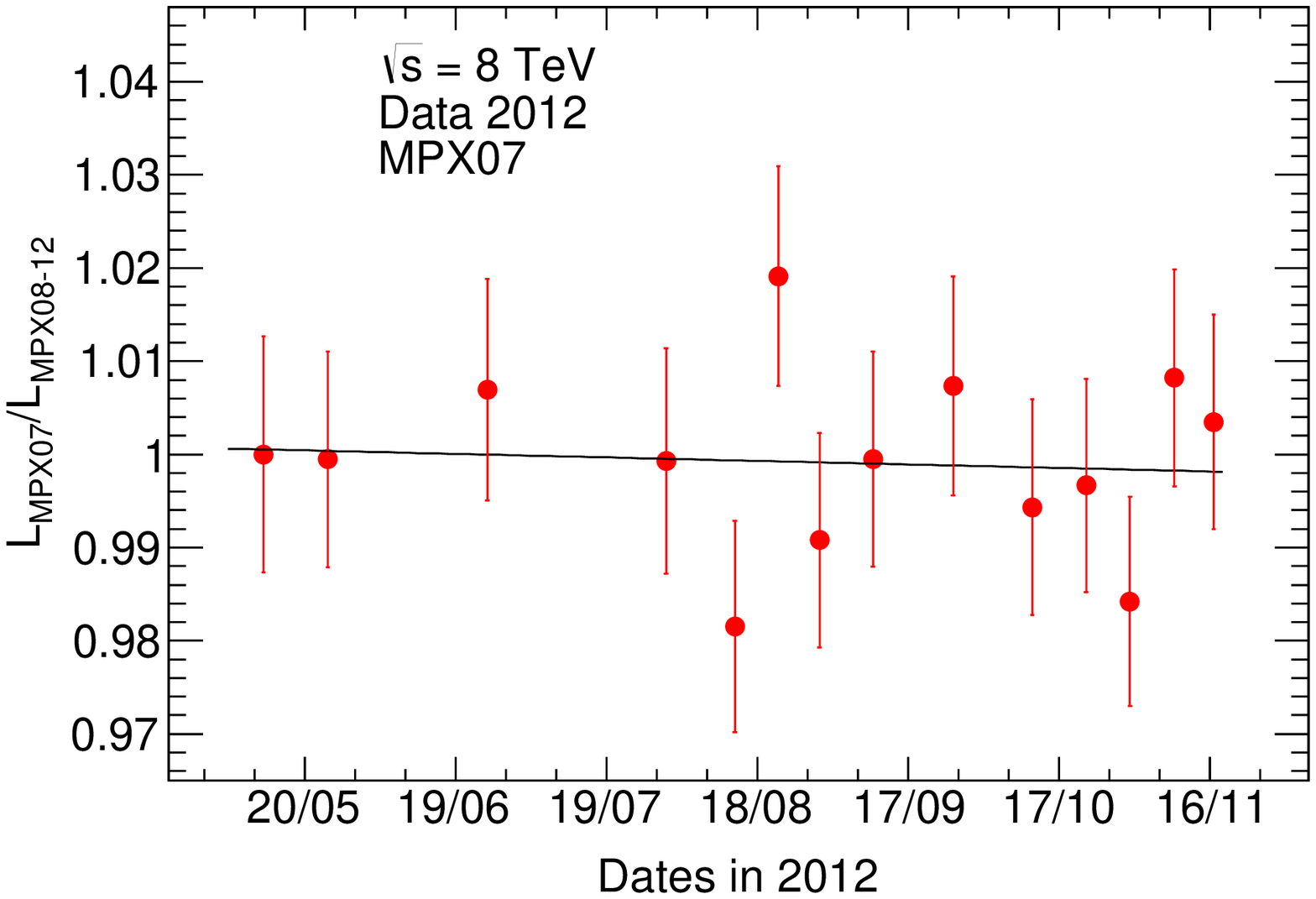}
\includegraphics[width=0.49\linewidth]{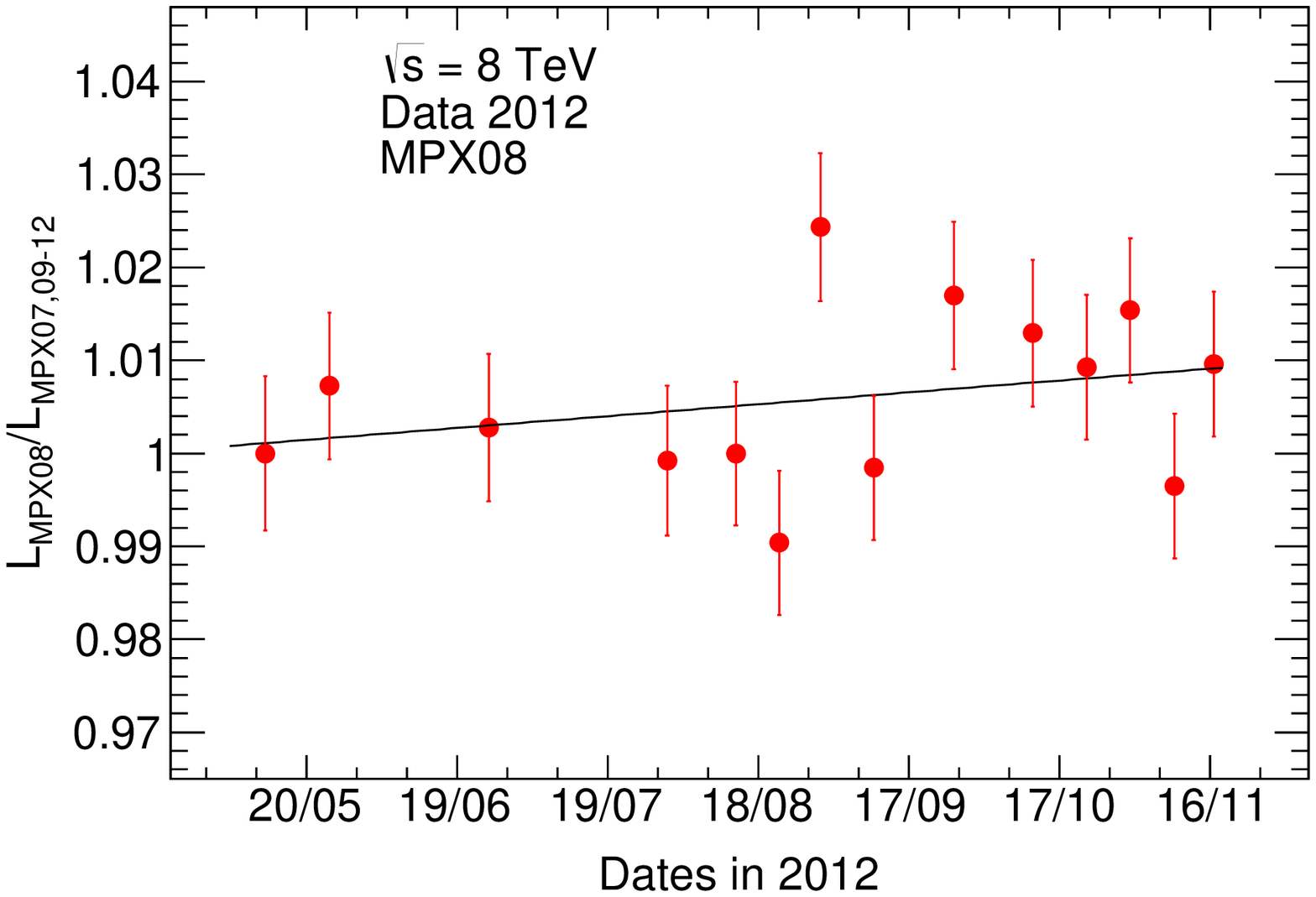}
\includegraphics[width=0.49\linewidth]{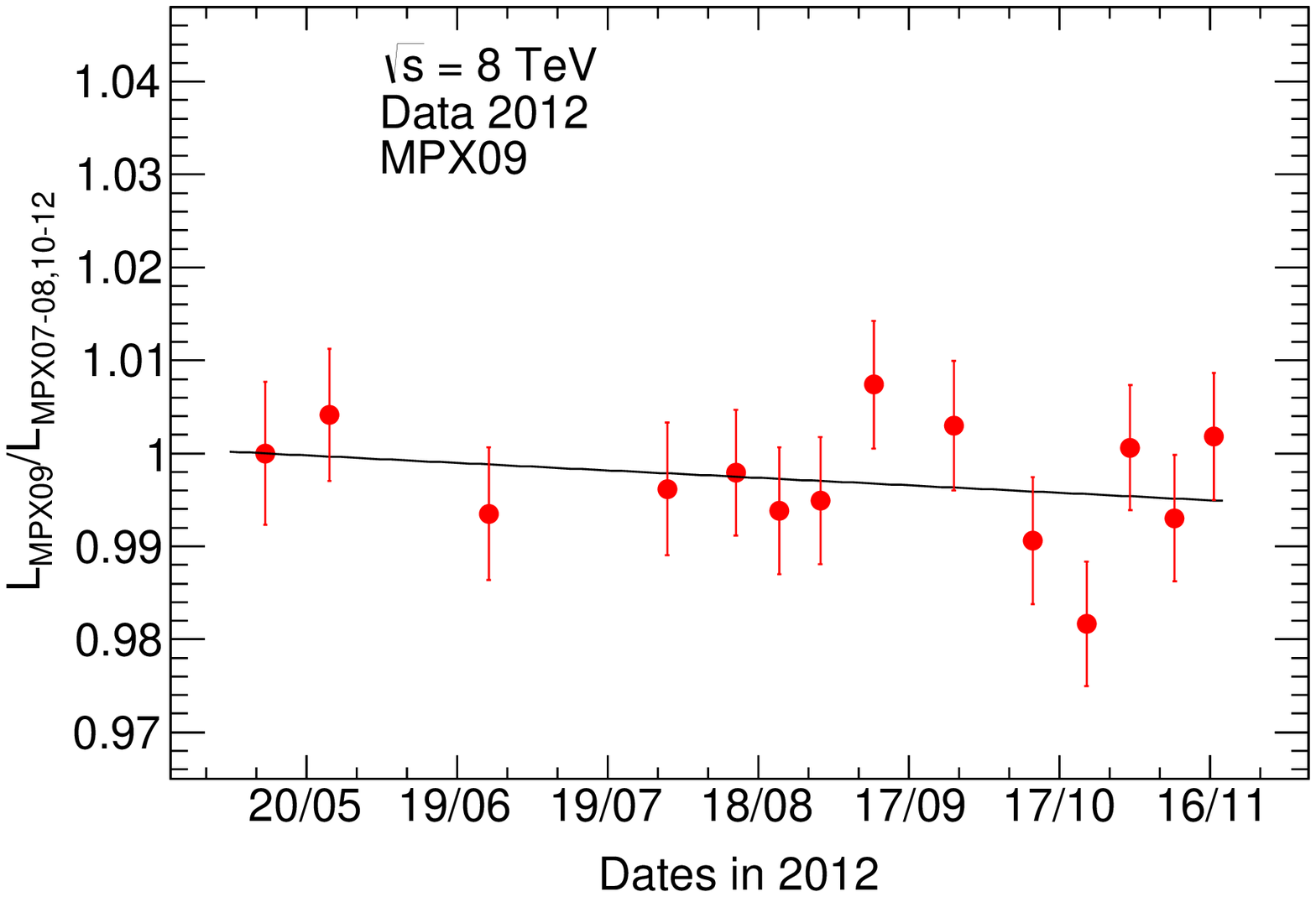}
\includegraphics[width=0.49\linewidth]{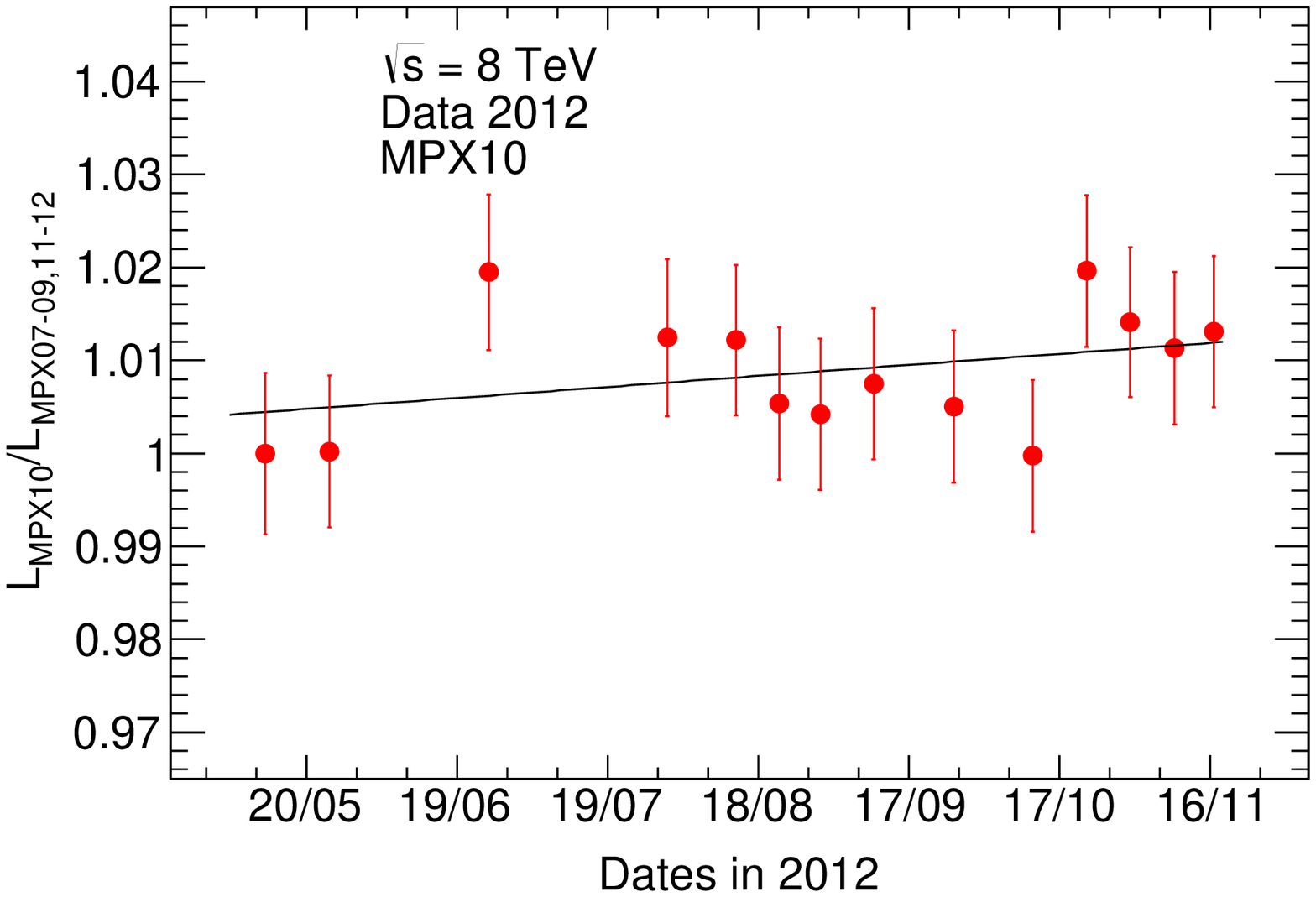}
\includegraphics[width=0.49\linewidth]{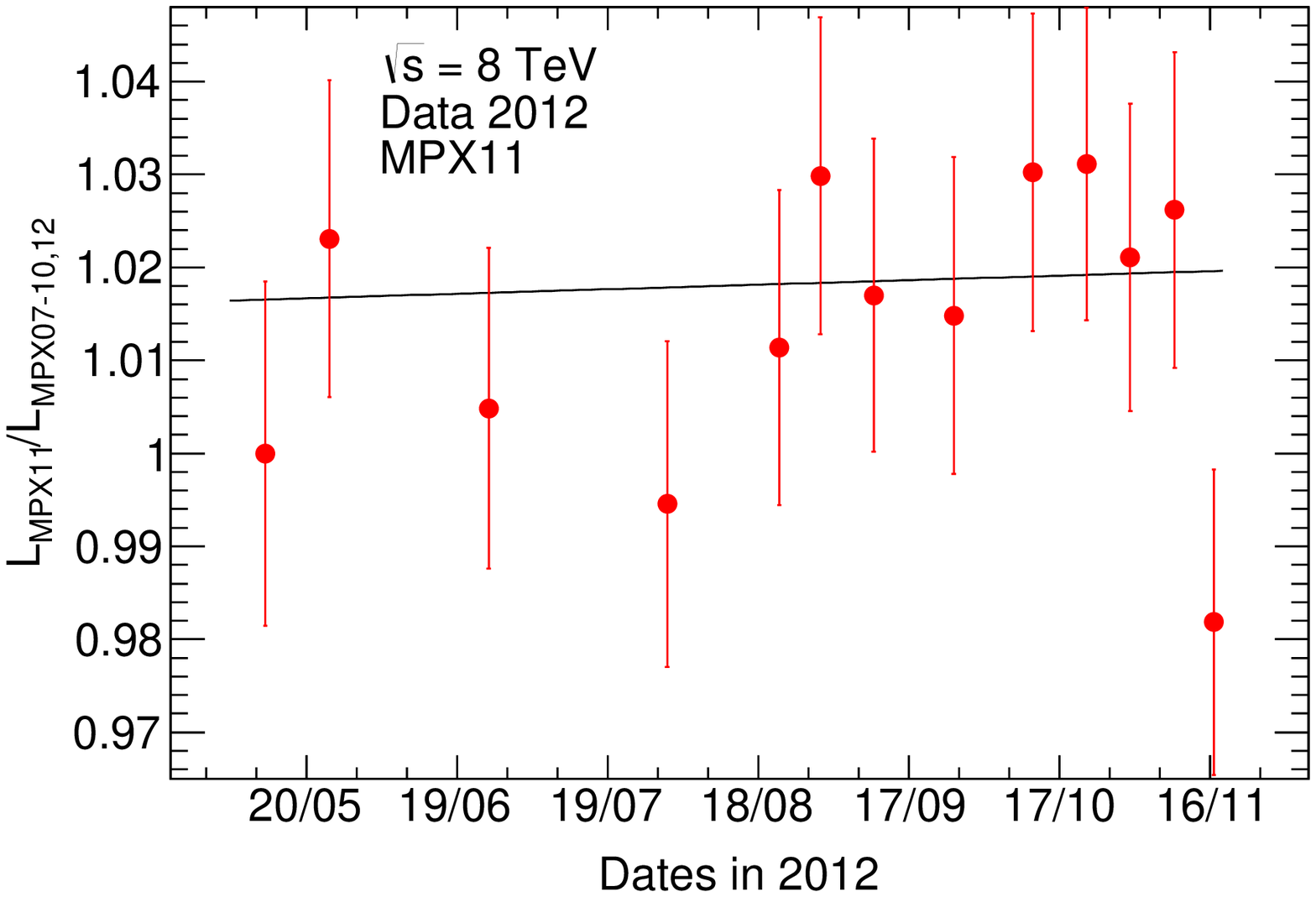}
\includegraphics[width=0.49\linewidth]{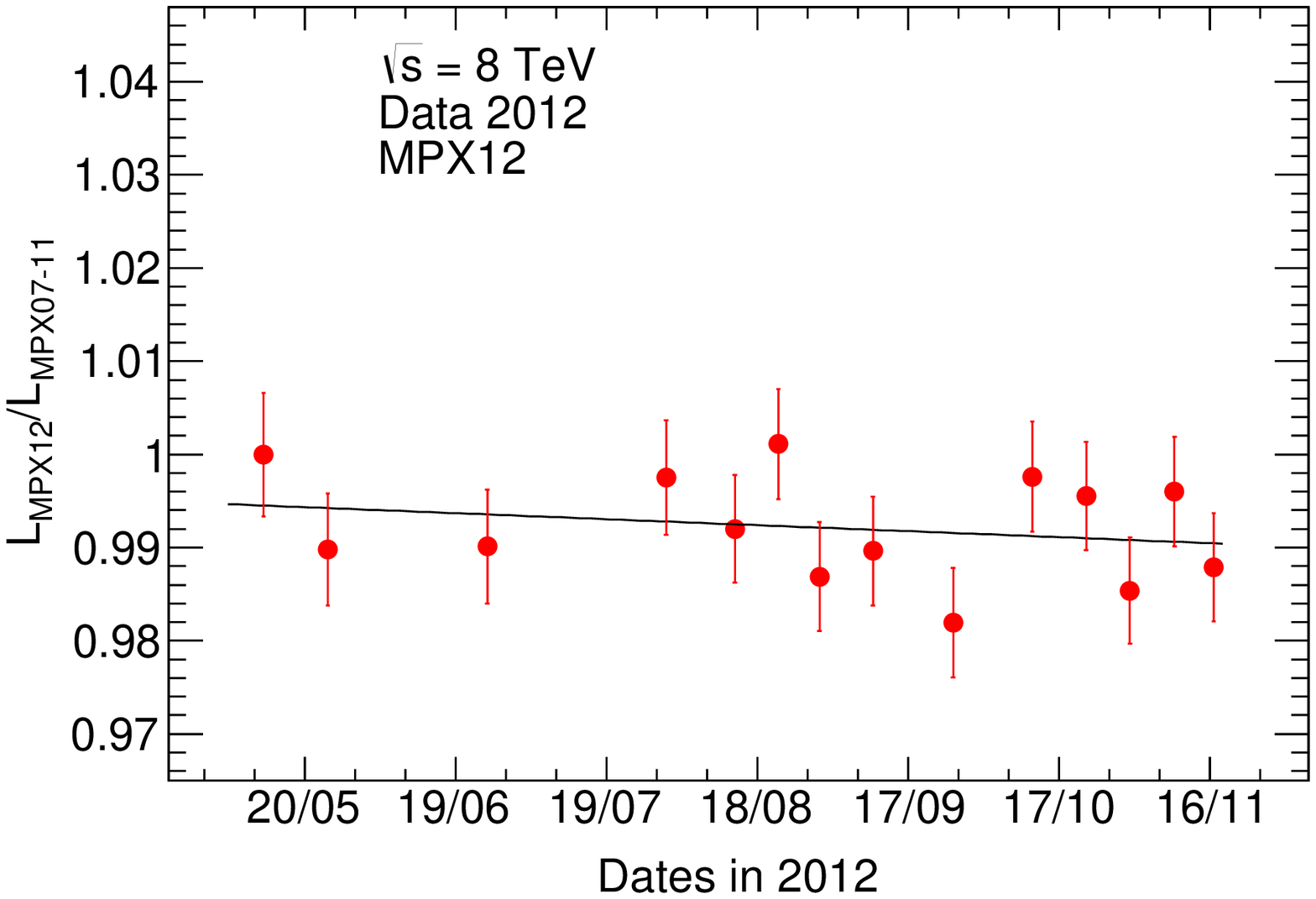}
\vspace*{-2mm}
\caption{Time history of the luminosity ratio MPX/MPXav for the six
         MPX devices with lowest particle flux (MPX07-12) 
         used for heavy blob (thermal neutron) counting.
         A linear fit is applied to determine the slopes.
         The 2012 data
         is divided 
         into 14 time periods with an equal number of LBs.
         The data is scaled such that the value of the first bin is unity.
         The statistical error bars are shown.
         LHC fills from May to November 2012.
}
\label{fig:internal_hb}
\end{figure*}

A linear fit is applied to determine the precision of the time history of 
the ratio between individual MPX devices with respect to the weighted average of all other devices,
shown in  Fig.~\ref{fig:internal_hb}.
Table~\ref{tab:MPX_SUM_neutrons} summarizes the slope values and the uncertainties
of the linear fits.
The variance of these slope measurements is 0.38 [\% per 200 days]$^2$.
The resulting standard deviation of 0.62 [\% per 200 days] is used as an estimation of the 
systematic uncertainty.
The $\rm \chi^2/ndf$ is close to unity, thus statistical uncertainties describe 
the fluctuations.

In the analysis of HB counting in the $^6$LiF-covered detector region
the statistical uncertainties are dominant since the HB count rate is rather 
small (a few HB per frame).

In summary, the long-term time-stability from June to November 2015 of the HB analysis 
is better than 1\% comparing single MPX devices with the weighted average of all other 
MPX devices.

\begin{table*}[hp]
\caption{\label{tab:MPX_SUM_neutrons}
Slope of time history of the luminosity ratio MPX/MPXav for the six
         MPX devices with lowest particle flux (MPX07-12)
         used for heavy blob (thermal neutron) counting.
The $\sigma$slope value results from the statistical uncertainty $1/\sqrt{N_{\rm HB}}$ per time period.
The $\chi^2/{\rm ndf}$ values of the linear fits from Fig.~\ref{fig:internal_hb}
are also shown.
.
}
\begin{center}
\renewcommand{\arraystretch}{1.3}
\begin{tabular}{cccccc}
\hline\hline
MPX  & Slope           & $\sigma$Slope    & Slope & $\sigma$Slope  & $\chi^2/{\rm ndf}$ \\
     & ($10^{-10}\,{\rm s}^{-1}$) & ($10^{-10}\,{\rm s}^{-1}$) & (\%/200d)& (\%/200d)  & \\ \hline
07   & -1.45           & 6.27             & -0.251   &  1.078      & 9.42 / 12\\
08   &  4.93           & 4.22             &  0.851   &  0.726      & 17.0 / 12\\
09   & -3.10           & 3.77             & -0.536   &  0.648      & 11.3 / 12\\
10   &  4.57           & 4.40             &  0.789   &  0.757      & 7.57 / 12\\
11   &  1.87           & 9.15             &  0.324   &  1.573      & 19.5 / 12\\
12   & -2.48           & 3.22             & -0.429   &  0.553      & 11.7 / 12\\
\hline\hline
\end{tabular}
\vspace*{-1cm}
\end{center}
\end{table*}


\section{Relation between Hits and Clusters}
\label{sec:statistics}
The relation between the number of hits and clusters is investigated
in order to determine the statistical uncertainty in luminosity from hit counting.
The definition of six cluster types is based on different shapes
observed. They are dots, small blobs, curly tracks, heavy blobs, heavy tracks and 
straight tracks~\cite[Sec. 2.2]{analysisRadiaField:2013}.

\begin{figure*}[htp]
 \vspace*{-1cm}
\centering
\includegraphics[width=0.49\linewidth]{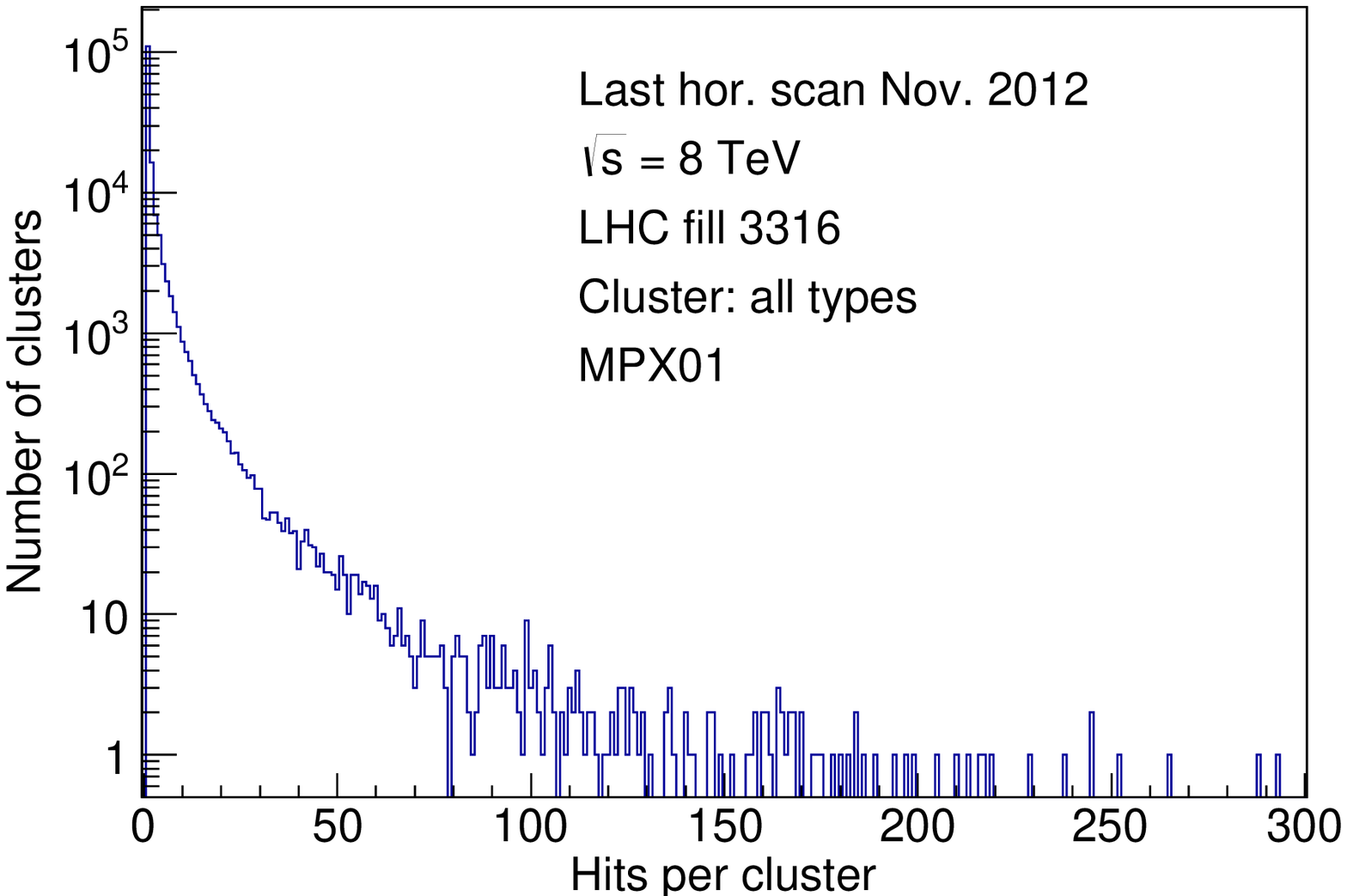}
\includegraphics[width=0.49\linewidth]{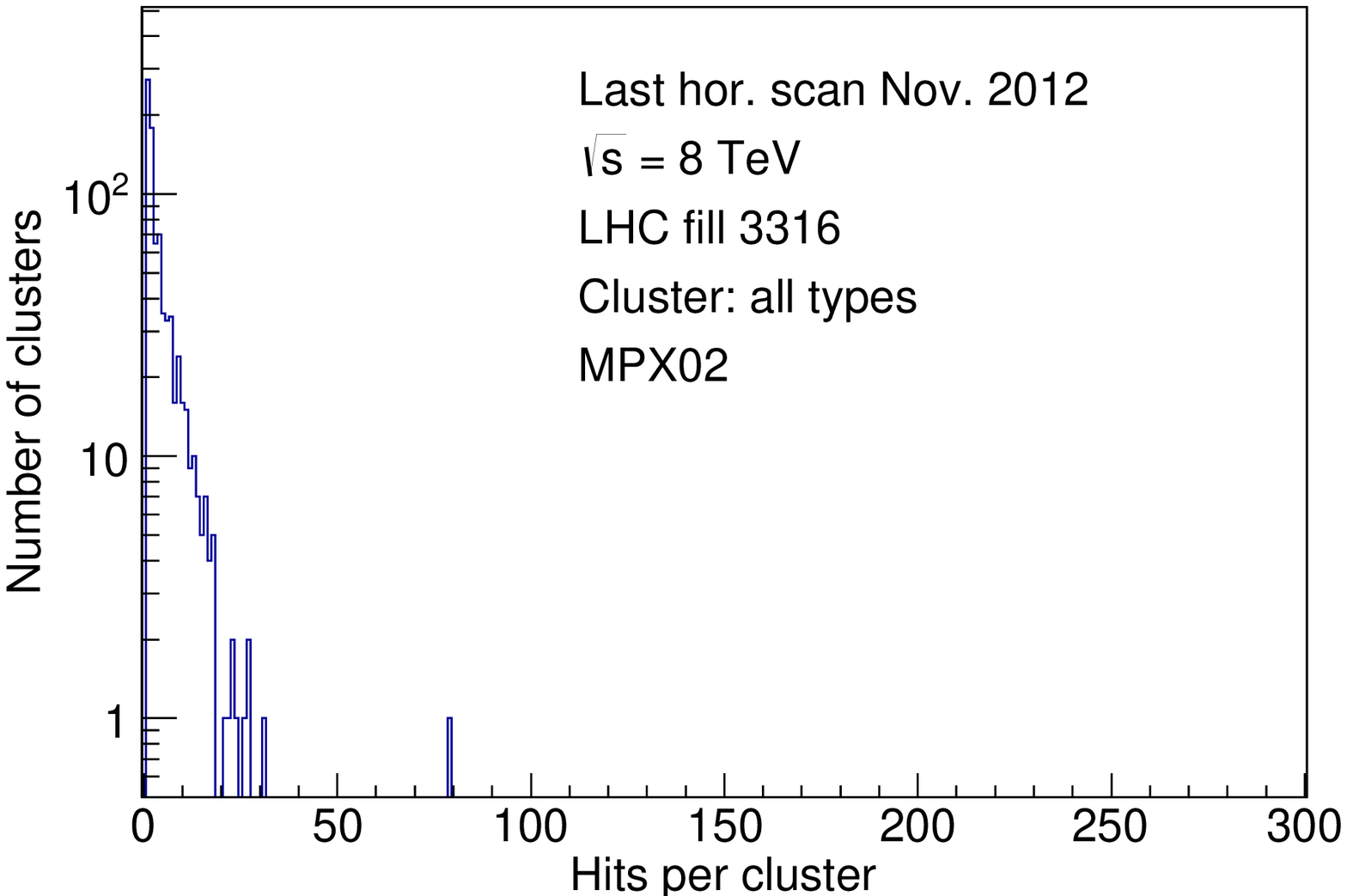}
\includegraphics[width=0.49\linewidth]{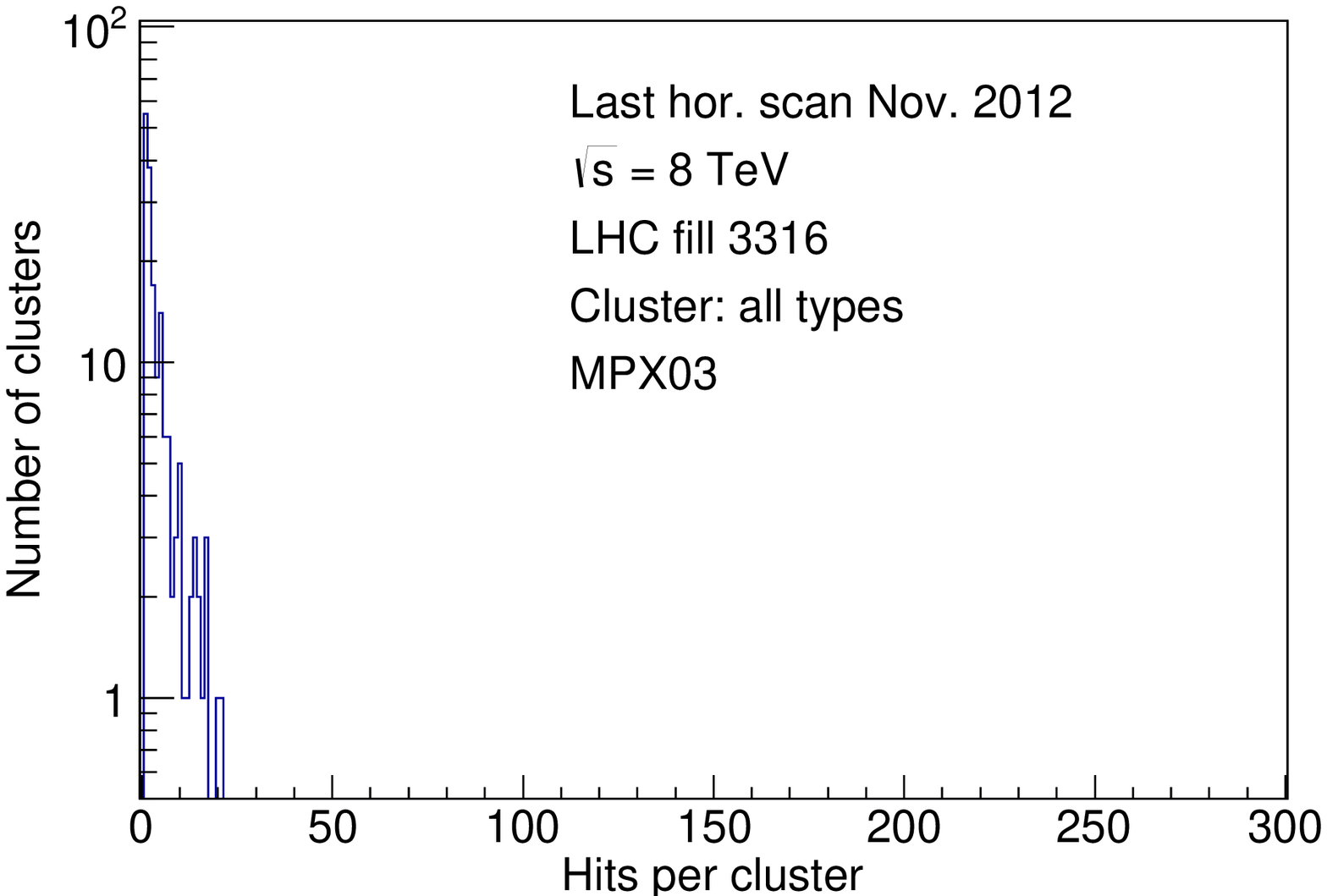}
\includegraphics[width=0.49\linewidth]{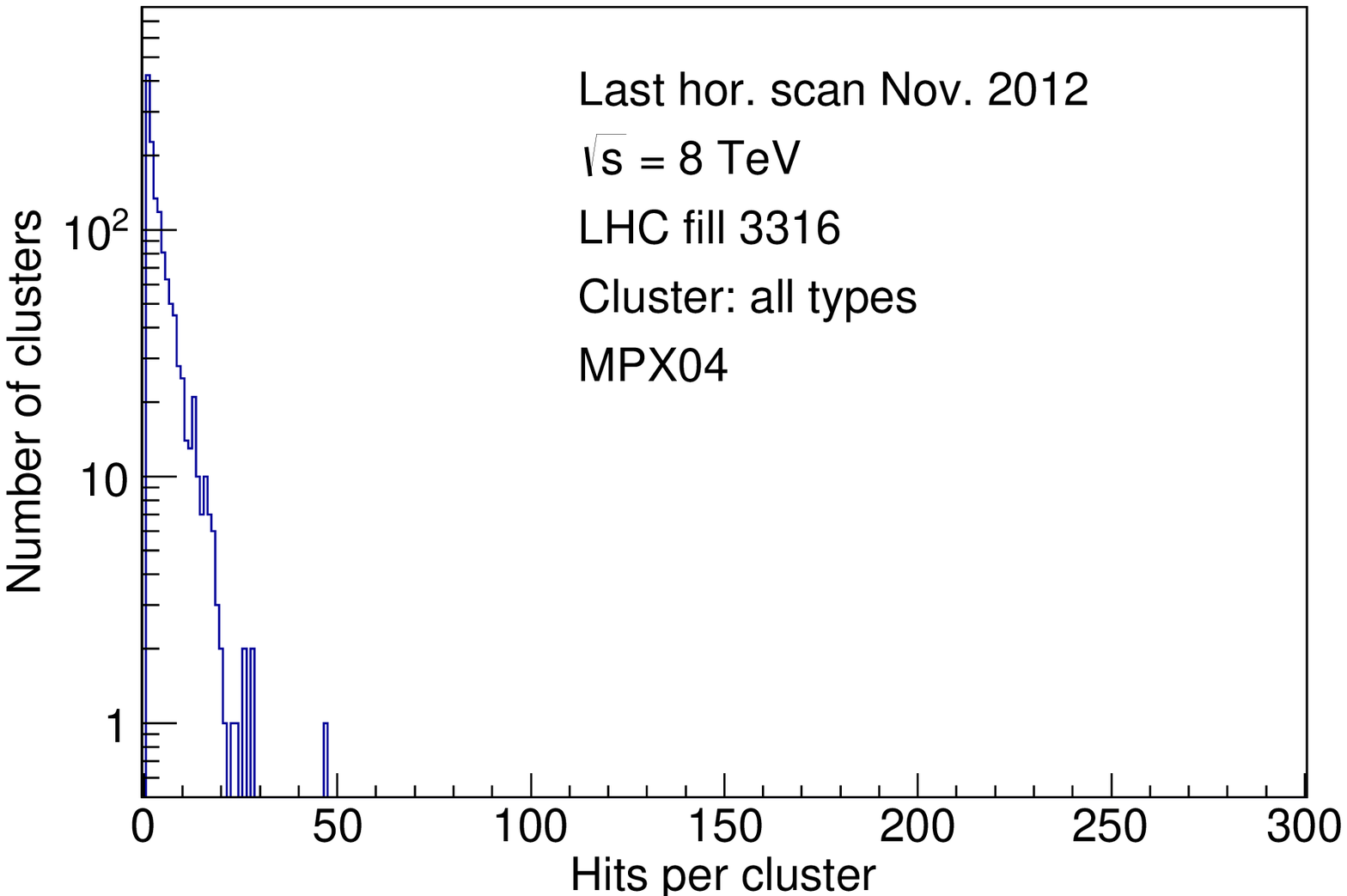}
\includegraphics[width=0.49\linewidth]{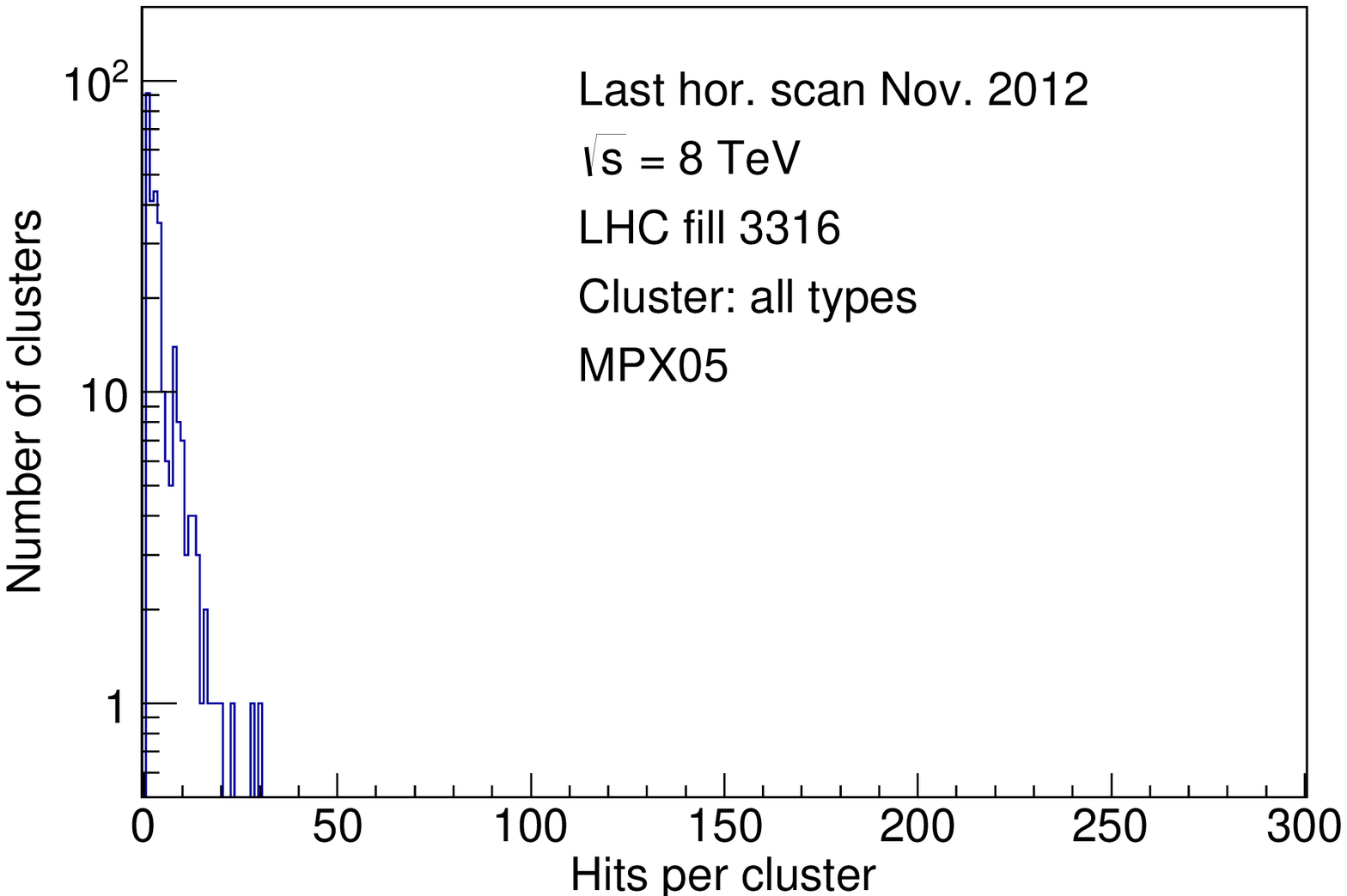}
\includegraphics[width=0.49\linewidth]{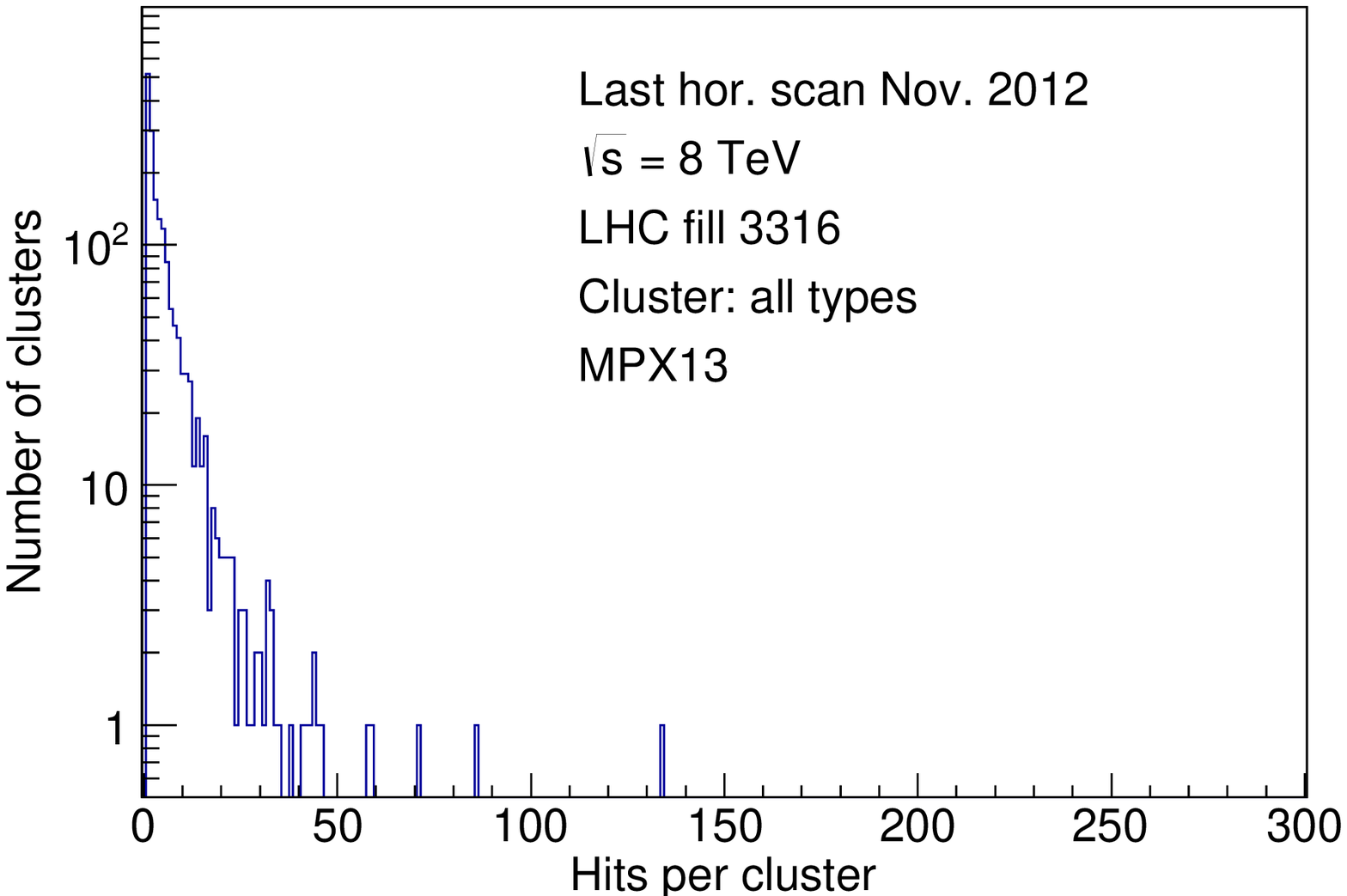}

\caption{Number of hits per cluster combined for six cluster types. 
Data is taken by MPX01-05 and MPX13 during the last horizontal van der Meer scan 
of November 2012 (LHC fill 3316).}
\label{fig:fill3316}
\end{figure*}

During physics data-taking MPX01 operates in counting mode. However, during 
the so-called van der Meer scans 
the occupancy of the device is sufficiently low for tracking particles.
As an example the last horizontal van der Meer scan of November 2012 (LHC fill 3316) is analysed
to determine the ratio between hits and clusters.
The data covers a time period of 1186 seconds, in which 103 frames were taken.
The total number of clusters is 155822.
Figure~\ref{fig:fill3316} shows the number of hits 
per cluster 
for the six MPX devices with the highest cluster rates
without distinguishing cluster types.
In summary, the ratio $N_{\rm hit}/N_{\rm cl} =413051/155822 = 2.65
$ is
smallest for MPX01 and largest for MPX13 with $N_{\rm hit}/N_{\rm cl} = 8257/1658 = 4.98$.
Table~\ref{tab:stat} lists the number of clusters, hit/cluster ratios and RMS 
values.
Assuming that one cluster is created by one particle, this ratio corresponds to the 
hit rate per interacting particle.
The fluctuations in the number of particles, not the number of hits, 
contribute to the statistical uncertainty of the luminosity measurement.

\begin{table*}[bpht]
\small
  \caption{Number of clusters, average ratio hits per cluster and 
           RMS values of the hits per cluster distributions, shown in Fig.\ref{fig:fill3316}. 
           Data is taken by MPX01-05 and MPX13 during the last horizontal van der Meer scan 
           of November 2012 (LHC fill 3316).
           For comparison the corresponding values are given for 
           one of the vertical van der Meer scans of July 2012 (LHC fill 2855).
           In this study, no noisy pixels are removed for the hit counting or
           cluster counting, the smaller ratio in November for MPX01 
           can be explained 
           by the increase in the number of noisy pixels during the year.}
 \centering
\renewcommand{\arraystretch}{1.3} 
    \begin{tabular}{cccccccccc}
\hline\hline
&MPX                & 01    & 02    & 03    & 04    & 05    & 13     \\\hline
&$N_{\rm cl}$       & 155822& 817   & 170   & 1294  & 285   & 1658   \\
NOV.&$N_{\rm hit}/ N_{\rm cl}$  
                   & 2.65  & 4.13  & 4.04  & 4.08  & 4.20  & 4.98   \\ 
&RMS                & 6.72  & 5.05  & 4.24  & 4.22  & 4.64  & 7.32   \\\hline
&$N_{\rm cl}$       & 61068 & 811   & 160   & 1191  & 310   & 1470   \\
JULY&$N_{\rm hit}/ N_{\rm cl}$  
                   & 5.04  & 4.95  & 4.65  & 4.08  & 3.51  & 4.67   \\ 
&RMS                &11.35  & 6.54  & 5.89  & 4.27  & 3.72  & 6.12   \\
\hline\hline
\end{tabular}
\label{tab:stat}
 \vspace*{-1cm}
\end{table*}


\section{Van der Meer Scans}
\label{sec:vdM_scans}

Van der Meer (vdM) scans are used for absolute luminosity 
calibration at the LHC~\cite{vdm}.
The vdM scan technique was pioneered at CERN in the 1960s to determine the luminosity calibration 
in a simple way.
It involves scanning the LHC beams through one another to determine the size of the beams at their 
point of collision. The scans are performed to determine the horizontal and vertical widths of the beams.
These width measurements are then combined with information on the number of circulating protons, 
allowing the determination of an absolute luminosity scale. 
Several ATLAS and CMS sub-detectors are used for vdM scans~\cite{improvedLumiDet:2013,cms:2013}
since the luminosity calibration is very important for physics analyses. 

The study of the MPX data taken during LHC vdM scans in April, July and November 2012 focuses 
on the horizontal and vertical width determination of the LHC proton beams.
The data used in this vdM scan study was taken with the MPX device with the highest count rate.
This study demonstrates that the operation of the MPX network is fully adapted to the low-luminosity 
regime of vdM scans and the high-luminosity regime of routine physics running.
Owing to the low statistics in heavy blob (thermal neutron) counting, only the hit counting mode
can be used for the vdM data analysis.

The beams are typically scanned transversely across each other in 25 steps. 
During each step, the beam orbits are left untouched (``quiescent beams") and the 
luminosity remains constant for approximately 29\,s. 
The beam separation is then incremented by several tens of microns 
(``non-quiescent beams") over an interval of several seconds, 
during which the luminosity varies rapidly and the luminosity measurements are unreliable. 
Since the MPX exposure (acquisition) time is about 5\,s per frame, followed by a 6\,s readout period,  
two frames typically occur within each quiescent-beam scan step. 
Occasionally, the MPX devices need to reload their configuration files, 
in which case the dead time can be as long as 30\,s. Therefore, 
only one frame is recorded in some scan steps.

\begin{figure}[htp]
\vspace*{-2mm}
\centering
\includegraphics[width=\linewidth]{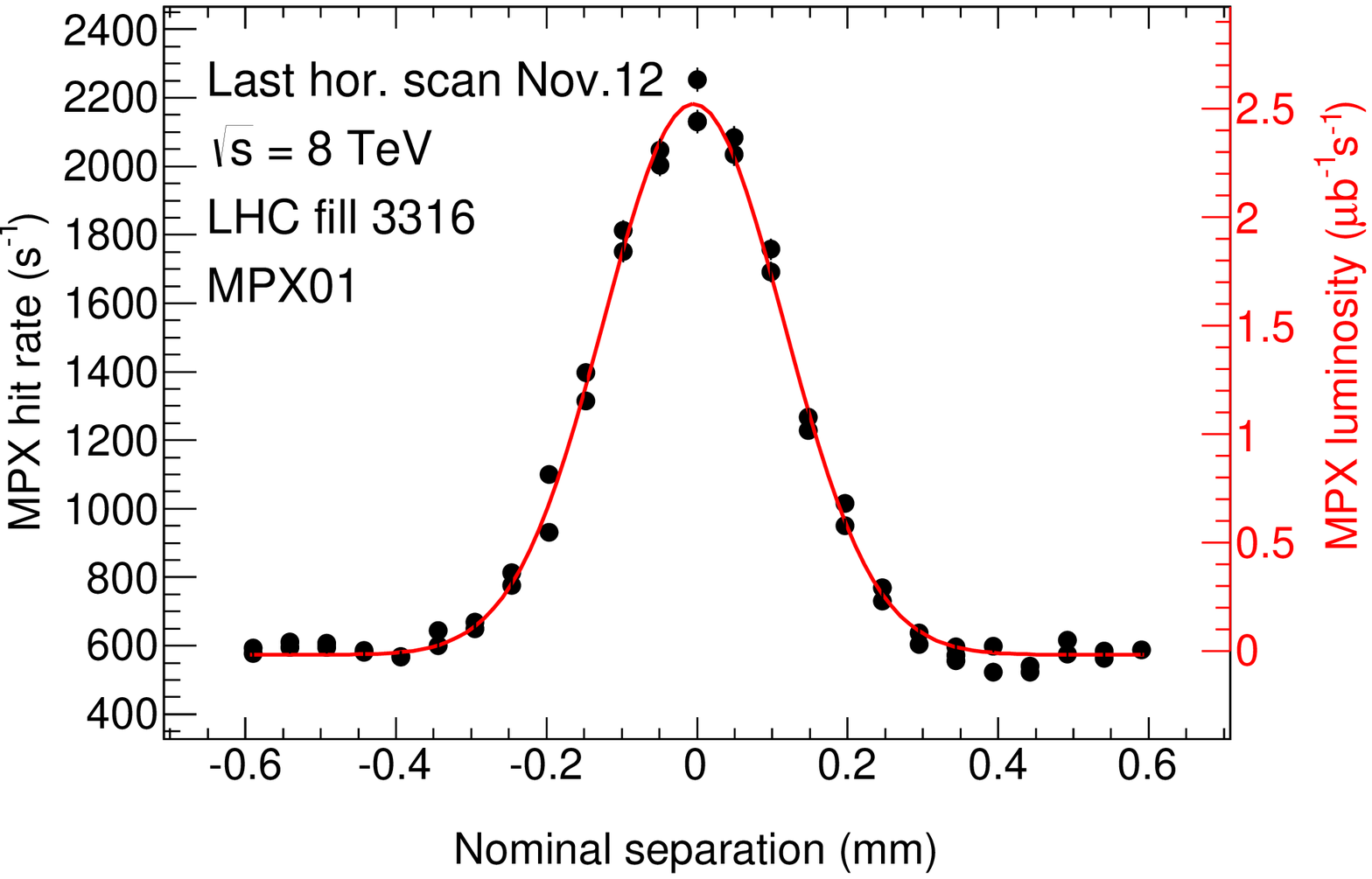}
\vspace*{-5mm}
\caption{Luminosity from hit counting as a function of nominal beam separation
         measured with MPX01 
         during the last horizontal vdM scan of November 2012. 
Each data point shows the measured instantaneous luminosity 
before background subtraction and averaged over one acquisition time.
Because the acquisition time
is significantly shorter than the duration of a scan step, 
there can be more than one MPX sampling per scan step. 
The MPX samplings that partially or totally overlap with non-quiescent 
scan steps (varying beam separation) are not shown. 
The fit function is the sum of a single Gaussian 
(representing the proper luminosity in this scan) 
and a constant term that accounts for instrumental 
noise and single-beam background.
The MPX normalization uses this horizontal and a vertical 
beam width from LHC vdM fill 3316.
}
\label{fig:mpx01HitLumi2}
\end{figure}

The beam separation dependence of the measured MPX luminosity is 
well represented by the sum of a single Gaussian and a 
constant (Fig.~\ref{fig:mpx01HitLumi2}).
The statistical uncertainty for each MPX frame, calculated from the number of hits,
is scaled up by a factor $\sqrt{2.65}$ to account for the ratio between hits and clusters,
as explained before in  Sec.~\ref{sec:statistics}.
In this approach it is assumed that one particle interacting with the MPX device creates one cluster
and it is the number of particles which lead to the statistical uncertainty.

\begin{figure}[hbtp]
\centering
\includegraphics[width=\linewidth]{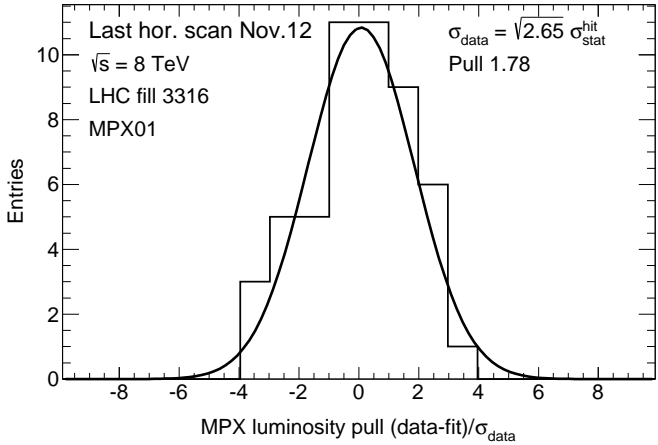}
\vspace*{-0.5cm}
\caption{Pull distribution defined as (data-fit)/$\sigma_{\rm data}$,
        where $\sigma_{\rm data} = \sqrt{R} \cdot \sigma_{\rm stat}^{\rm hit}$ and
        the ratio $R=N_{\rm hit}/N_{\rm cl} =2.65$ for MPX01. 
        The data shown in Fig.~\ref{fig:mpx01HitLumi2} is used.
        LHC fill 3316.
}
\label{fig:vdmpull}
\vspace*{-2mm}
\end{figure}

The precision of the MPX01 device can be determined with respect to the expected statistical 
precision. For this study, the pull (data-fit)/$\sigma_{\rm data}$ is calculated (Fig.~\ref{fig:vdmpull})
for the last horizontal vdM scan in November 2012, where  
$\sigma_{\rm data} = \sqrt{R} \cdot \sigma_{\rm stat}^{\rm hit}$ and $R=2.65$.
The sigma of the pull distribution is 1.78, which indicates that additional uncertainties are present 
beyond the determined statistical uncertainties.

The data shows that the sensitivity of the MPX01 device is sufficient to measure the 
luminosity down to about 
$\rm 0.5\,\micro b^{-1} s^{-1}$, 
four orders of magnitude smaller than the luminosity typical of routine physics data-taking.

The luminosity can be calculated as:
\begin{equation}
L_{\rm MPX} = N_{\rm b} N_{\rm p1} N_{\rm p2} f/(2\pi \Sigma_{\rm x}\Sigma_{\rm y}),
\end{equation}
where 
$N_{\rm b}$ is the number of bunch crossings producing collisions per machine revolution, 
$N_{\rm p1}$ and $N_{\rm p2}$ are the average bunch populations (number of protons) 
in beam~1 and beam~2, respectively,
$f$ is the machine revolution frequency (11245.5~Hz), and
$\Sigma_{\rm x}$ ($\Sigma_{\rm y}$) are the convoluted horizontal (vertical) beam sizes.

The specific luminosity is defined as:
\begin{equation}
L_{\rm specific} =  L_{\rm MPX} / (N_{\rm b} N_{\rm p1} N_{\rm p2}) = f/(2\pi \Sigma_{\rm x}\Sigma_{\rm y}).
\end{equation}
Table~\ref{tab:vdm} summarizes the scan results for all 2012 vdM scans registered with the MPX01 
device.

\begin{table}[htbp]
\small
  \caption{MPX01 van der Meer (vdM) scan results for 2012 data. 
           The fit results for the bunch-averaged horizontal $\Sigma_{\rm x}$ 
           and vertical $\Sigma_{\rm y}$ convoluted beam sizes are given.
           The units of the specific luminosity, $L_{\rm specific}$, are
           $\rm 10^{27}\,cm^{-2}\,s^{-1}\,(10^{11}\,protons)^{-2}$.
}
\vspace*{-2mm}
 \centering
 \renewcommand{\arraystretch}{1.5} 
    \begin{tabular}{cccccr}
\hline\hline
          & Fill      & Scan     & $\Sigma_{\rm x}$ & $\Sigma_{\rm y}$& $L_{\rm specific}$           \\
          & number    &          & ($\micron$)   & ($\micron$)  &   \\ \hline
    \multirow{3}[7]{*}{APRIL} & \multirow{3}[7]{*}{2520} 
                &  1   & 25.94 & 33.07 &2086 \\
          &     &  2   & 26.04 & 32.65 &2105 \\
          &     &  3   & 26.70 & 33.64 &1992 \\\hline
    \multirow{4}[8]{*}{JULY} & \multirow{4}[8]{*}{2855} 
                &  4   & 120.5 & 127.6 & 116  \\
          &     &  5   & 124.7 & 127.9 & 112  \\
          &     &  6   & 123.8 & 128.2 & 112  \\
          &     &  8   & 120.6 & 127.1 & 116  \\\hline
    \multirow{4}[7]{*}{NOV.} & \multirow{3}[6]{*}{3311} 
                & 10   & 121.4 & 128.2 & 114   \\
          &     & 11   & 129.3 & 129.6 & 106   \\
          &     & 14   & 139.9 & 131.0 & 97   \\
          & 3316& 15   & 119.3 & 133.6 & 112   \\
\hline\hline
    \end{tabular}%
  \label{tab:vdm}%
\vspace*{-3mm}
\end{table}%

In this paper,
the last horizontal and vertical November 2012 scans are used 
for the absolute luminosity calibration. 
These two scans are well described by a single Gaussian. 
The horizontal scan (Fig.~\ref{fig:mpx01HitLumi2})
has $\chi^2/{\rm ndf}=136/47$ and 
a similar value is obtained for the vertical scan $\chi^2/{\rm ndf}=114/44$.
These ratios indicate that in addition to the statistical uncertainty 
(augmented by the factor $\sqrt{2.65}$) systematic uncertainties are also present.
The widths of the horizontal and vertical nominal beam separations and 
their uncertainties are $(119.3\pm1.6)\,\micron$ and $(133.6\pm1.9)\,\micron$, respectively.
The LHC parameters for fill 3316 are~\cite{lhc}:
\begin{itemize}
\item Number of bunches: 29
\item Average number of protons (in units $10^{11}$) per bunch in beam~1 and in beam~2: 
      $25.3/29=0.872$ and $25.7/29=0.886$, respectively.
\end{itemize}
Thus, the resulting luminosity is 
$L_{\rm MPX} = 2.515\,{\rm \micro b^{-1}s^{-1}}$. 

The corresponding number of MPX hits at the peak is 
determined from a Gaussian fit plus a constant background.
The fit provides $(1609.34\pm0.03)$ hits/s at the peak above the background.
Thus, the normalization factor $n_{\rm f}$ between the MPX01 hit rate and 
the instantaneous LHC luminosity is 
\begin{equation}
n_{\rm f}=\frac{\rm 2.515\,\micro b^{-1}s^{-1}}{\rm 1609.34\,hit\,s^{-1}}=
1.5628\cdot 10^{-3}~{\rm \micro b^{-1} / hit}.
\end{equation}

The normalization factor for the absolute luminosity 
is only approximate since the MPX acquisition time is much 
longer than the bunch spacing. 
Therefore, the bunch-integrated luminosity averages 
over the different bunch profiles.

The uncertainty on $n_{\rm f}$ due to the bunch-integration by the MPX data-taking has been estimated
by simulating 29 overlapping Gaussian distributions, corresponding to 29 colliding bunches, 
with varying individual widths. The simulated bunches vary in width in equal distances up to $\pm 25$\%.
The envelope of the summed Gaussians is fitted and the resulting width and height are compared with the 
nominal value without variations.
The width uncertainty is 0.36\% and height uncertainty is 0.49\%.
The fit is repeated 100 times with different sets of random numbers to test the reproducibility.
For the determination of luminosity uncertainty it is assumed that the horizontal and vertical
width uncertainties are correlated, thus the luminosity uncertainty is 0.72\%.
Furthermore, it is assumed that the widths and height (fitted hit rate) uncertainties are correlated, 
thus the uncertainty on the normalization factor is $0.7\%+0.5\%=1.2\%$.

Although further uncertainties could arise from non-Gaussian shapes, this study shows that the 
Gaussian approximation of the sum of Gaussians is quite robust and the luminosity approximation 
by bunch integration is a sensible approach.
No attempt is made for a precise determination of the total uncertainty which would require a
dedicated study~\cite{improvedLumiDet:2013}.

In summary,
Fig.~\ref{fig:mpx01HitLumi2} shows the hit rate and the absolute luminosity determined 
from the scan widths. The resulting normalization factor is used throughout this paper.

\section{LHC Luminosity Curve and MPX Short-Term Precision}
\label{sec:short-term}
The MPX network precisely measures the LHC luminosity as a function of time. 
As a proof of principle it is demonstrated that the MPX network has the capability to study the 
underlying mechanisms of the rate of reduction of LHC luminosity.

The LHC luminosity reduction is mainly caused by beam-beam interactions 
(burning-off the proton bunches) and beam-gas (single bunch) interactions by the protons of the circulating beams 
with remaining gas in the vacuum pipe.
The particle loss rate due to proton burn off in collision
is proportional to the number of protons in the second power
since protons are lost in both colliding bunches. In the case that the protons in a beam collide 
with remaining gas in the vacuum pipe, the particle loss rate is proportional to the number 
of protons in the beam. The loss rate of protons $N$ in the colliding beam is thus governed by:
\begin{equation}
-dN/dt = \lambda_{\rm bb}N^2/N_0+\lambda_{\rm g}N,
\label{eq:rate}
\end{equation}
where $N_0$ is the initial number of protons, and $\lambda_{\rm bb}$ and $\lambda_{\rm g}$ 
are constants related to beam-beam and beam-gas interactions, respectively.
This equation has a known solution:
\begin{equation}
N(t) = \frac{N_0 {\rm e}^{-\lambda_{\rm g}t}}{1+ \frac{\lambda_{\rm bb}}{\lambda_{\rm g}}(1-{\rm e}^{-\lambda_{\rm g}t})},
\label{eq:exponent}
\end{equation}
with two well-known border cases:
\begin{equation}
N(t) = N_0 {\rm e}^{-\lambda_{\rm g}t}~ {\rm for}~ \lambda_{\rm bb}\ll \lambda_{\rm g}~{\rm and},
\label{eq:exponenta}
\end{equation}
\begin{equation}
N(t) = \frac{N_0}{1+ \lambda_{\rm bb} t}~{\rm for}~\lambda_{\rm g}\ll \lambda_{\rm bb}.
\label{eq:exponentb}
\end{equation}
In the following we will be interested in the time dependence of the luminosity and
of the average number of interactions per bunch crossing $\mu$.
By definition $\mu$ is proportional to the luminosity $L$. 
Since both of these
quantities are proportional to $N^2$ we expect the time dependence of $\mu$ to be described 
by: 
\begin{equation}
\mu(t) = \frac{\mu_0 {\rm e}^{-2\lambda_{\rm g}t}}{[1+ \frac{\lambda_{\rm bb}}{\lambda_{\rm g}}(1-{\rm e}^{-\lambda_{\rm g}t})]^2}.
\label{eq:exponent3a}
\end{equation}

A fit is applied to the data using eq.~(\ref{eq:exponent3a}). 
The uncertainty on the fit result is evaluated by  
several independent measurements of the MPX network.

The LHC fill 3236, taken on 28-29 October 2012, has been investigated in this study. 
The MPX luminosity is converted to an average interaction per bunch crossing by:
\begin{equation}
\mu=L\cdot \sigma_{\rm inel}/(k\cdot f), 
\end{equation}
where $k=1368$ colliding bunches, $f=11245.5$~Hz and the inelastic cross-section $\sigma_{\rm inel} = 73$~mb.
The fill was chosen since it has a large $\mu$ range from about $\mu = 35$ at the beginning 
to about $\mu=8$ at the end.

When studying the MPX measurements of the LHC luminosity, structures resulting
from LHC parameter tuning can be noted.
These beam tuning adjustments change the shape of the 
luminosity decrease, and are not described by eq.~(\ref{eq:exponent3a}).
It is noted that in the first half of a fill these tunings were 
frequent (about one every hour) 
while in the second half of a fill, adjustments of the beam were rarely made. 
Therefore, only the second half of a fill is used for this study and
the range $\mu = 15$ to $\mu=8$ is used for fitting the data distribution.

\begin{figure*}[hbp]
\vspace*{-5mm}
\centering
\includegraphics[width=0.49\linewidth]{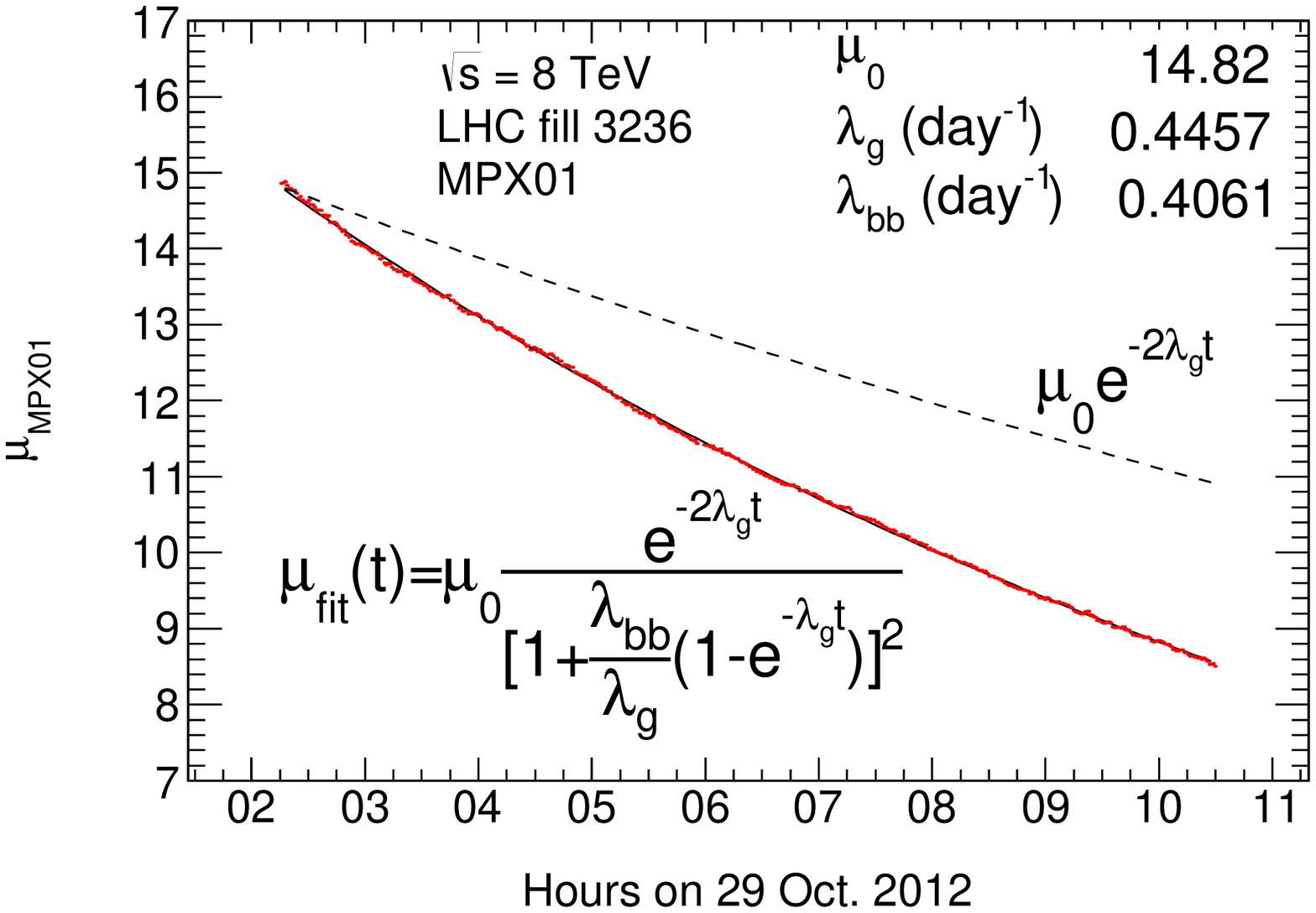}
\includegraphics[width=0.49\linewidth]{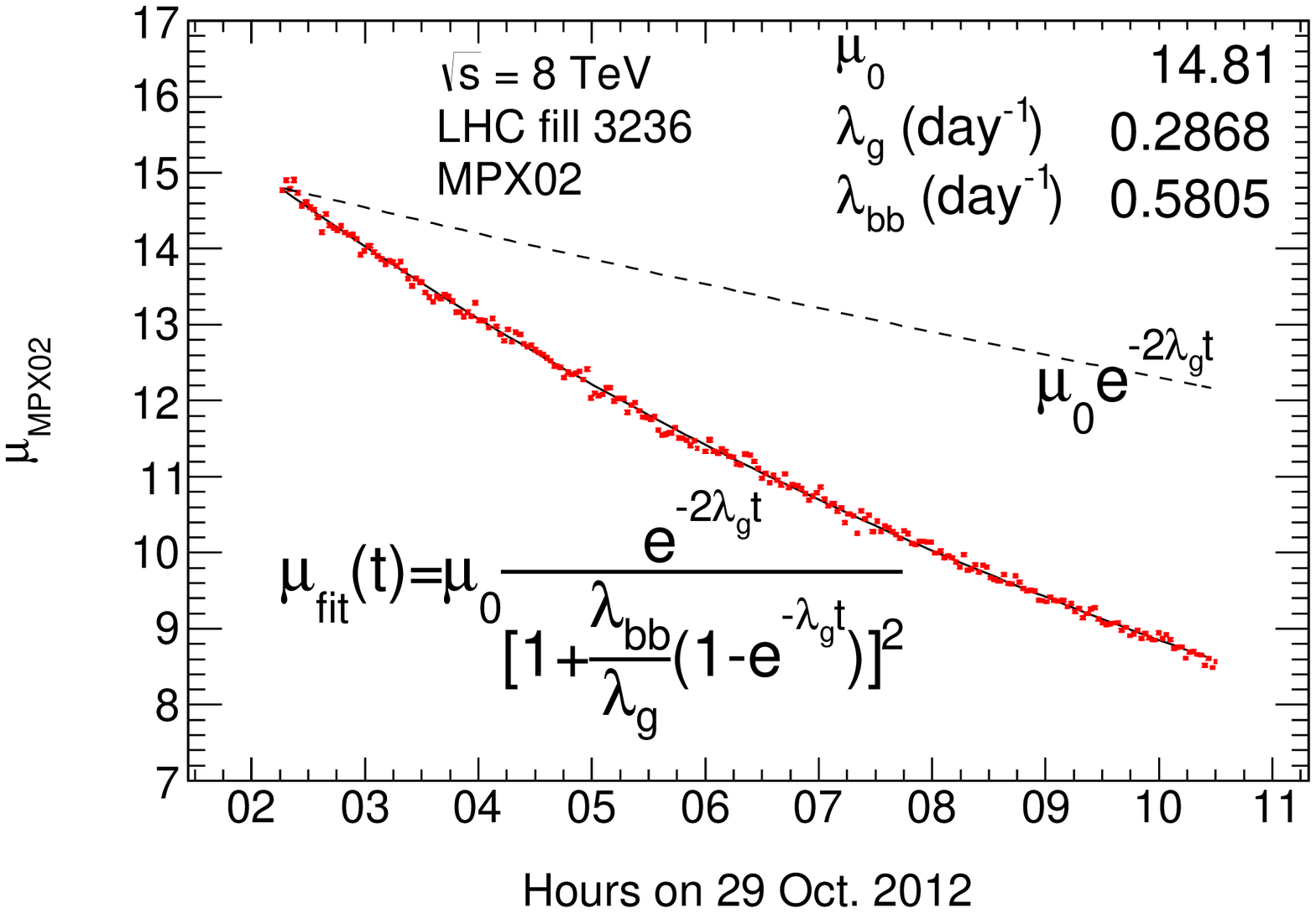}
\includegraphics[width=0.49\linewidth]{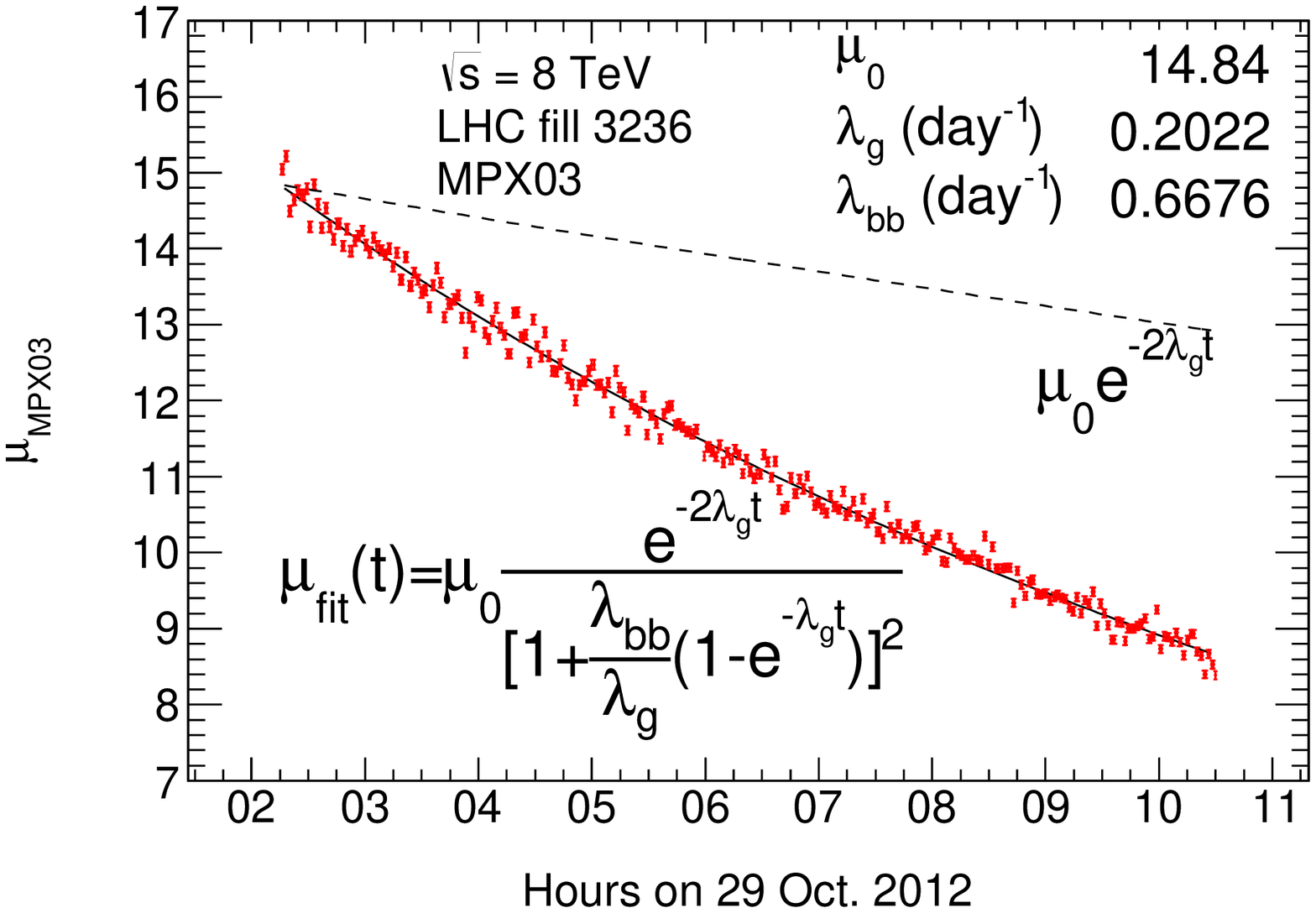}
\includegraphics[width=0.49\linewidth]{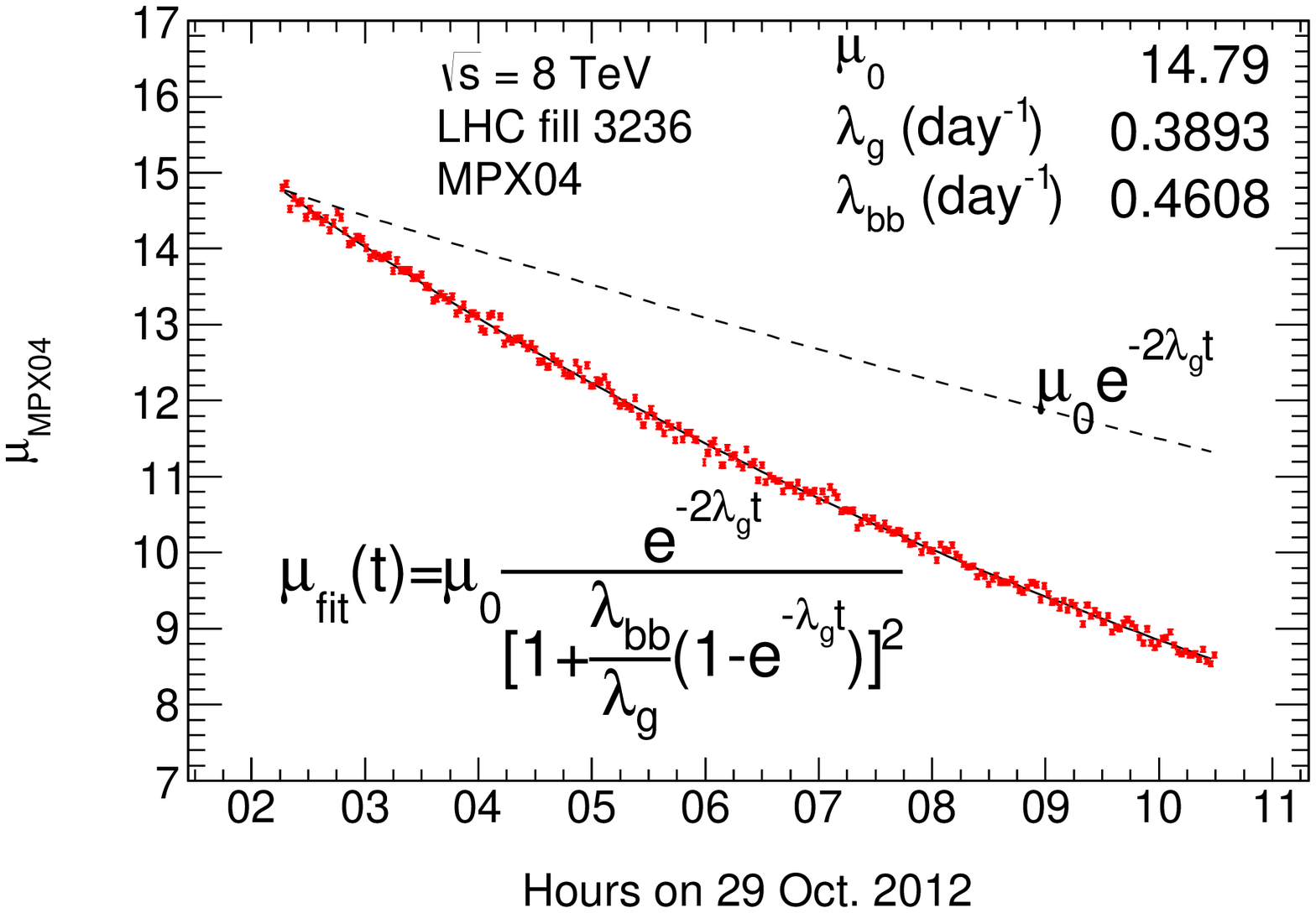}
\includegraphics[width=0.49\linewidth]{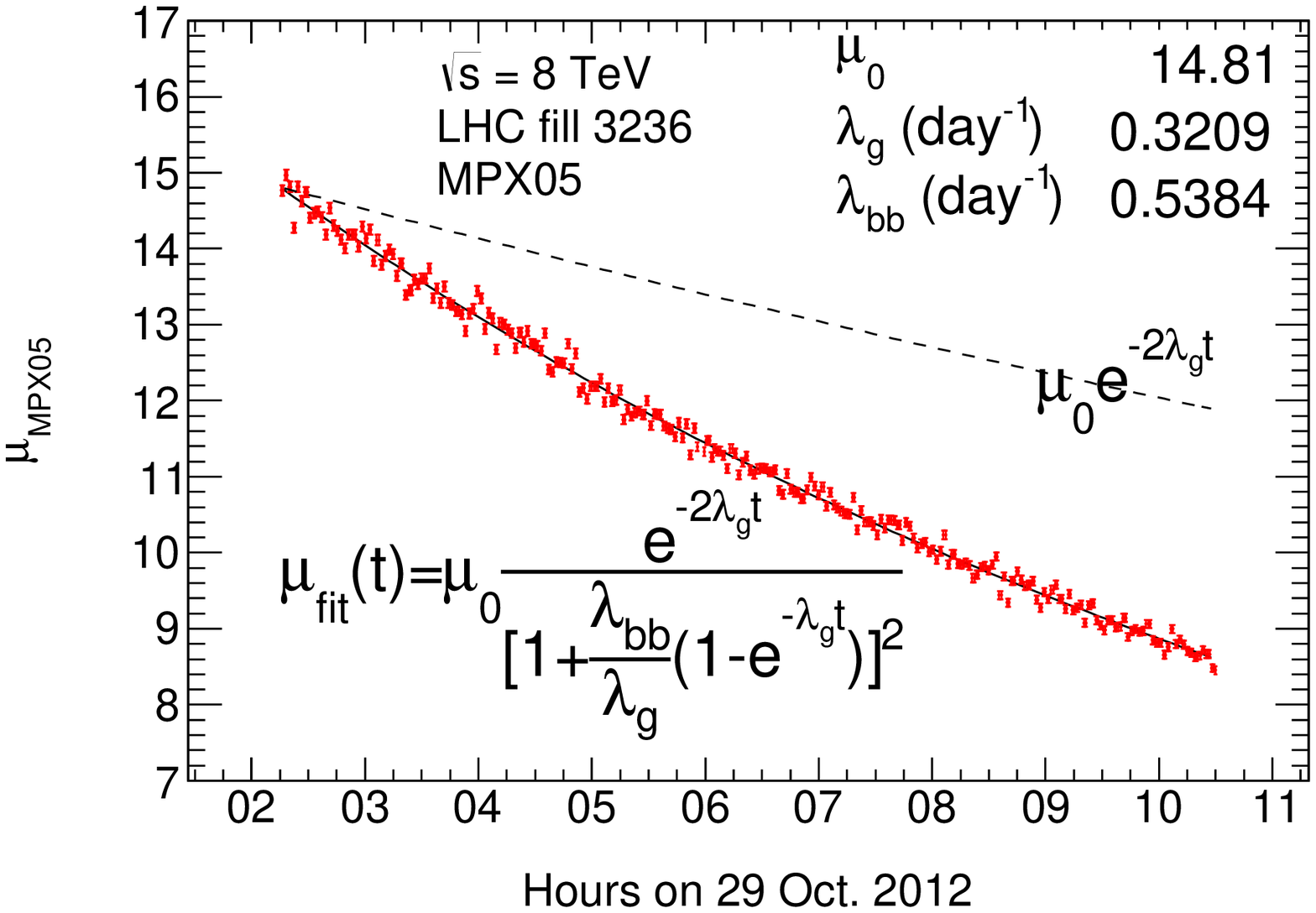}
\includegraphics[width=0.49\linewidth]{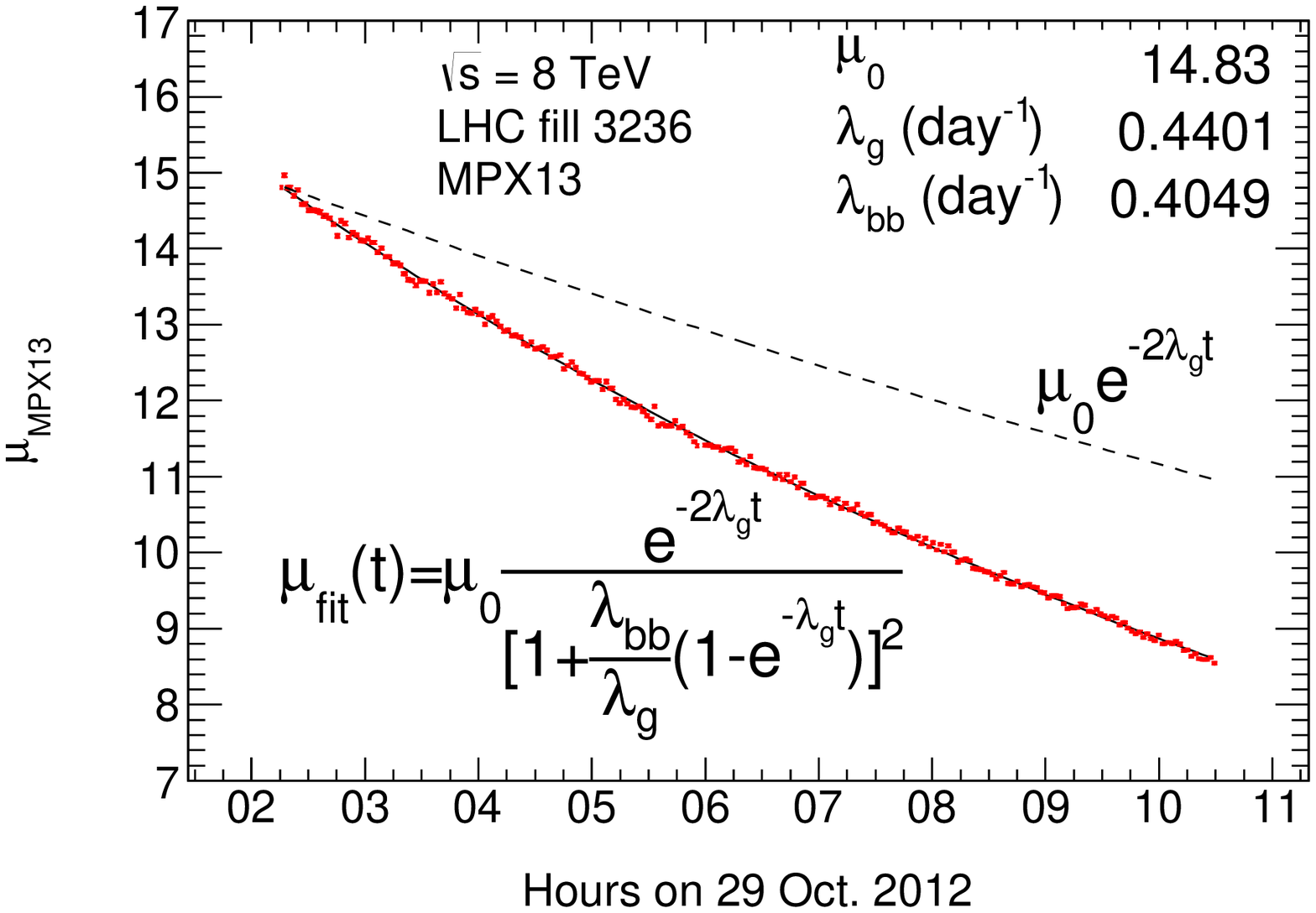}
\vspace*{-2mm}
\caption{Average number of interactions per bunch crossing  
         as a function of time seen by MPX devices with highest particle 
         flux (MPX01-05 and MPX13).
         The distribution is approximately described by a function as 
         given in the figure.
         The parameters are defined in the text.
         The statistical uncertainties per data point are indicated.
         They depend on the hit statistics scaled by a factor $N_{\rm hit}/N_{\rm cl}$
         given in Table~\ref{tab:stat}.
         In addition to the fit result described by the solid line, the dashed 
         line shows the result of beam-gas interactions alone 
         $\mu_0\exp(-2\lambda_{\rm g}t)$, 
         with $\mu_0$ and $\lambda_{\rm g}$ obtained from the three parameter fit.
         LHC fill 3236.
}
\label{fig:fit} 
\end{figure*}

The fits 
are shown in Fig.~\ref{fig:fit} individually for MPX01-05 and MPX13.
The fits give the initial number of interactions per bunch crossing $\mu_0$, 
$\lambda_{\rm bb}$ and $\lambda_{\rm g}$,  summarized in Table~\ref{tab:fit}.

The fits with both parameters
$\lambda_{\rm bb}$ and $\lambda_{\rm g}$ describes the data significantly better 
compared to using only one of the two.

\begin{table*}[htbp]
\small
  \caption{Fit values of $\mu_0$, $\lambda_{\rm bb}$ and $\lambda_{\rm g}$
           for the MPX devices with highest particle flux 
           for statistical uncertainties only, and 
           for systematic uncertainties only, constant in time, such that $\chi^2/{\rm ndf}=1$.
           LHC fill 3236.}
 \centering
\renewcommand{\arraystretch}{1.3} 
    \begin{tabular}{ccccccccc}
\hline\hline
MPX                & 01    & 02    & 03    & 04    & 05    & 13   & Mean & Standard deviation\\\hline
\multicolumn{9}{c}{Statistical uncertainties only} \\
$\mu_0$            & 14.82 & 14.81 & 14.84 & 14.79 & 14.81 & 14.83&14.82 & 0.02   \\
$\lambda_{\rm bb}$  
($10^{-6}\,$s$^{-1}$)        & 4.70  & 6.72  & 7.72  & 5.33  & 6.23  & 4.68&5.89   & 1.18   \\ 
$\lambda_{\rm g}$ 
($10^{-6}\,$s$^{-1}$)        & 5.16  & 3.32  & 2.34  & 4.50  & 3.71  & 5.09&4.02   & 1.10   \\ \hline
\multicolumn{9}{c}{Systematic uncertainties only}  \\
$\mu_0$            & 14.82 & 14.81 & 14.85 & 14.79 & 14.82 & 14.83&14.82 & 0.02   \\
$\lambda_{\rm bb}$  
($10^{-6}\,$s$^{-1}$)        & 4.98  & 7.00  & 8.22  & 5.41  & 6.37  & 4.94&6.15   & 1.30   \\ 
$\lambda_{\rm g}$ 
($10^{-6}\,$s$^{-1}$)        & 4.91  & 3.06  & 1.91  & 4.44  & 3.59  & 4.87&3.80   & 1.18   \\
\hline\hline
\end{tabular}
\label{tab:fit}
\vspace*{-0.9cm}
\end{table*}

\begin{figure}[tbp]
\centering
\includegraphics[width=\linewidth]{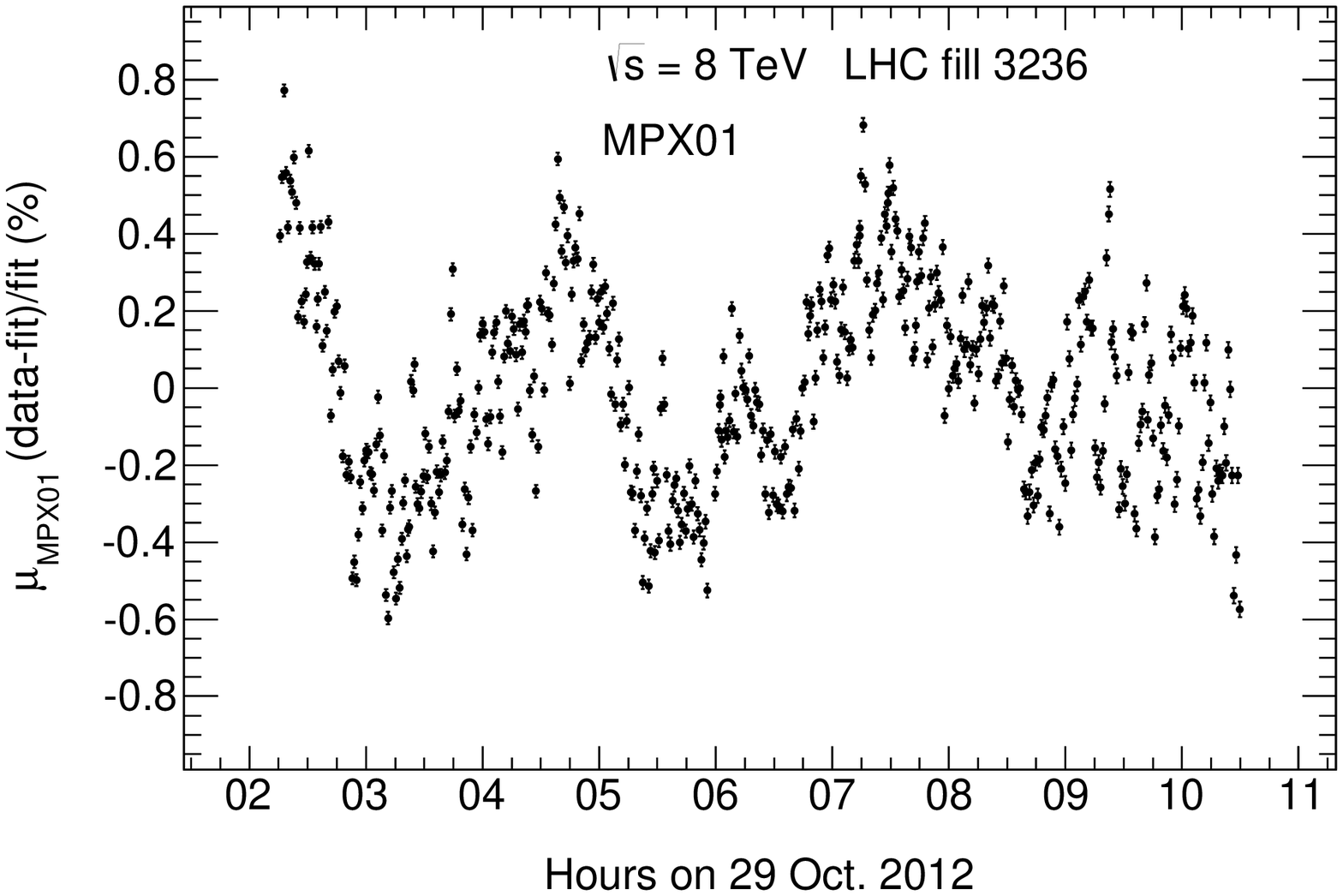}
\vspace*{-0.6cm}
\caption{Relative difference between data and fitted average 
         number of interactions per bunch crossing as a function
         of time seen by MPX01.
         The relative deviations between data and fit have an RMS of 0.3\%.
         The statistical uncertainties $\Delta\mu/\mu$ per data point are indicated and
         vary from $0.0094 \sqrt{2.65}$\% to $0.0125\sqrt{2.65}$\%
         where the factor $2.65$ is the averaged ratio of hits per 
         interacting particle.
         The apparent structure is discussed in the text.
         LHC fill 3236.
}
\label{fig:fiterr}
\vspace*{-0.2cm}
\end{figure}

Figure~\ref{fig:fiterr} shows the difference between the fit and the data for MPX01.
The observed structures could be attributed to small LHC luminosity fluctuations
not described by the fit function.
However, the origin of these fluctuations cannot be determined from the MPX data alone,
and thus they are conservatively attributed to MPX systematics.
The size of these fluctuations estimated as departure of the data from the fitted curve, 
amounts to an RMS of approximately 0.3\%,
corresponding to $\Delta\mu^{\rm fct}_{\rm sys} = 0.03$ for $\mu=10$. 

For all high-statistics devices used (MPX01-05 and MPX13), the uncertainties of 
the fits are dominated by systematic effects.
The fit of the MPX01 data, for example, has a 
$\chi^2/{\rm ndf}=4.4\cdot 10^7/499$ much larger than one.
Therefore, the fits are repeated with a constant systematic uncertainty for each MPX data point
such that $\chi^2/{\rm ndf}=1$.
These fit parameters are also summarized in Table~\ref{tab:fit}.
The $\lambda$ mean values of the beam-beam and beam-gas interactions are
\begin{equation}
\rm
\lambda_{\rm bb} = (6.2 \pm 1.3) \cdot 10^{-6}\,s^{-1}
\label{eq:lbb}
\end{equation}
and
\begin{equation}
\rm
\lambda_{\rm g}  = (3.8 \pm 1.2) \cdot 10^{-6}\,s^{-1},
\end{equation}
where the given standard deviations are calculated as the square root of the variance 
from the numbers in Table~\ref{tab:fit} for constant systematic uncertainties.

The fit results indicate that the LHC luminosity reduction is predominantly reduced by the 
beam-beam interactions since a larger value of $\lambda$ corresponds to a shorter lifetime. 
In addition to the interactions between the proton beams and the remaining gas in the vacuum pipe,
there could be other processes which, assuming that they also depend linearly on the number of protons, 
are incorporated in the value of $\lambda_{\rm g}$.  

In the following, the expected mean lifetime of inelastic beam-beam interactions is calculated
and compared with the experimental results.
The mean lifetime from inelastic beam-beam interactions is given by~\cite{LHC94}:
\begin{equation}
\label{eq:tbb}
t_{\rm bb}^{\rm inel} = kN_0/(N_{\rm exp}L_0 \sigma_{\rm inel}),
\end{equation}
where $k$ is the number of bunches, 
$N_0$ is the initial number of protons per bunch 
($kN_0 = 2.2\cdot 10^{14}$ protons~\cite{lhc}).
The initial luminosity is 
$L_0 = 7360\,{\rm \micro b^{-1} s^{-1}}$~\cite{lhc},
the number of experiments is $N_{\rm exp} = 2$ (ATLAS~\cite{atlasCollaboration:2013} 
and CMS~\cite{cmsCollaboration}).
We obtain $t_{\rm bb}^{\rm inel} = 2.05\cdot 10^5\,{\rm s}$ and thus
\begin{equation}
\lambda_{\rm bb}^{\rm inel} = 1/t_{\rm bb}^{\rm inel} = 4.87\cdot 10^{-6}\,{\rm s}^{-1}.
\end{equation}

We note that $\lambda_{\rm bb}^{\rm inel}$ 
depends on the initial luminosity
and the initial number of protons, thus on the starting value of $\mu_0=15$ for the fit. 
Since $L \propto N^2$
we can write $ \lambda_{\rm bb}^{\rm inel}  \propto \sqrt{L_0} \propto \sqrt{\mu_0} $. 
Thus for the lower initial luminosity in the fit, we expect a longer lifetime from
beam-beam interactions and therefore a smaller 
\begin{equation}
\lambda^{15}_{\rm bb} =\sqrt{15/35}\cdot 4.87\cdot 10^{-6}\,{\rm s}^{-1}= 3.20 \cdot 10^{-6}\,{\rm s}^{-1}.
\label{eq:inel}
\end{equation}

We observe that the fitted  $\lambda_{\rm bb}$ value of eq.~(\ref{eq:lbb}) 
is larger than  $\lambda^{15}_{\rm bb}$ calculated using 
the inelastic cross-section only, given in eq.~(\ref{eq:inel}). This suggests,
as expected, that the  proton-proton interaction cross-section leading to the proton loss from the 
beam is indeed larger than the inelastic cross-section.
In addition to inelastic hadronic scattering,
hadronic diffractive, hadronic elastic and Coulomb scattering contribute to the proton burn-off.

\begin{figure}[htbp]
\centering
\vspace*{-0.1cm}
\includegraphics[width=\linewidth]{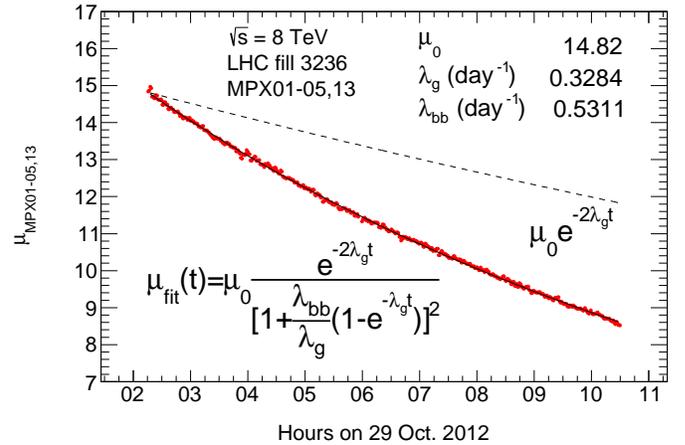}
\vspace*{-0.6cm}
\caption{Average number of interactions per bunch crossing 
         as a function of time using combined data from MPX01-05 and MPX13.
         The distribution is approximately described by the function  
         given in the figure.
         The parameters are defined in the text.
         In addition to the fit result described by the thin line, the dashed 
         line shows the result of beam-gas interactions alone 
         $\mu_0\exp(-2\lambda_{\rm g}t)$,
         with $\mu_0$ and $\lambda_{\rm g}$ obtained from the three parameter fit.
         LHC fill 3236.}
\label{fig:fitcombi}
\end{figure}

Another fit is performed using the combined data from MPX01-05 and MPX13
as a consistency check of the previously described analysis procedure which used 
the average of the fit parameters from the individual MPX devices.
The fit using the combined data is shown in Fig.~\ref{fig:fitcombi}. 
It gives, as expected,  almost identical fit parameters to the ones obtained as the average values 
of the previous procedure.

The ratio of the data in the first half and the extrapolated fit curve is studied
since the performed fits were based only on data from the second half of the LHC fill 3236. 
It was found that data at the beginning of the fill is up to 30\% above the fit curve.
This could indicate that the luminosity decreases more quickly than expected 
from beam-beam (burn-off) and beam-gas interactions alone 
at the beginning of a fill, possibly due to 
non-linear effects with small emittance and short-length bunches. 
Furthermore, faster reductions in collision rates at the beginning of the fill could arise 
from denser bunches.

An intrinsic uncertainty in the MPX luminosity measurements results from
the bunch integration of the MPX network since the colliding bunches in the LHC 
contribute with different intensities.

\begin{figure}[htbp]
\centering
\vspace*{-0.5cm}
\includegraphics[width=\linewidth]{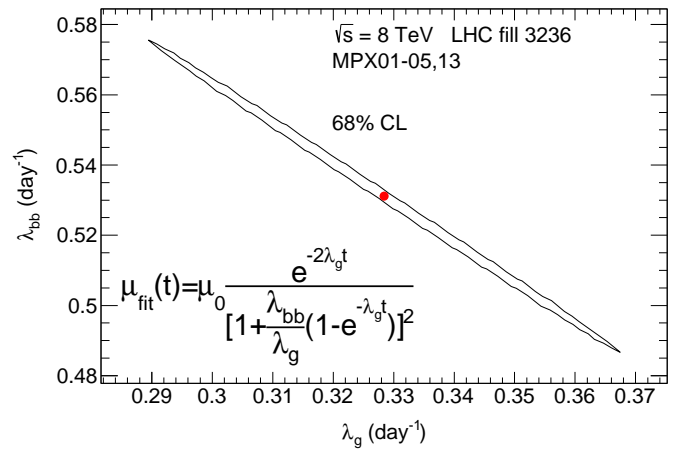}
\vspace*{-0.6cm}
\caption{Contour curve of $\lambda_{\rm g}$ and $\lambda_{\rm bb}$, given at 68\% CL,
         for the fit results of the average number of interactions per bunch crossing  
         as a function of time using 
         combined data from MPX01-05 and MPX13, shown in Fig.~\ref{fig:fitcombi}.
         The dot in the center of the contour indicates the fit values.
         LHC fill 3236.}
\label{fig:contour}
\end{figure}

The contour plot of $\lambda_{\rm g}$ and $\lambda_{\rm bb}$, given at 68\% CL 
in Fig.~\ref{fig:contour}, shows a strong anti-correlation. 
The individual uncertainties are taken as constant (giving equal
weight to the individual MPX devices in the combined data) since the uncertainty of the MPX 
data is systematically dominated. 
The constant uncertainty is scaled to $\Delta\mu = 0.0448$ yielding $\chi^2/{\rm ndf}=1$. 

\begin{figure*}[hbp]
\centering
\includegraphics[width=0.49\linewidth]{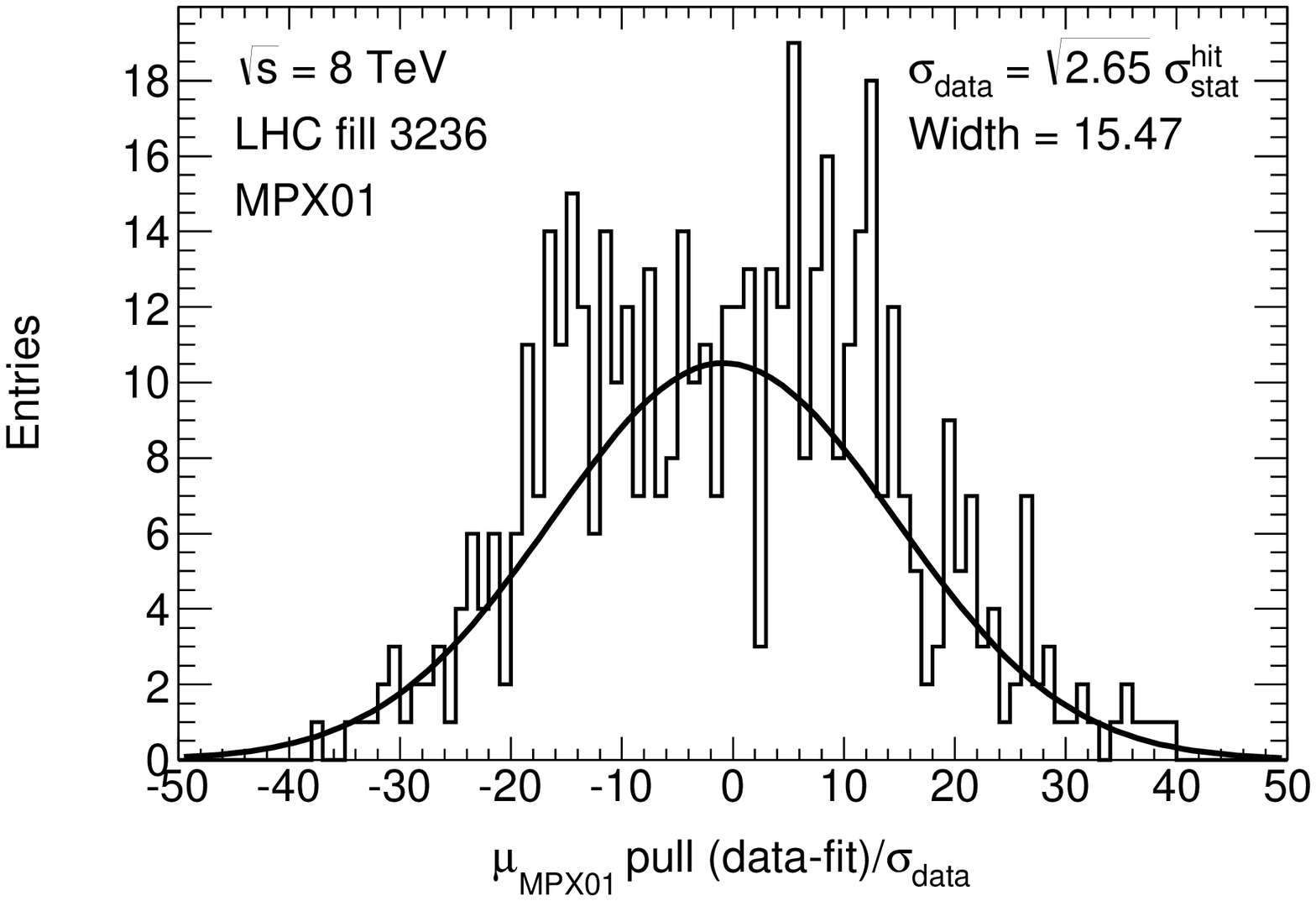}
\includegraphics[width=0.49\linewidth]{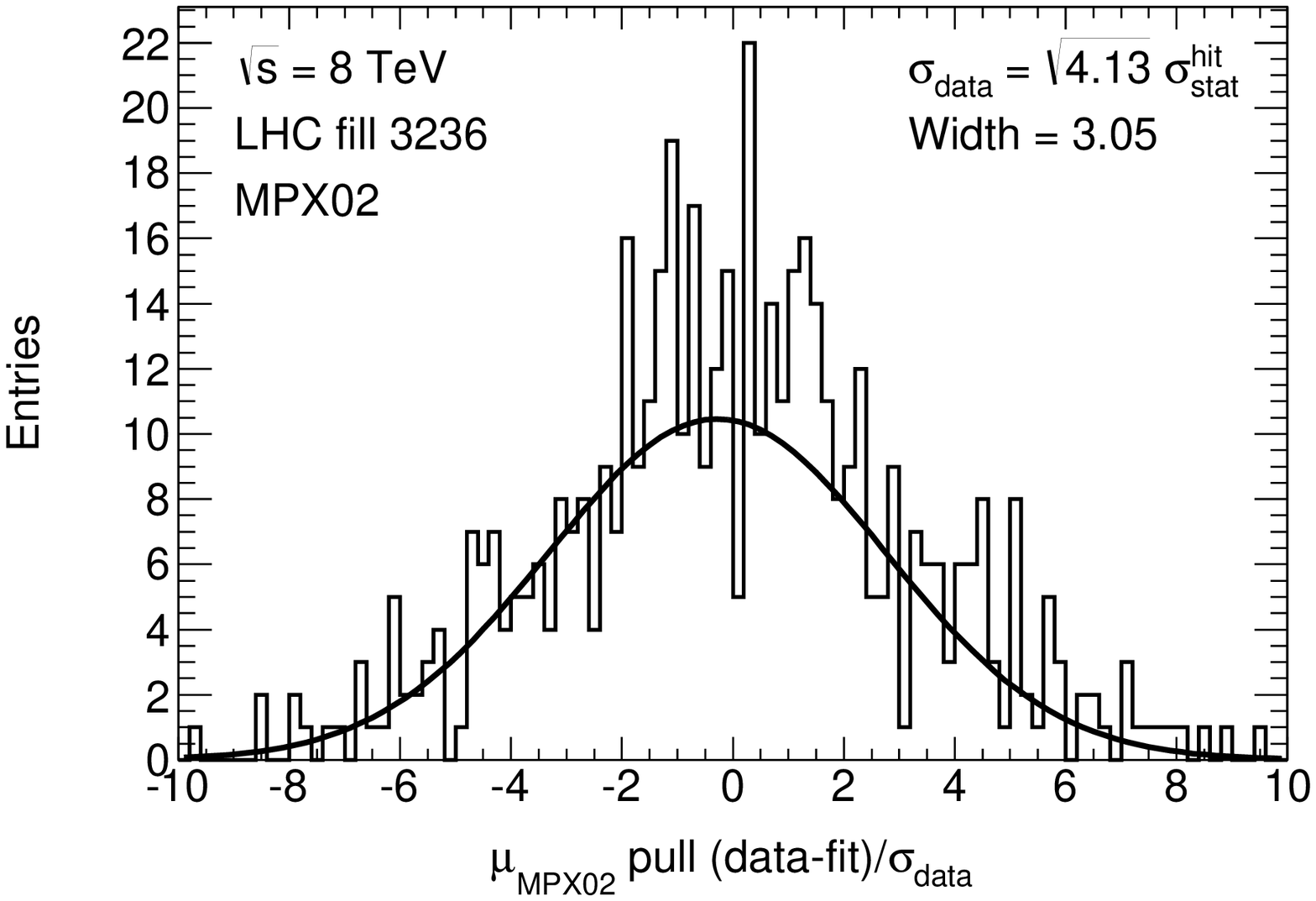}
\includegraphics[width=0.49\linewidth]{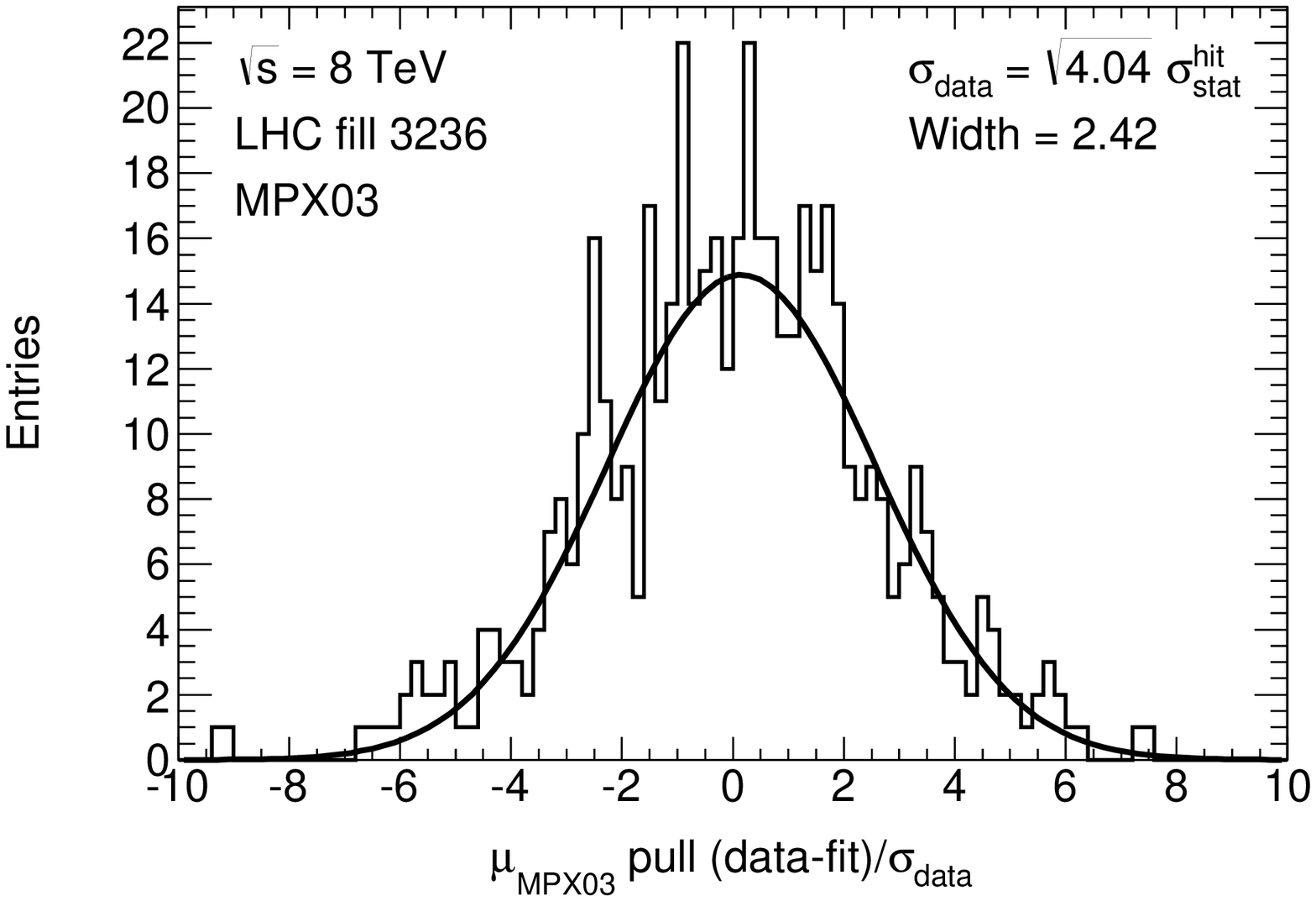}
\includegraphics[width=0.49\linewidth]{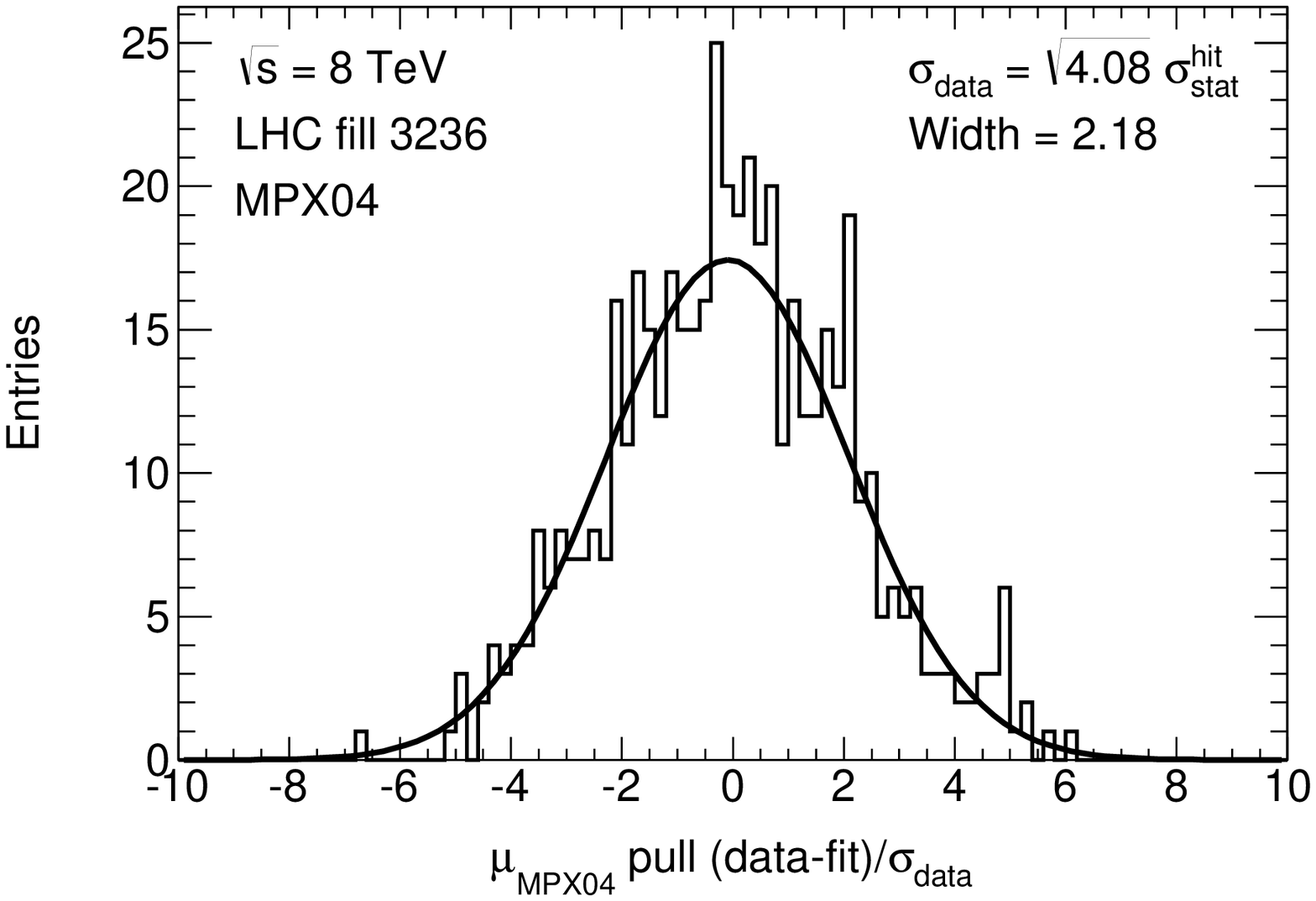}
\includegraphics[width=0.49\linewidth]{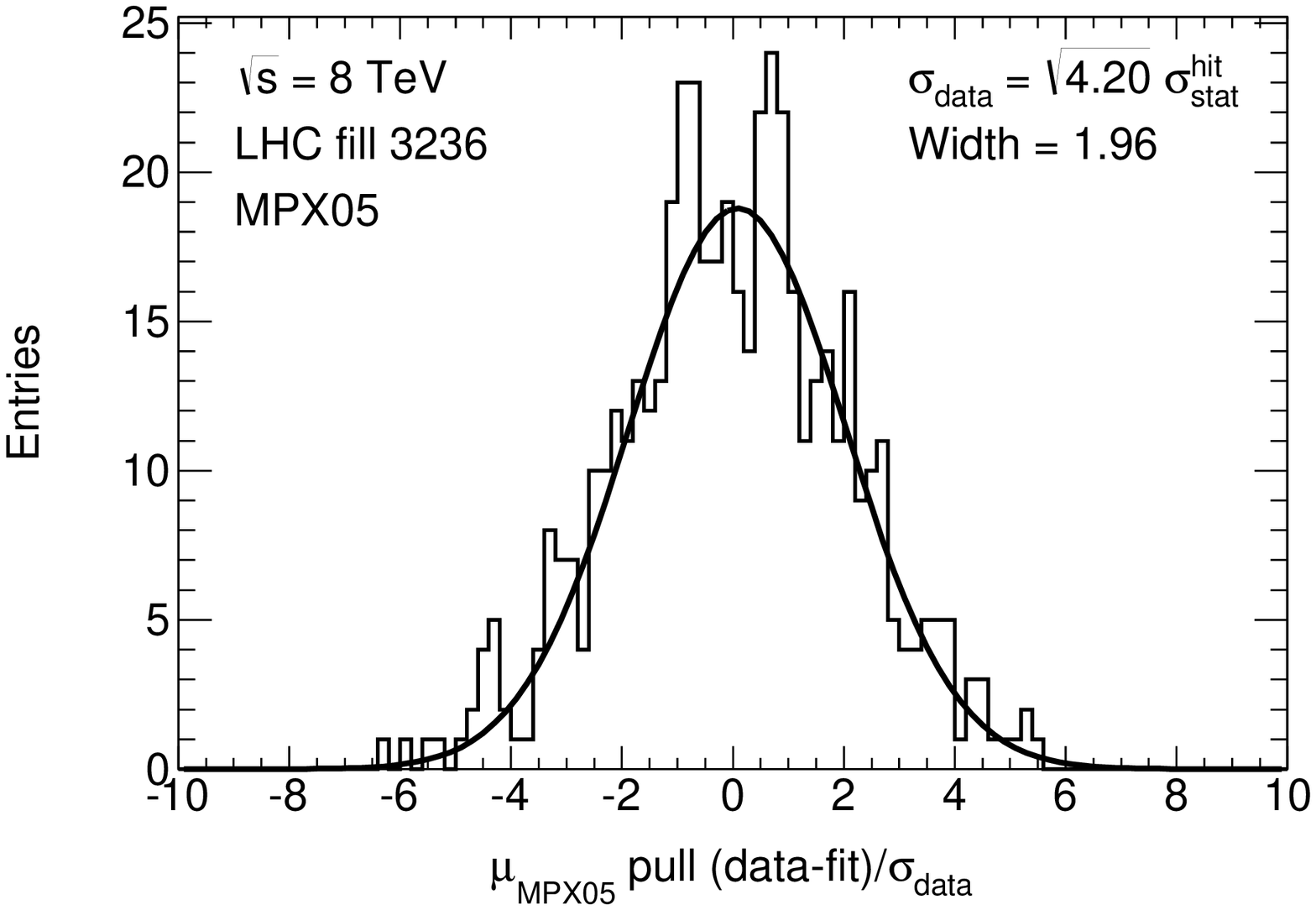}
\includegraphics[width=0.49\linewidth]{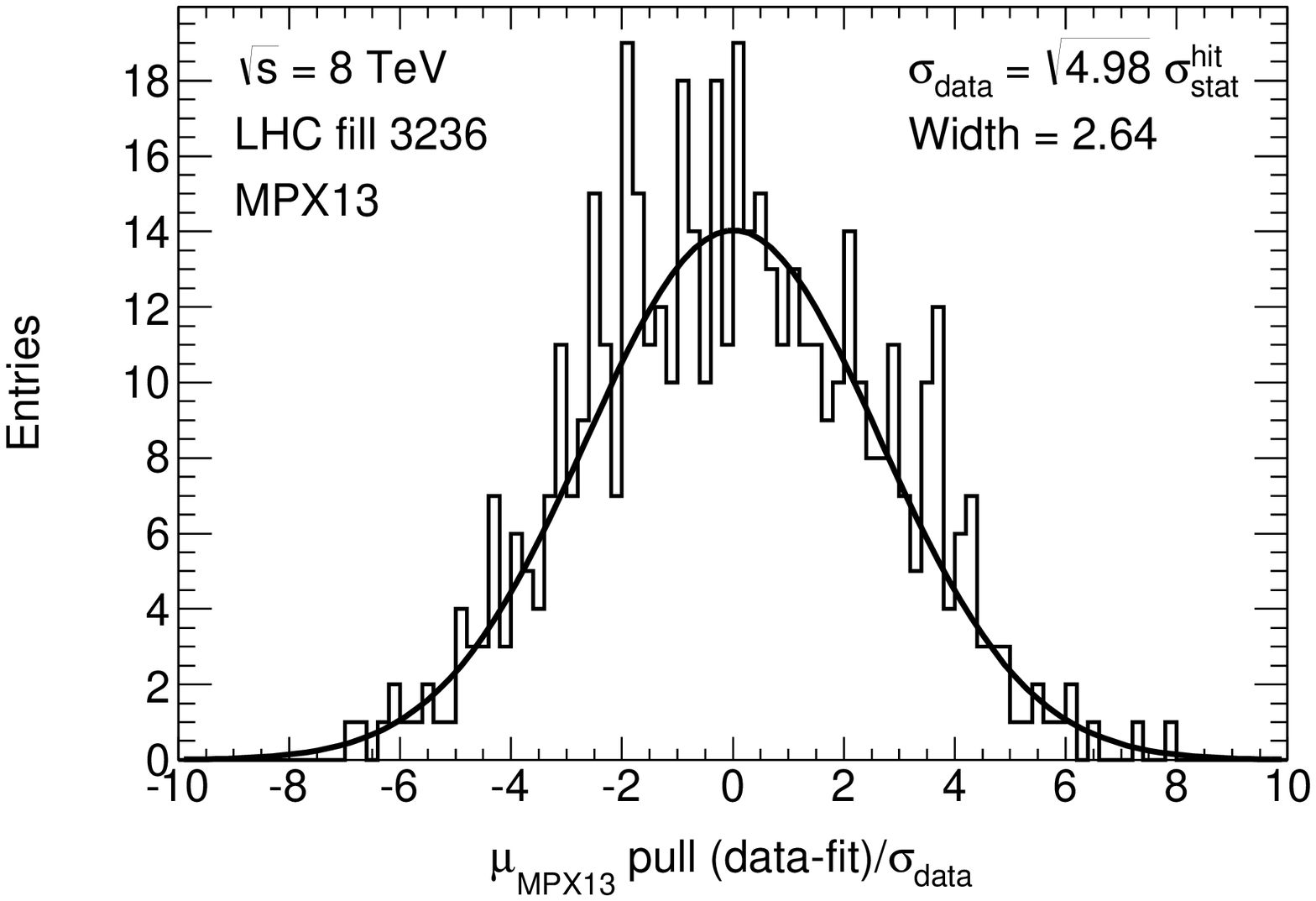}
\caption{Pull distributions defined as (data-fit)/$\sigma_{\rm data}$,
        where $\sigma_{\rm data} = \sqrt{R} \cdot \sigma_{\rm stat}^{\rm hit}$.
        The ratio $R=N_{\rm hit}/N_{\rm cl}$ is determined for each MPX
        device separately. LHC fill 3236.
}
\label{fig:pull}
\end{figure*}

This study of the LHC luminosity reduction allows us to investigate the contribution
of statistical and systematic uncertainties for each MPX device without relying 
on comparisons with other luminometers, only using MPX01 data due to its large data statistics.
The pull, (data-fit)/$\sigma_{\rm data}$,
is studied for each high-statistics MPX device individually with
$\sigma_{\rm data} = \sqrt{R} \cdot \sigma_{\rm stat}^{\rm hit}$.
The $R=N_{\rm hit}/N_{\rm cl}$ value is given in Table~\ref{tab:stat} for each
MPX device.
Figure~\ref{fig:pull} shows the pull distribution for MPX01-05 and MPX13.
As expected from the observed structures in Fig.~\ref{fig:fiterr},
the pull value 15.47 (defined as the width of a Gaussian fit) is large for MPX01.
The pull values vary between 2 and~3 for MPX02-05 and MPX13, given in Table~\ref{tab:pull}.
The uncertainty tends to be statistical in nature since the pull distributions 
are well described by Gaussians. 
While the hit statistics of MPX02-05 and MPX13 vary more
than a factor 10 (Table~\ref{tab:stat}) the pull is almost device independent. Thus,
the systematic uncertainty in addition to the hit statistics seems to also be  
of statistical nature.
Therefore, the total uncertainty (statistical and systematic) can be described
by scaling the statistical uncertainties such that the pull is unity. 

Comparing the pull value of MPX01 with those of the other MPX devices, 
the statistical precision of MPX01 is high enough that in addition to the 
intrinsic MPX uncertainties (scaling with the MPX hit statistics),
variations in the LHC luminosity which are not described by the fit function become visible,
shown in Fig.~\ref{fig:fiterr}.
This is corroborated by the observation that the pull distributions as a function of 
time show a structure only for MPX01.

\begin{figure*}[hbp]
\centering
\includegraphics[width=0.49\linewidth]{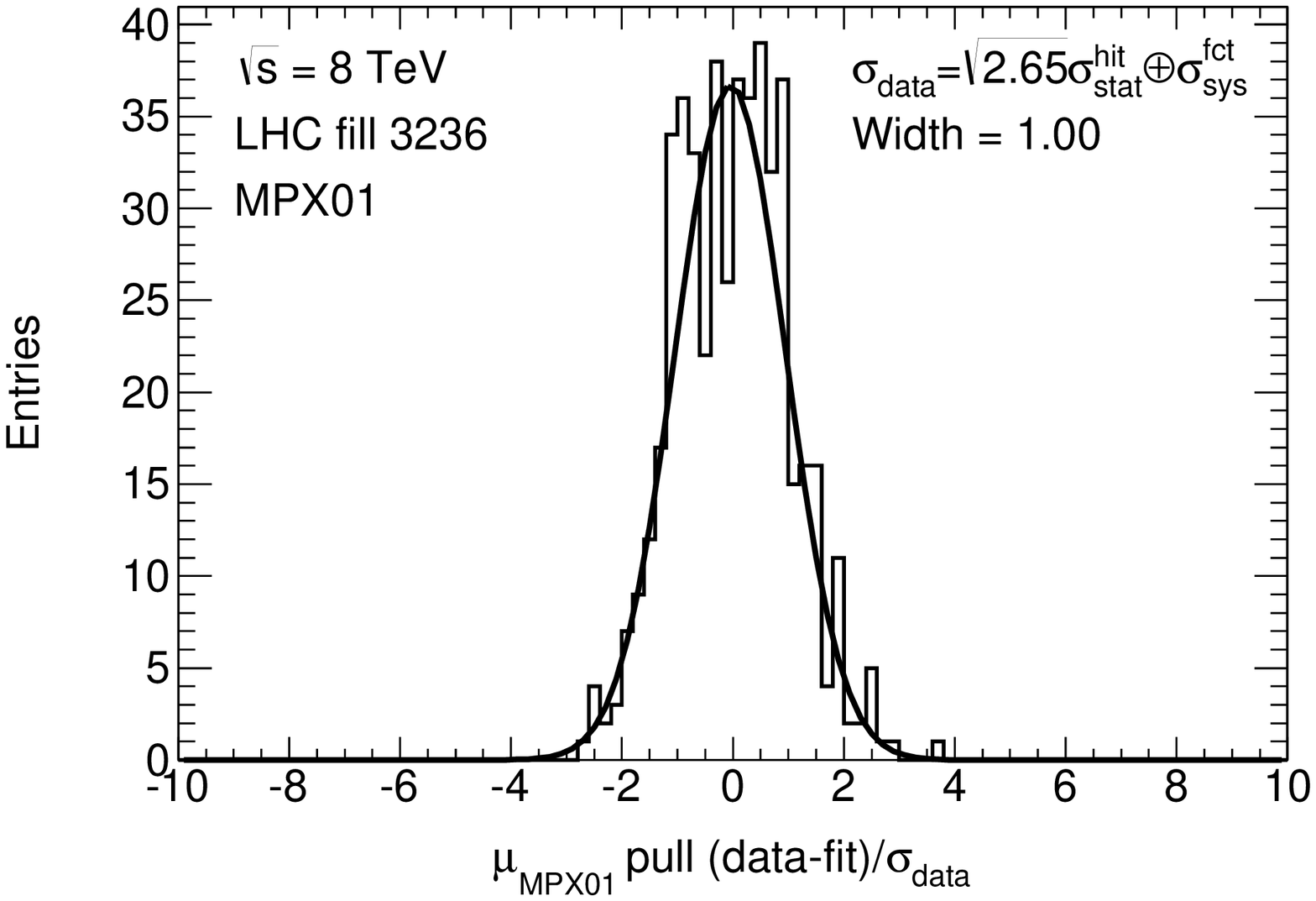}
\includegraphics[width=0.49\linewidth]{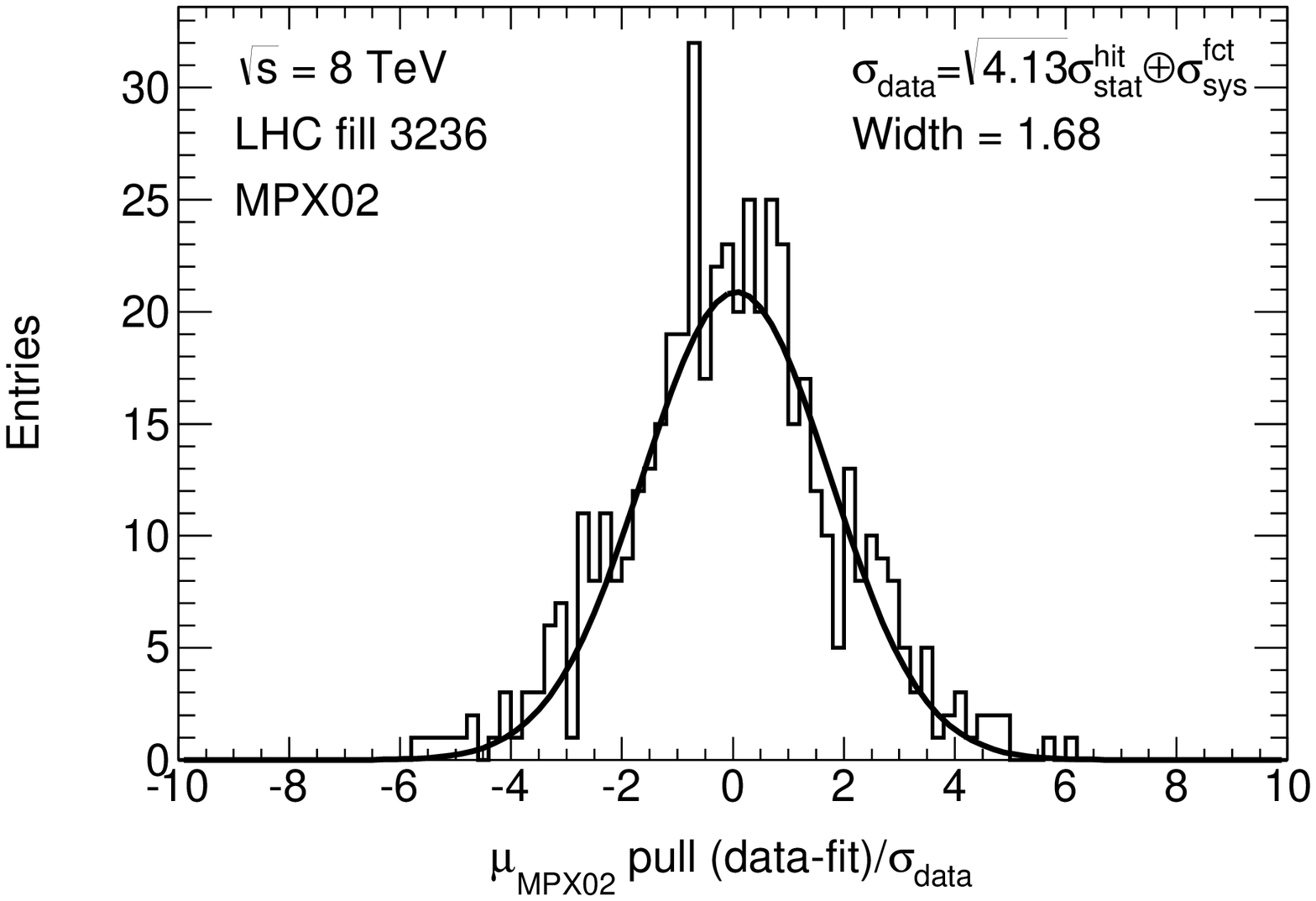}
\includegraphics[width=0.49\linewidth]{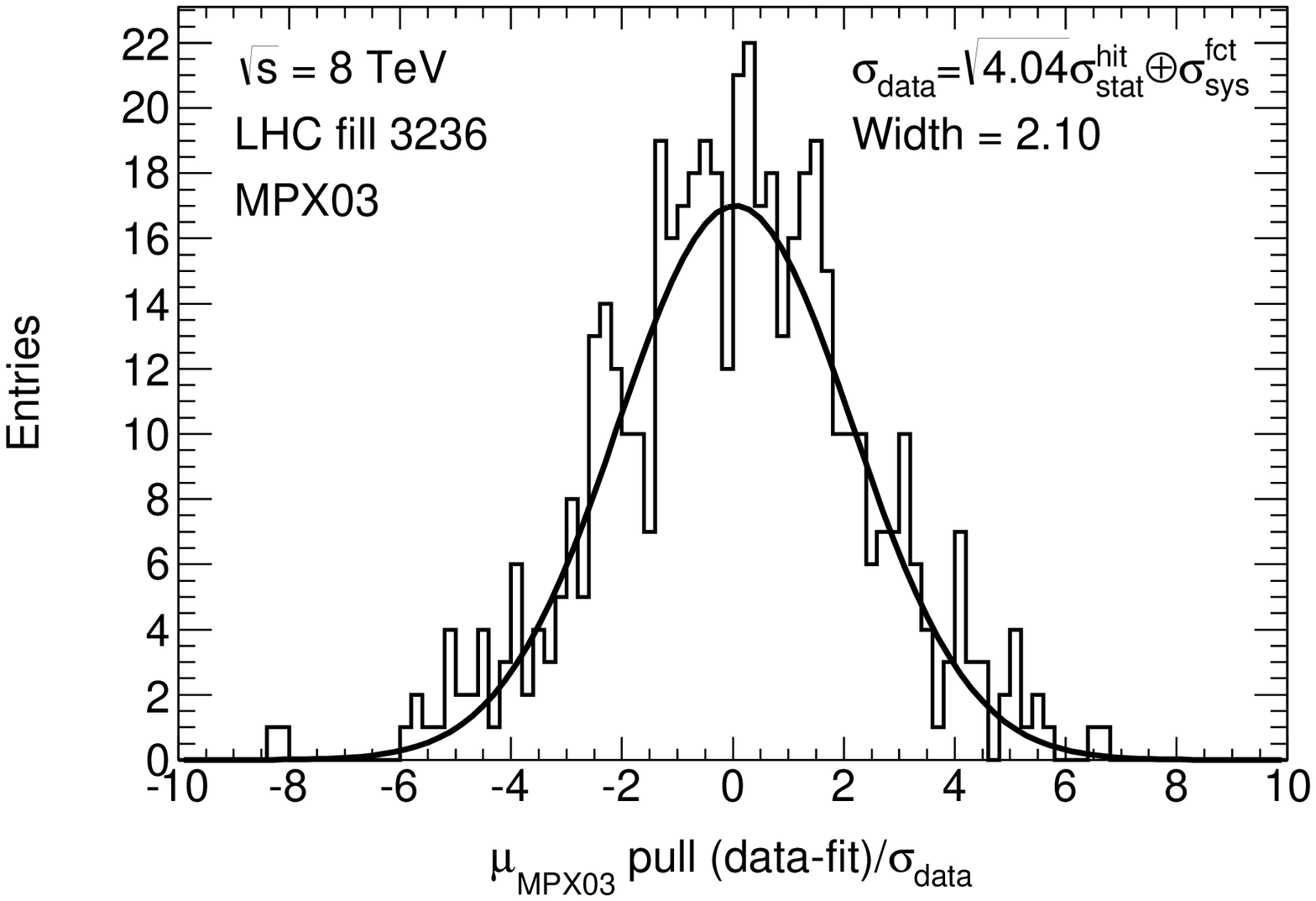}
\includegraphics[width=0.49\linewidth]{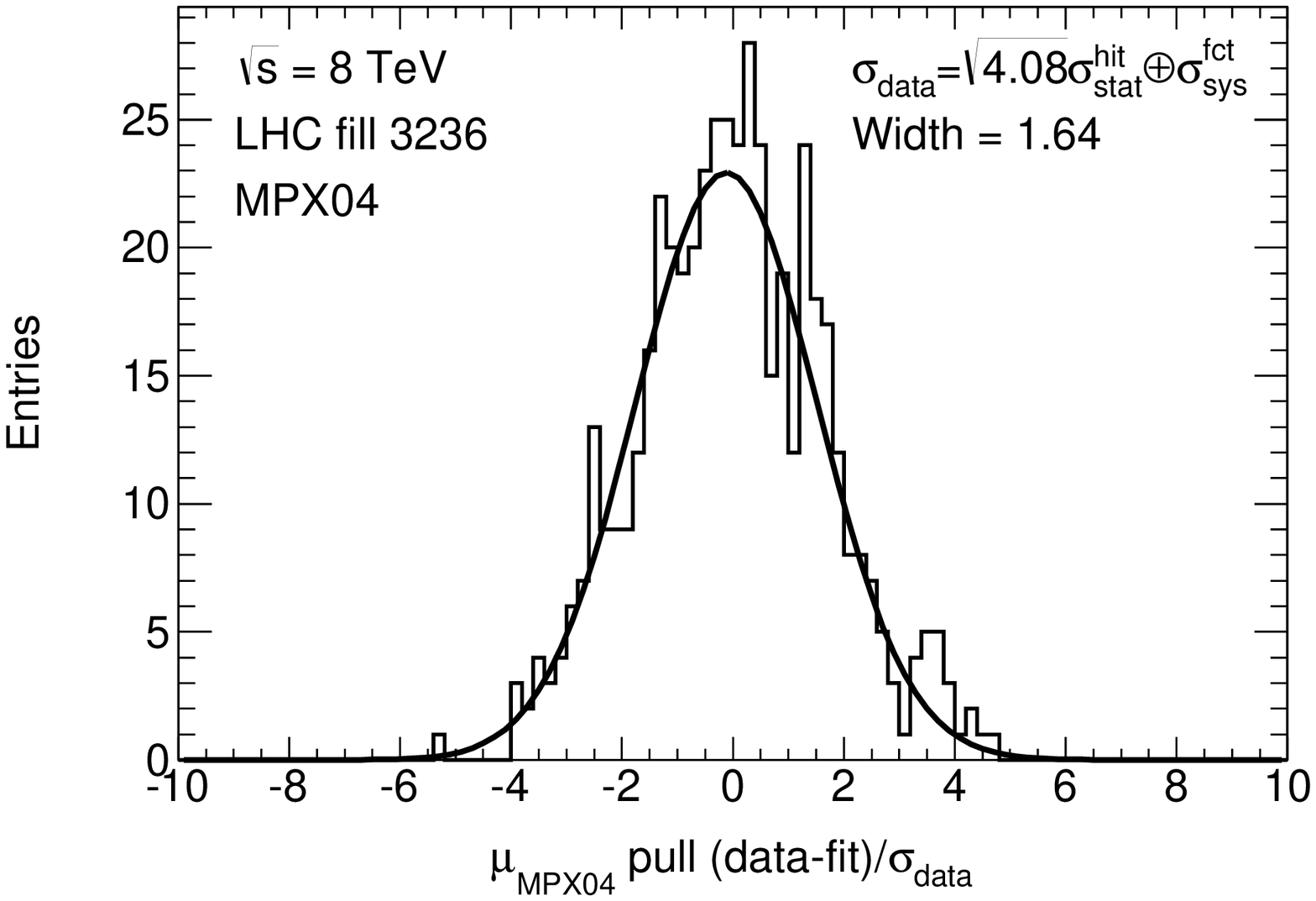}
\includegraphics[width=0.49\linewidth]{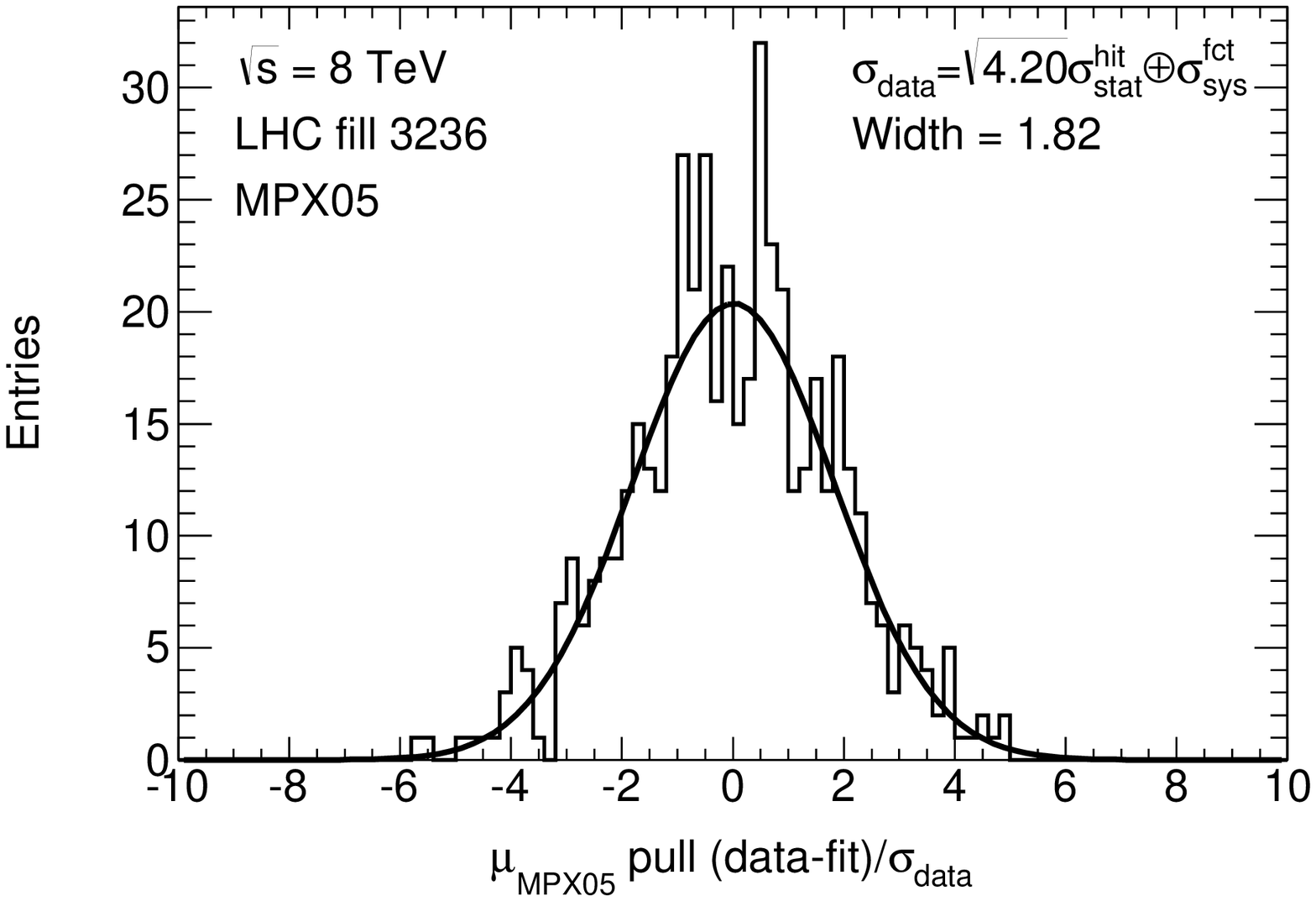}
\includegraphics[width=0.49\linewidth]{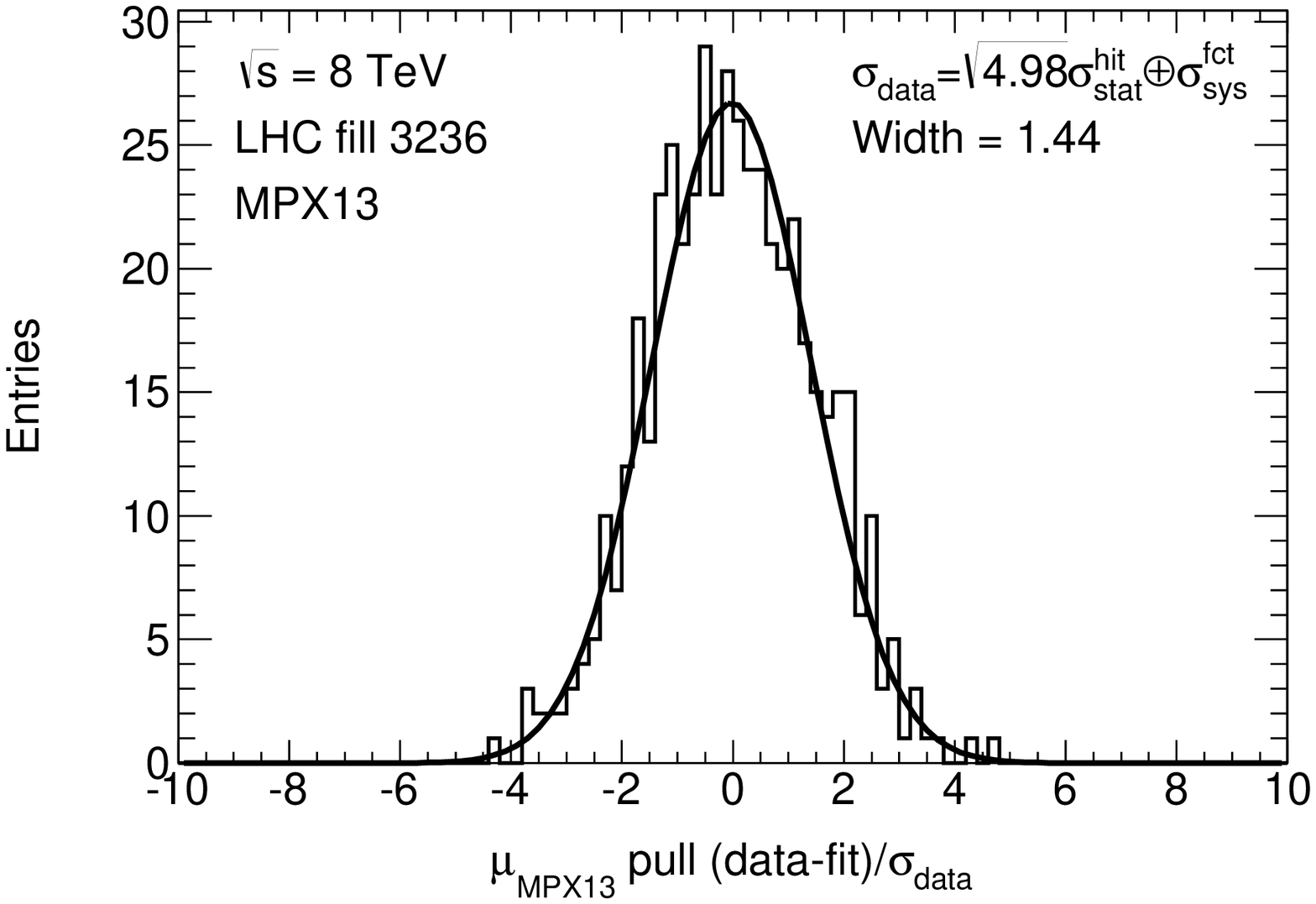}
\caption{Pull distributions defined as (data-fit)/$\sigma_{\rm data}$,
        where $\sigma_{\rm data} = \sqrt{R} \cdot \sigma_{\rm stat}^{\rm hit} 
        \oplus \sigma_{\rm sys}^{\rm fct}$.
        The ratio $R=N_{\rm hit}/N_{\rm cl}$ is determined for each MPX
        device separately and takes into account the systematic uncertainty from the 
        fluctuations not described by the fit function.
        LHC fill 3236.
}
\label{fig:pull2}
\end{figure*}

The fluctuations of luminosity not described by the fit function (Fig.~\ref{fig:fiterr})
are interpreted as systematic uncertainty. 
Therefore, an additional uncertainty corresponding to the RMS of the data deviations from the fit 
function,
$\Delta\mu^{\rm fct}_{\rm sys} = 0.03$, is added in quadrature to the
statistical uncertainty.
Figure~\ref{fig:pull2} shows the resulting pull distributions for MPX01-05 and MPX13,
and Table~\ref{tab:pull} summarizes the fit results.
The width of the pull distribution is close to unity as MPX01 was used to derive 
the systematic uncertainty.  
For the other MPX devices the pull varies between 1.44 and 2.10. 
Thus, as for the vdM scan, the fluctuations are about a factor two larger than expected 
from hit statistics alone for physics data-taking which is about 3000 times larger in luminosity. 

The effect of the varying hit/cluster ratio on the statistical 
evaluation is studied with a simple Monte Carlo simulation. 
The goal is to determine whether using the average hit/cluster 
ratio (2.65 for MPX01), rather than the varying ratios frame-by-frame,
could increase the pull value.
This study cannot be done with recorded data since the pixel occupancy is
too large during physics data-taking when the pull value is determined.
First, a pull distribution has been simulated using a Gaussian distribution 
with unity width.
Then, the statistical uncertainties are reduced by a constant value $\sqrt{2.65}$. 
Alternatively, they are reduced by a varying factor $\sqrt{N_{\rm hit}/N_{\rm cl}}$
frame-by-frame reflecting the MPX01 distribution in Fig.~\ref{fig:fill3316}.  
Both resulting pull distributions are fitted with a Gaussian. The former gives a width of
$\sqrt{2.65}=1.6$, as expected, and the latter 1.2. 
Therefore, there is no increase of the width when using the varying hit/cluster ratios.
Compared to the initial Gaussian distribution,
the pull distribution with varying hit/cluster ratios
shows a higher peak and higher populated side-bands.
Thus, it can be excluded that the varying hit/cluster ratio is the source 
of the observed pull values for MPX02-05 and MPX13 ranging between 
1.44 and 2.10 (Table~\ref{tab:pull}).

The following observations could explain the about twice as large luminosity fluctuations
compared to the statistical expectations from the hit statistics alone:
\begin{itemize}
\item The average hit/cluster ratio, in particular for MPX01, could be larger as noisy pixels are
not excluded when the ratio is determined.
\item The assumption that one cluster corresponds to one particle bears an uncertainty involving the
cluster definition.
\item Some particles are reflected and could pass the MPX sensor more than once.
\item The showering of particles in the ATLAS detector material increases the number of particles passing
the MPX devices, thus the number of independent particles to be used in the 
statistical determination of the expected uncertainty would be smaller. The determination of this
effect would require a full simulation of the  material distribution of the ATLAS detector in front of
each of the 
MPX devices which is beyond the scope of this study. However, the same pull value of about two, observed 
for all six MPX devices used in the hit study, indicates that the effect is independent 
of the MPX device and its position.
\end{itemize}
These effects could lead to a smaller number of independent objects 
impacting the luminosity measurement. 
Thus, the statistical uncertainty is presumably larger than from hit counting alone.
A reduction of the number of independent objects by 
about a factor four increases  the statistical uncertainty by a factor two, 
and would reduce the pull distribution width to unity.

\begin{figure*}[htbp]
\centering
\includegraphics[width=0.49\linewidth]{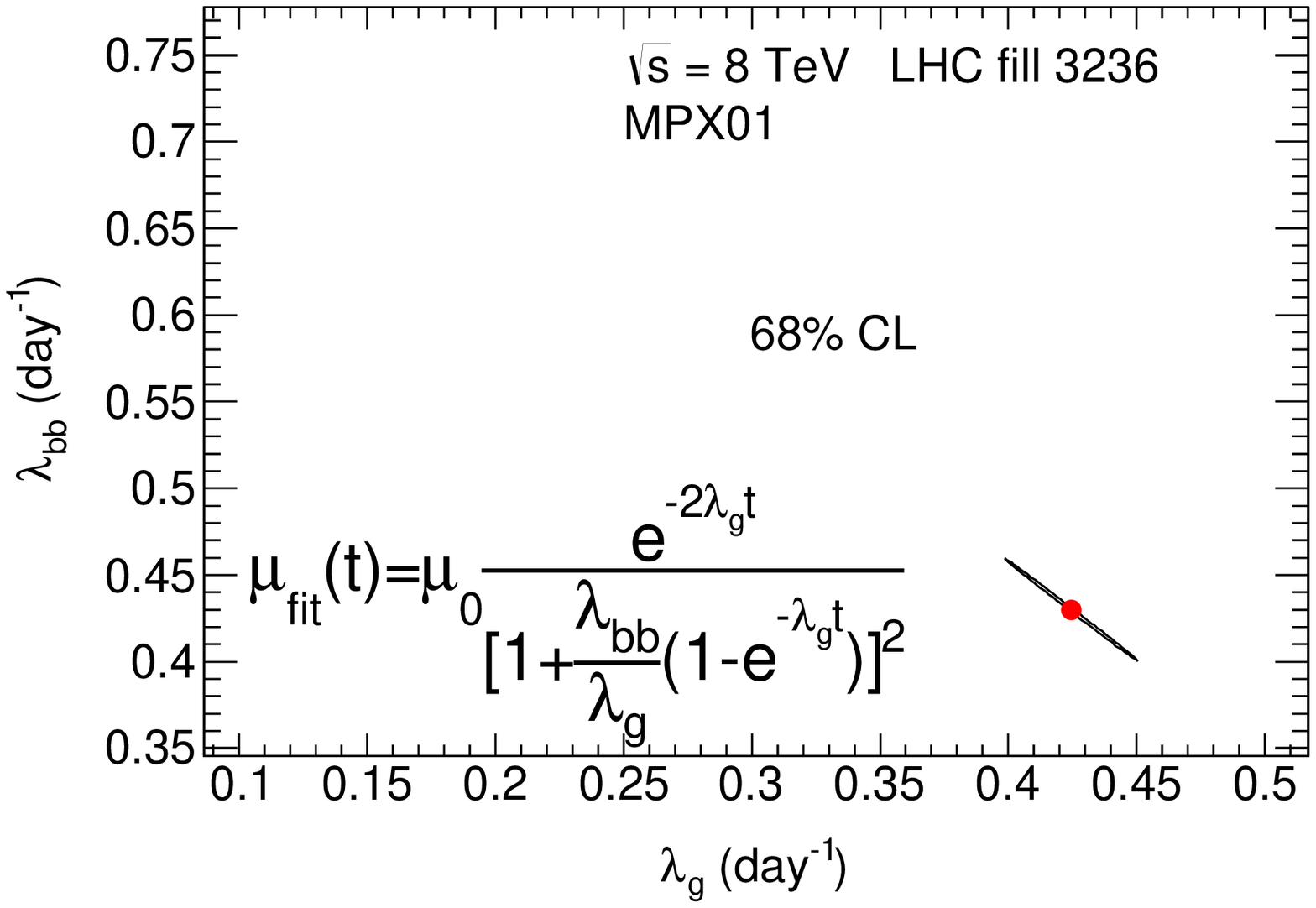}
\includegraphics[width=0.49\linewidth]{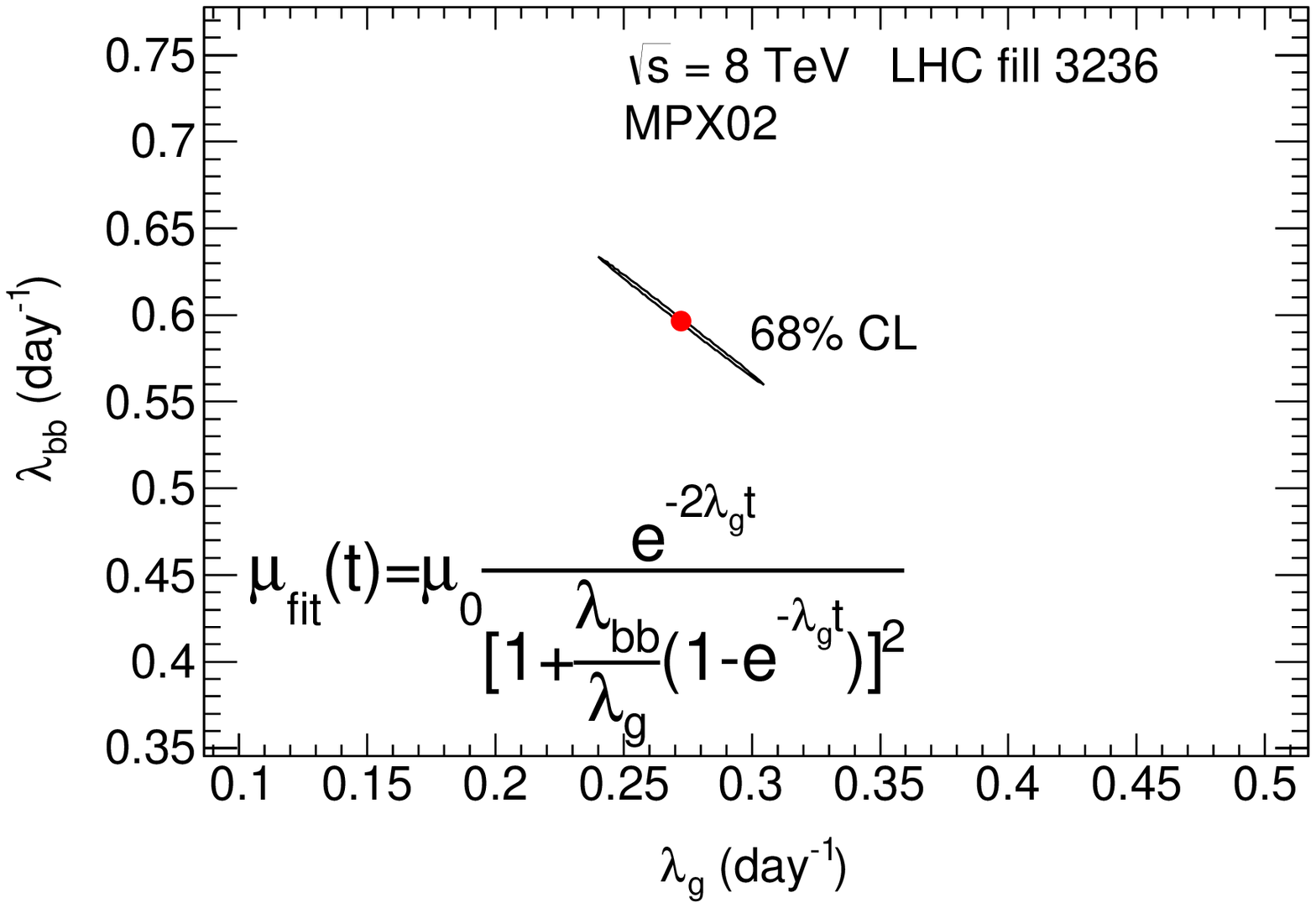}
\includegraphics[width=0.49\linewidth]{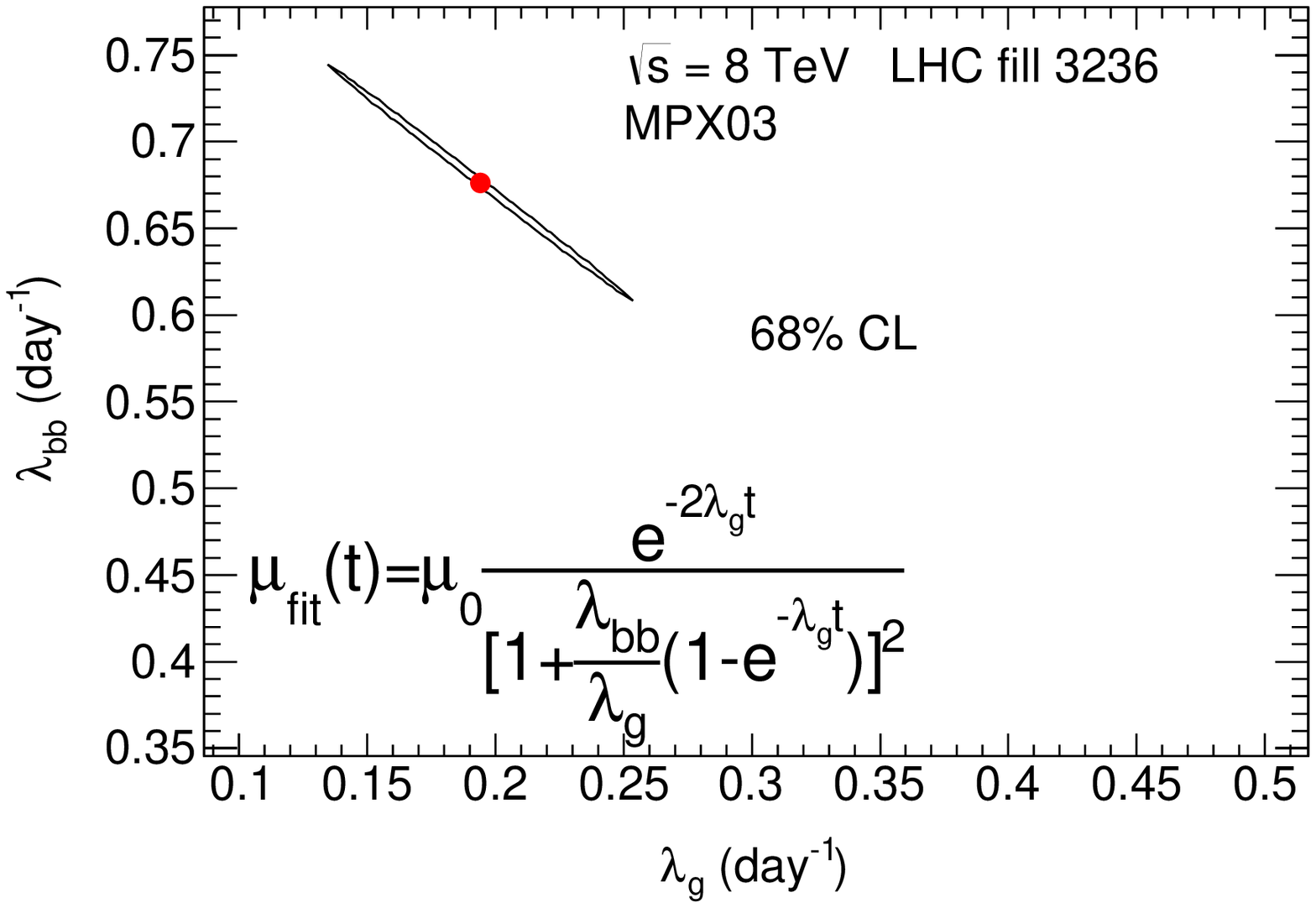}
\includegraphics[width=0.49\linewidth]{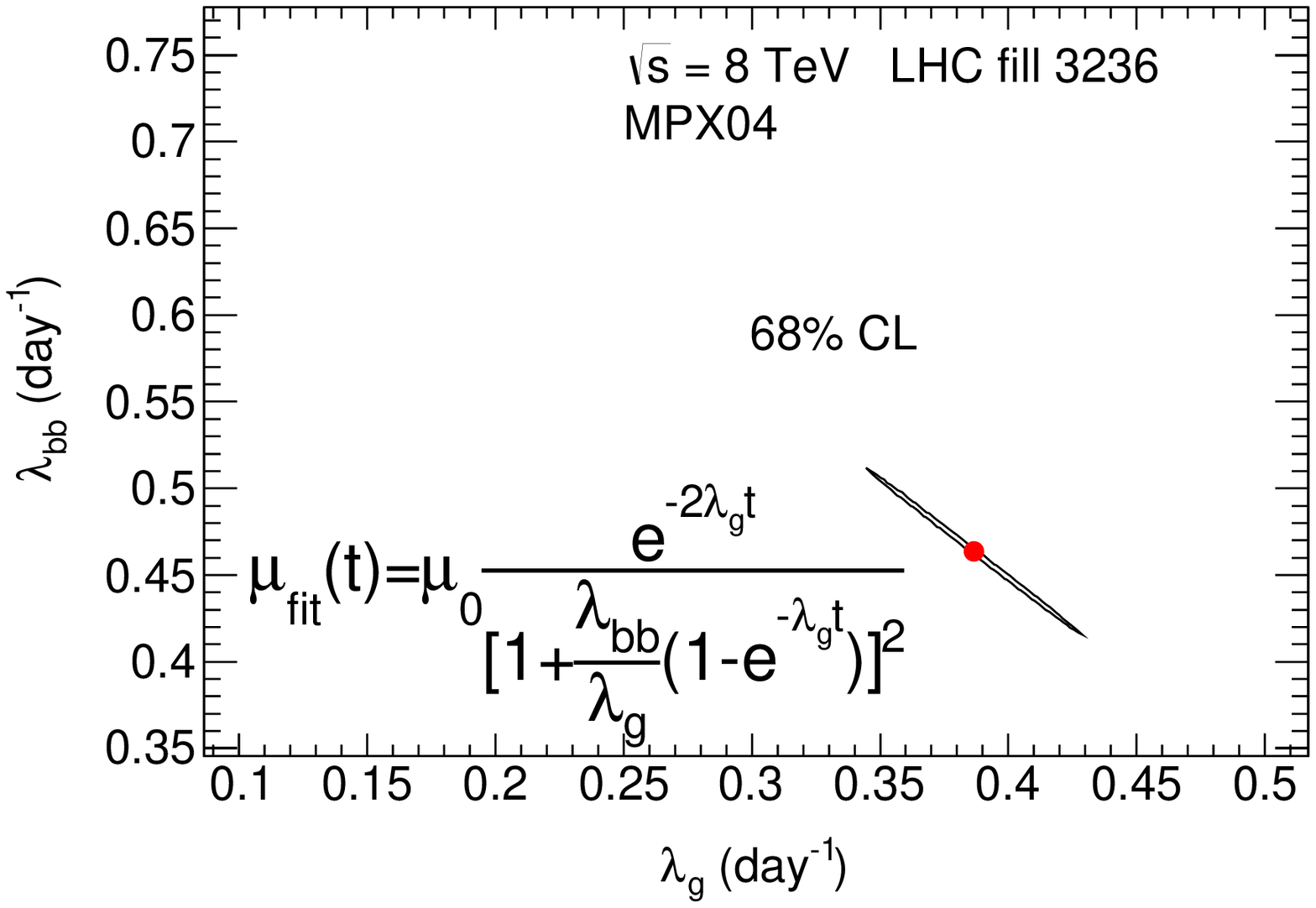}
\includegraphics[width=0.49\linewidth]{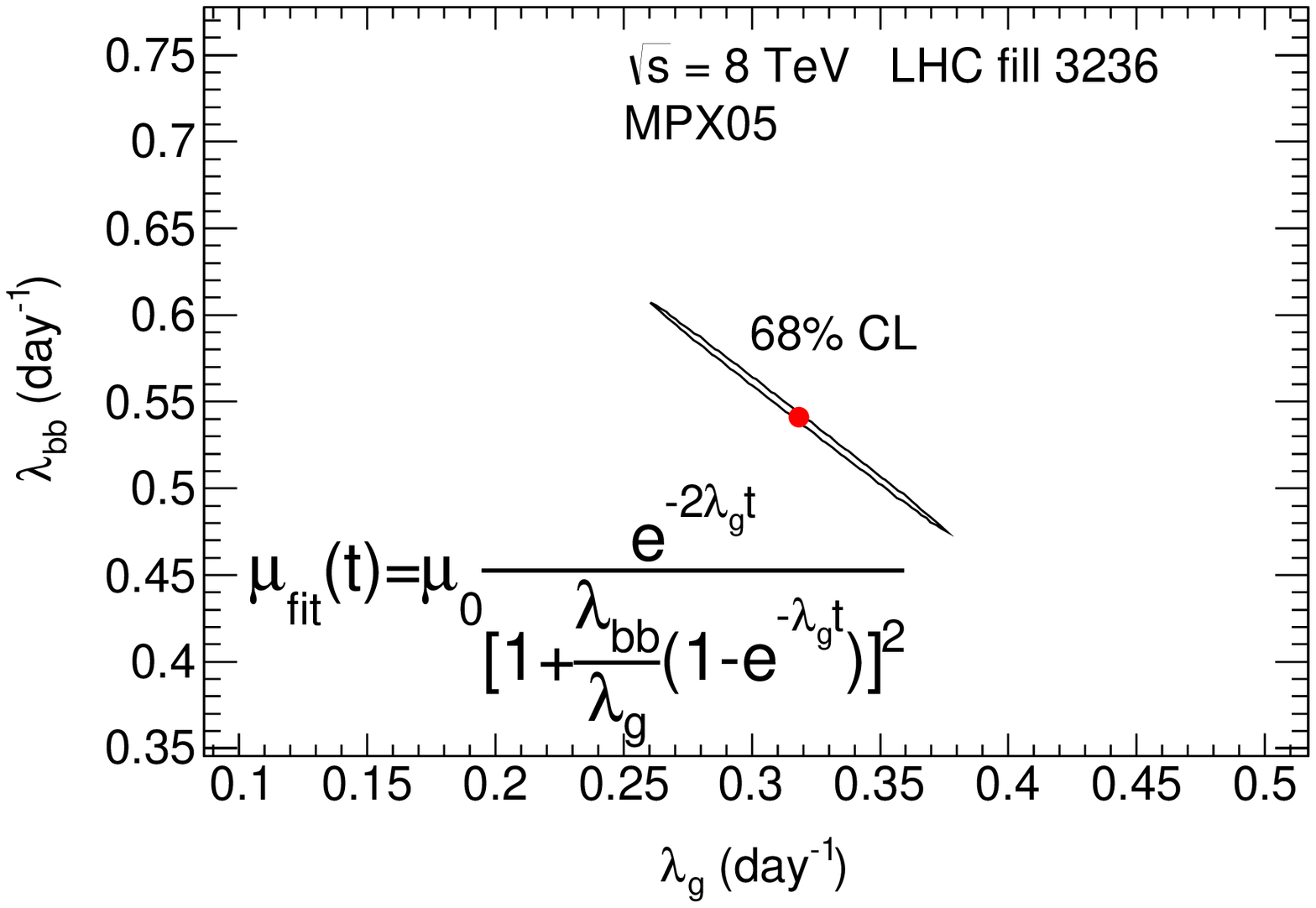}
\includegraphics[width=0.49\linewidth]{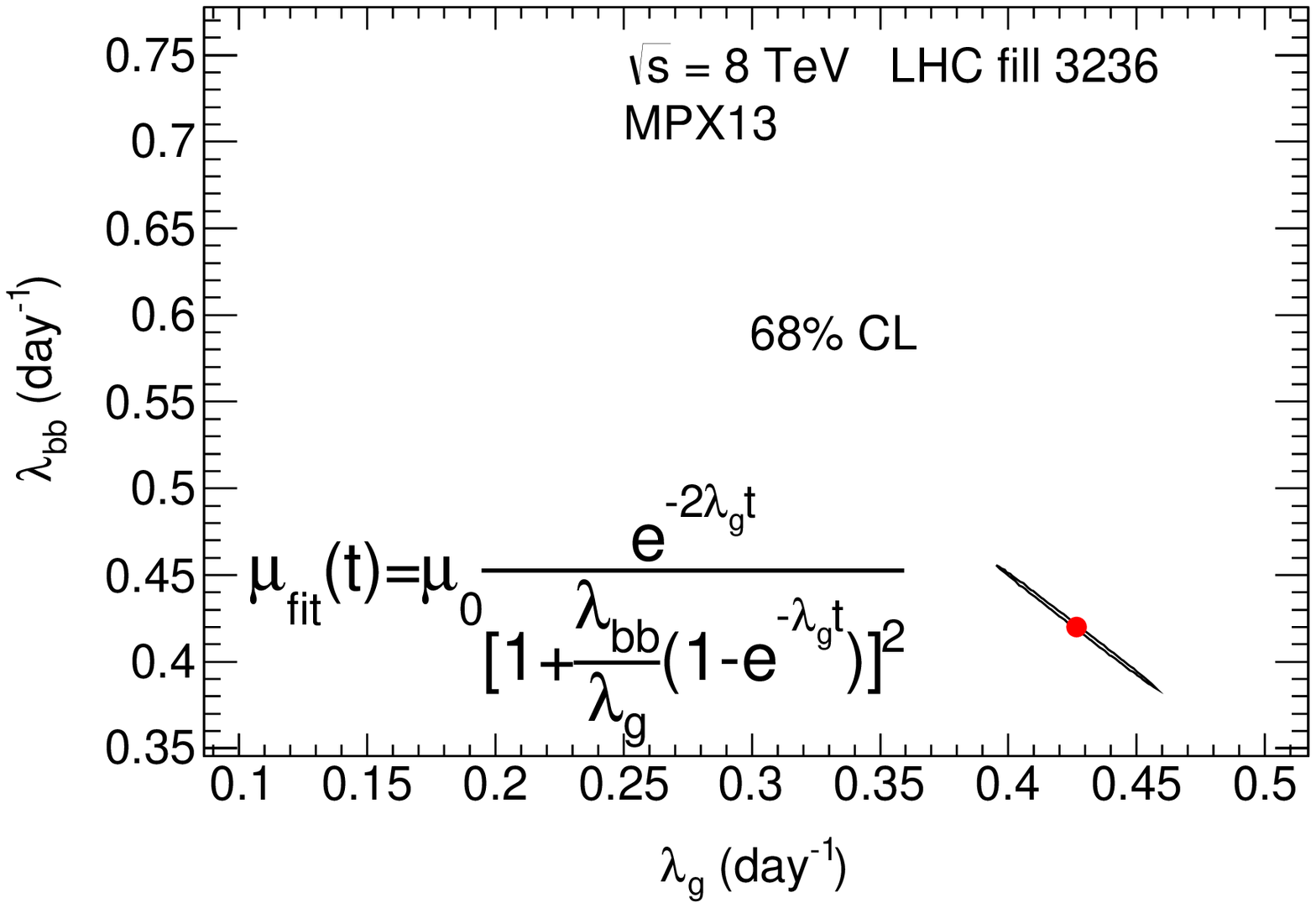}
\caption{Contour curves of $\lambda_{\rm g}$ and $\lambda_{\rm bb}$, given at 68\% CL,
         for the fit results of the average number of interactions per bunch crossing 
         as a function of time seen by MPX01-05 and MPX13.
         The dot in the center of the contour indicates the fit values.
         The hit statistical uncertainties and systematic 
         uncertainties from luminosity fluctuations not described by
         the fit function are added in quadrature, given in Table~\ref{tab:uncert}.
         The $\chi^2$ values are 528, 1811, 2666, 1415, 1720, and 1078, for 
         MPX01-05 and MPX13, respectively, for 499 degrees of freedom.
         LHC fill 3236.
        }
\label{fig:contour_ind}
\end{figure*}

\begin{figure*}[htbp]
\centering
\includegraphics[width=0.49\linewidth]{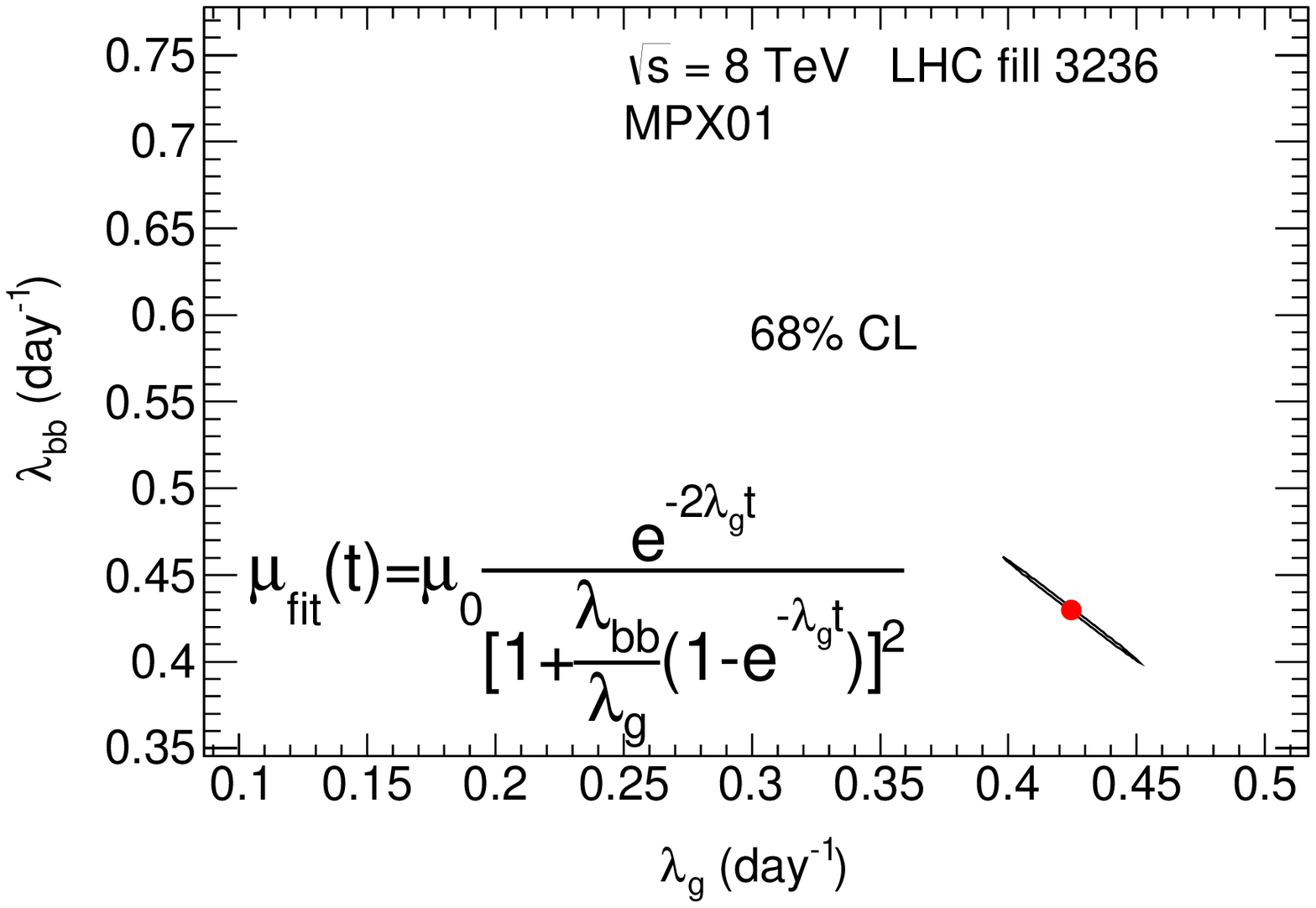}
\includegraphics[width=0.49\linewidth]{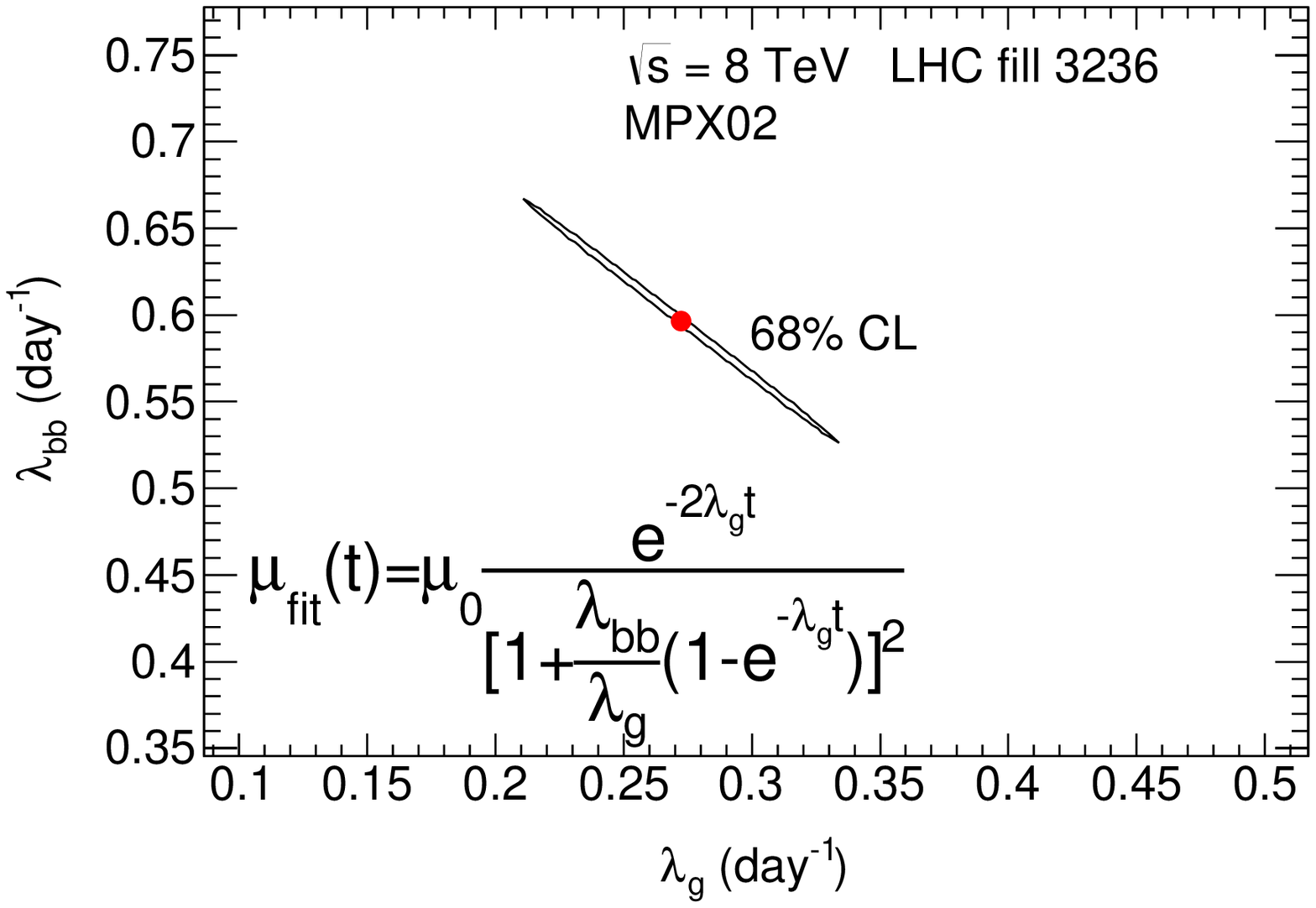}
\includegraphics[width=0.49\linewidth]{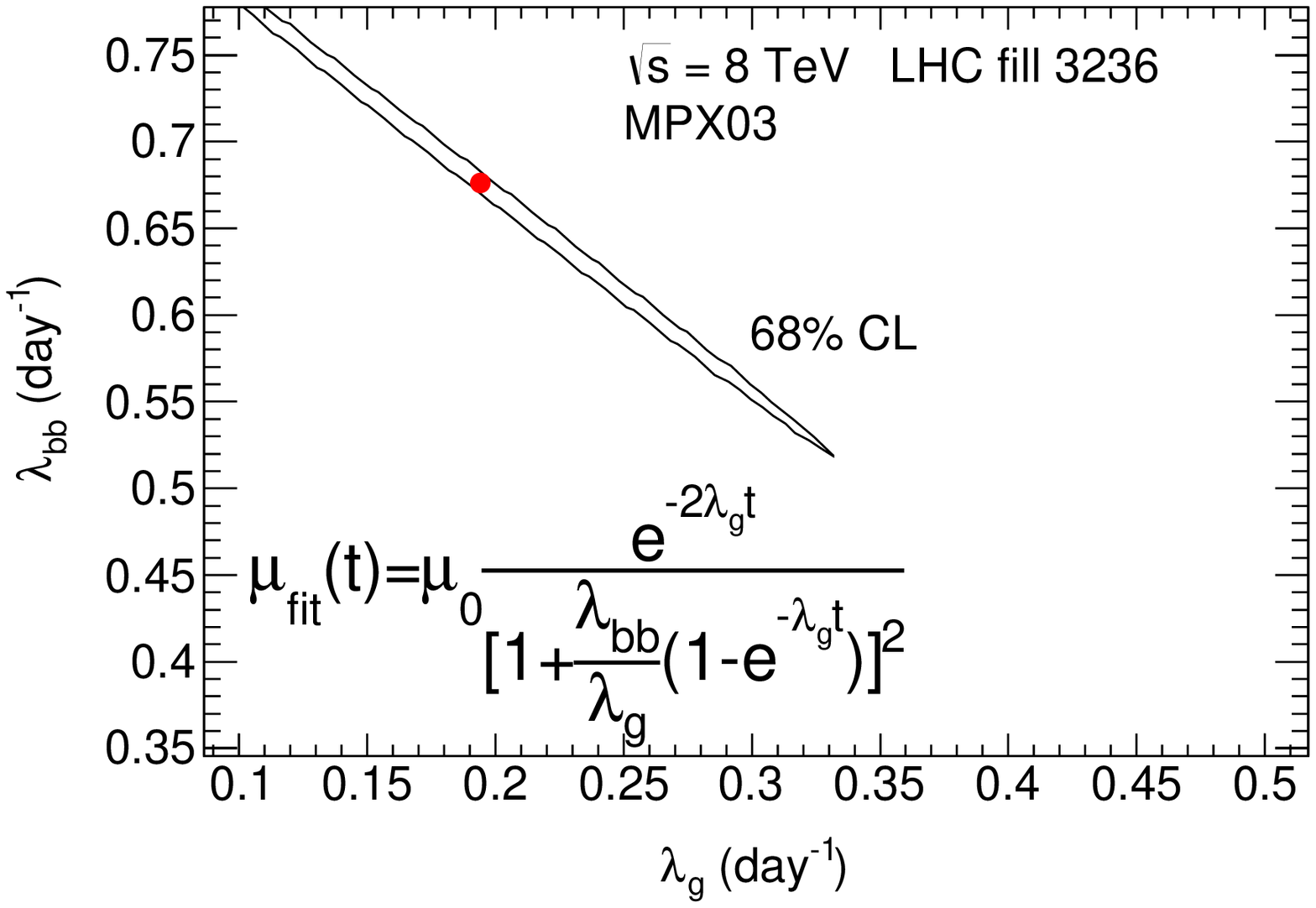}
\includegraphics[width=0.49\linewidth]{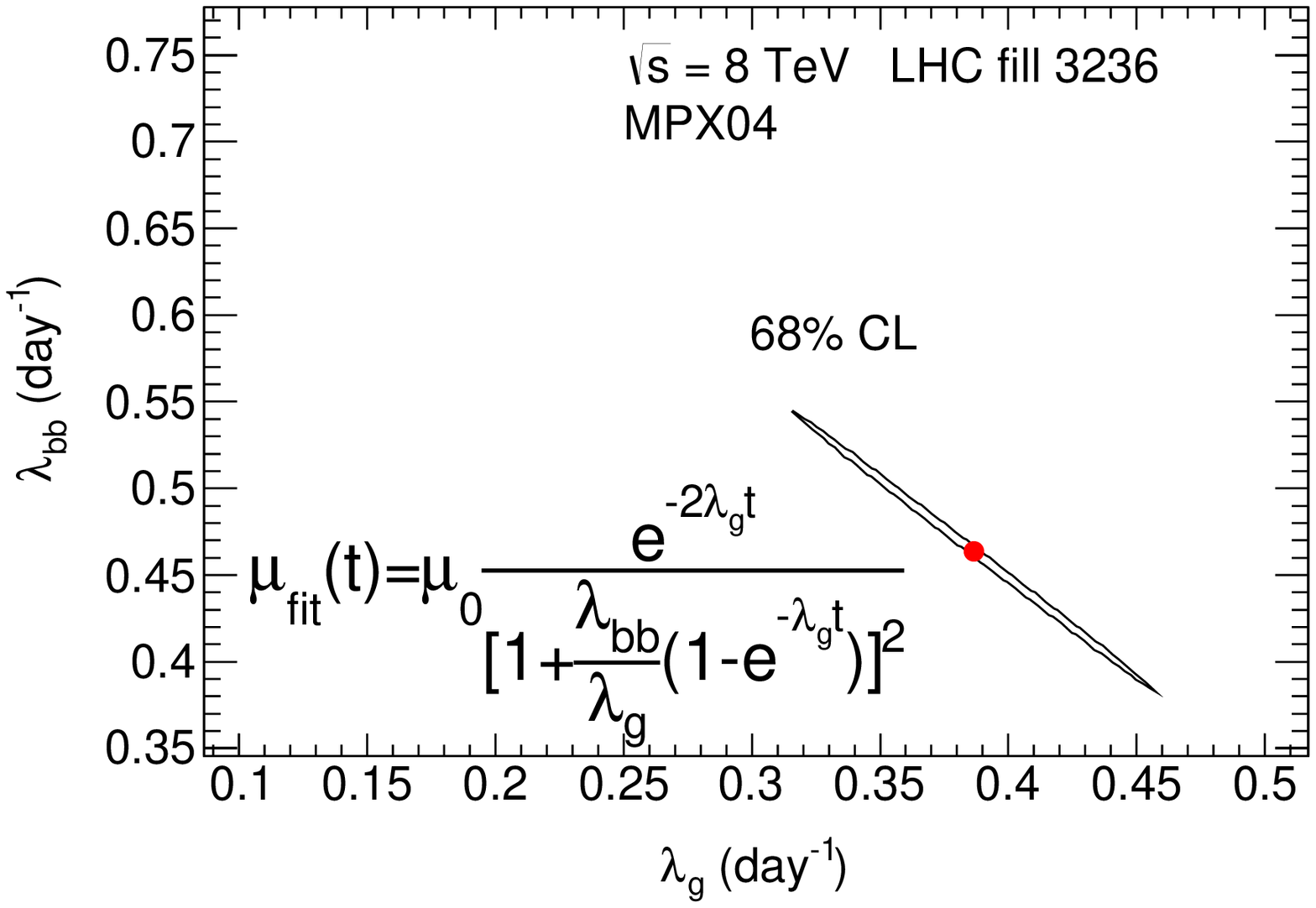}
\includegraphics[width=0.49\linewidth]{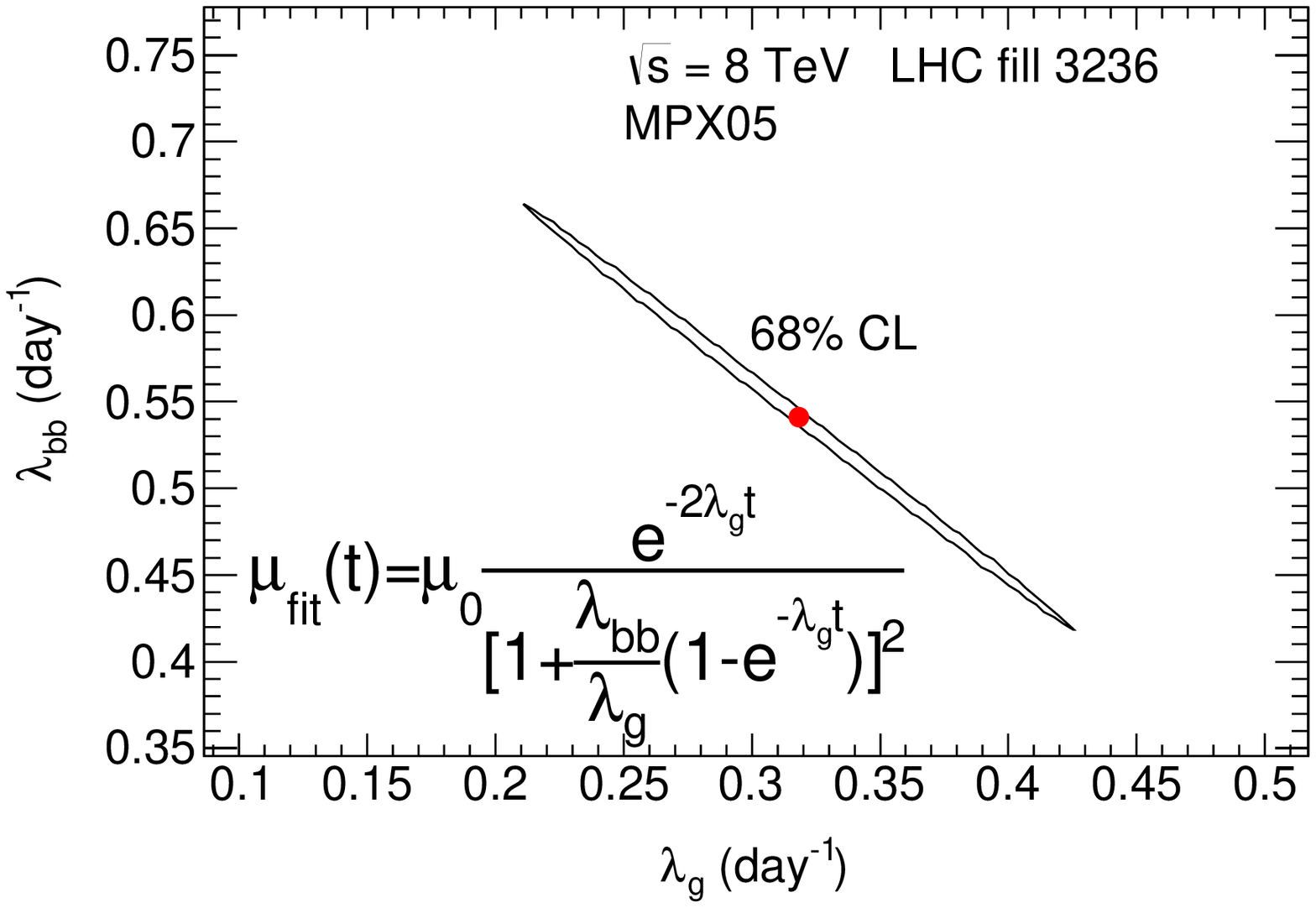}
\includegraphics[width=0.49\linewidth]{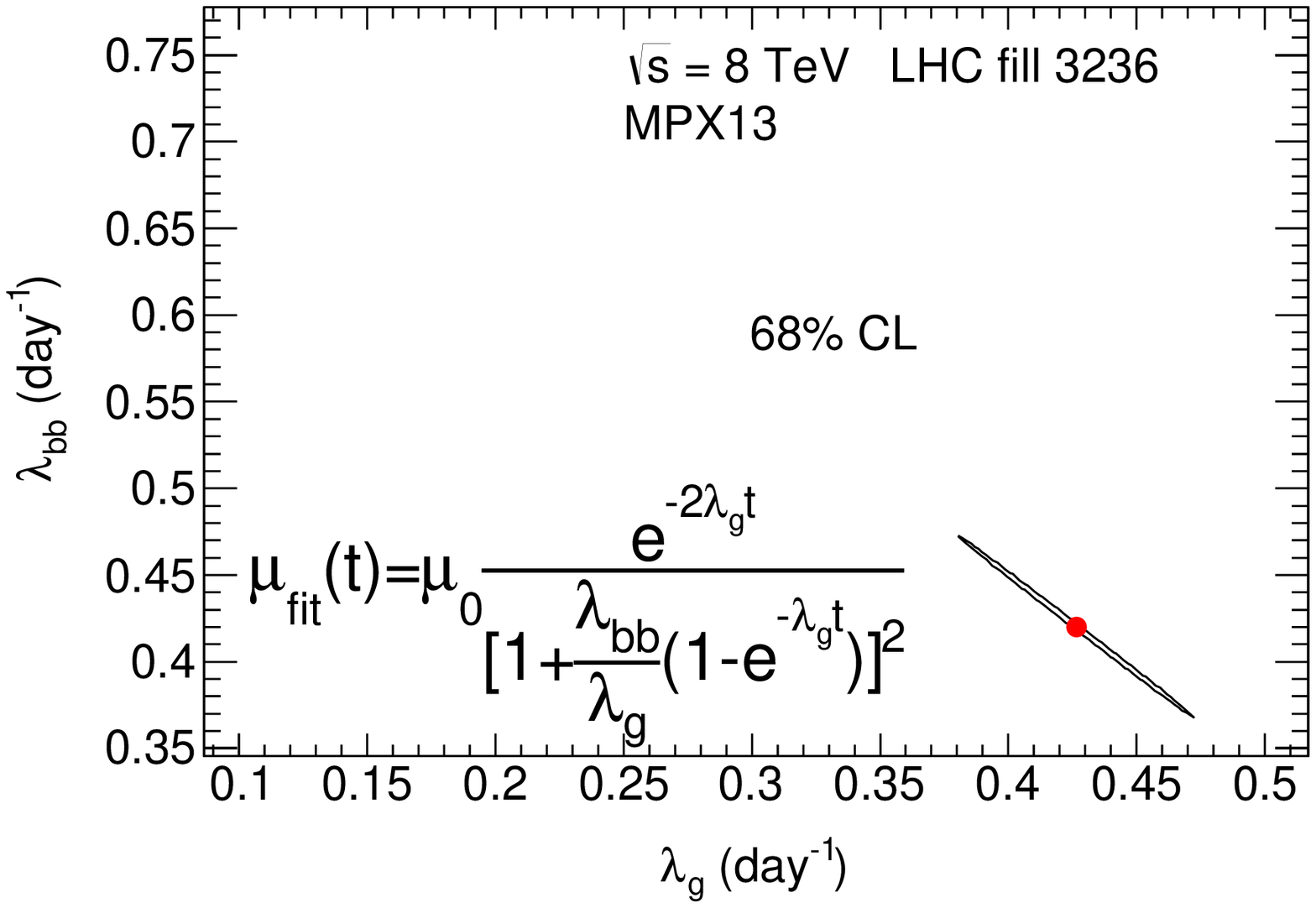}
\caption{Contour curves of $\lambda_{\rm g}$ and $\lambda_{\rm bb}$, given at 68\% CL,
         for the fit results of the average number of interactions per bunch crossing  
         as a function of time seen by 
         MPX01-05 and MPX13.
         The dot in the center of the contour indicates the fit values.
         The hit statistical uncertainties and systematic uncertainties from 
         luminosity fluctuations not described by the fit function are added 
         in quadrature, given in Table~\ref{tab:uncert}, and then scaled to 
         yield $\chi^2/{\rm ndf}=1$. LHC fill 3236.
        }
\label{fig:contour_chi_1}
\end{figure*}

\begin{figure}[htbp]
\centering
\vspace*{0.2cm}
\includegraphics[width=\linewidth]{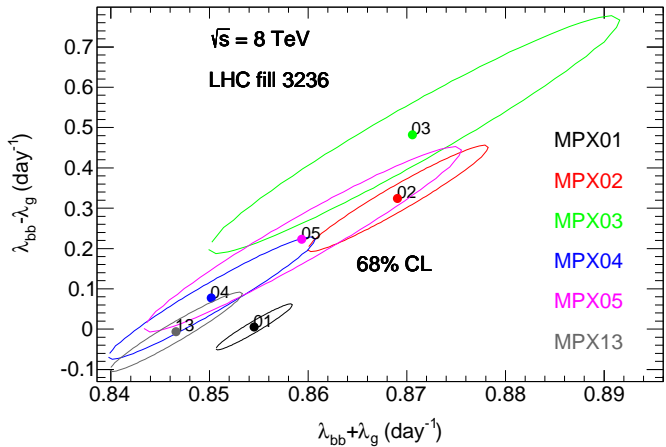}
\vspace*{-0.5cm}
\caption{Contour curves of 
  $(\lambda_{\rm bb} - \lambda_{\rm g})$
  and $(\lambda_{\rm bb} + \lambda_{\rm g})$, given at 68\% CL,
         for the fit results of the average number of interactions per bunch crossing  
         as a function of time seen by 
         MPX01-05 and MPX13.
         The dots in the center of the contours indicate the fit values.
         The hit statistical uncertainties and systematic uncertainties from 
         luminosity fluctuations not described by the fit function are added 
         in quadrature, given in Table~\ref{tab:uncert}, and then scaled to 
         yield $\chi^2/{\rm ndf}=1$.
         LHC fill 3236.
        }
\label{fig:contour_summary}
\vspace*{0.4cm}
\end{figure}

In summary, Fig.~\ref{fig:contour_ind} shows the contour plot of
$\lambda_{\rm g}$ and $\lambda_{\rm bb}$, given at 68\% CL, for the 
hit statistical uncertainties and systematic uncertainties from 
luminosity fluctuations not described by the fit function. They are added in 
quadrature, given in Table~\ref{tab:uncert}.
Figure~\ref{fig:contour_chi_1} shows the corresponding plots for $\chi^2/{\rm ndf}=1$.
The variations of the fit values are well described by the contours.
Owing to the strong anti-correlation between 
$\lambda_{\rm bb}$ and $\lambda_{\rm g}$ a transformation of the fit results is 
performed. 
The $(\lambda_{\rm bb} - \lambda_{\rm g})$ 
versus $(\lambda_{\rm bb} + \lambda_{\rm g})$ plane is used in order to 
illustrate the relative uncertainty between the fit values.
The comparative sensitivities of the MPX devices  at 68\% CL are summarised in
Fig.~\ref{fig:contour_summary}. 

\begin{figure}[htbp]
\centering
\includegraphics[width=\linewidth]{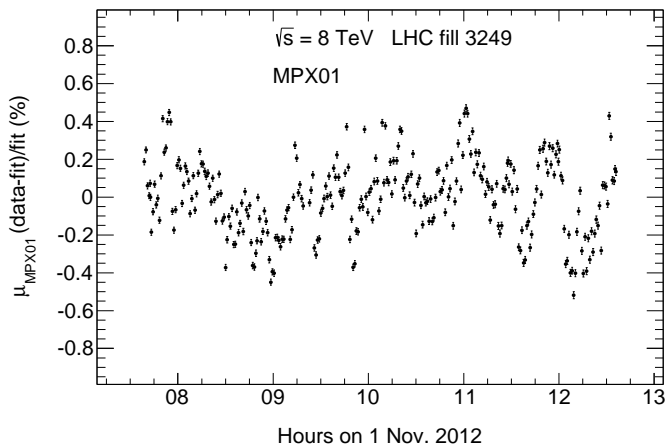}
\vspace*{-0.5cm}
\caption{Deviations between data and the fit of the average number of interactions per 
         bunch crossing as a function of time seen by MPX01.
         The relative deviations between data and fit have an RMS of 0.2\%.
         The statistical uncertainties $\Delta\mu/\mu$ per data point are indicated and
         vary from about $0.0099 \sqrt{2.65}$\% to about $0.0116 \sqrt{2.65}$\%
         where the factor $2.65$ is the averaged ratio of hits per 
         interacting particle.
         The apparent structure is similar to the one previously discussed in the text.
         LHC fill 3249.
}
\label{fig:fill3249res}
\end{figure}

The analysis performed with LHC fill 3236 data, has been 
repeated with 
LHC fill 3249 data, taken 31 October -- 1 November 2012, for $\mu=13.4$ to $\mu=9.8$. 
The MPX01 results from LHC fill 3249 lead to a relative precision below 
RMS 0.3\% (Fig.~\ref{fig:fill3249res}) in agreement with the results from
LHC fill 3236.
Table~\ref{tab:uncert3249} gives the relative statistical uncertainties, and 
lists the total uncertainties for  RMS 0.2\% (obtained in LHC fill 3249) and
RMS 0.3\% (obtained for LHC fill 3236).
The corresponding pull values for statistical and total uncertainties are given 
in Table~\ref{tab:pull3249}.

\clearpage

\begin{table*}[htbp]
\small
  \caption{Widths of the Gaussian fits to the pull distributions for MPX01-05 and MPX13.
          The widths are given for uncertainties resulting from the number of hits scaled
          by a factor $\sqrt{N_{\rm hit}/N_{\rm cl}}$ for each MPX device,
          given in Table~\ref{tab:stat}.
          The widths are also given for statistical and systematic 
          uncertainties added in quadrature. The systematic uncertainties 
          result from luminosity fluctuations not described by the fit function.
          Table~\ref{tab:uncert} indicates the sizes of the uncertainty applied.
          LHC fill 3236.
}
  
 \centering
\renewcommand{\arraystretch}{1.3} 
    \begin{tabular}{ccccccccc}
\hline\hline
MPX                              & 01    & 02    & 03    & 04    & 05    & 13     \\\hline
Width pull stat. uncert.       &15.47  & 3.05  & 2.42  & 2.18  & 1.96  & 2.64   \\
Width pull total uncert. (0.03 sys.)& 1.00  & 1.68  & 2.10  & 1.64  & 1.82  & 1.44   \\
\hline\hline
\end{tabular}
\label{tab:pull}
\end{table*}

\begin{table*}[htbp]
\small
  \caption{Statistical uncertainty ranges for the data points in Fig.~\ref{fig:fit}.
          The uncertainty resulting from the number of hits is scaled by a factor 
          $\sqrt{N_{\rm hit}/N_{\rm cl}}$,
          given in Table~\ref{tab:stat}. 
          The values are given for the beginning and end of the fit range.
          Also shown are the total uncertainties calculated adding in quadrature the statistical 
          and systematic uncertainties,
          where the systematic uncertainties result from luminosity fluctuations 
          not described by the fit function ($\sigma_{\rm sys}^{\rm fct} =0.03$),
          shown in Fig.~\ref{fig:fiterr}.
          LHC fill 3236.
}
 \centering
\renewcommand{\arraystretch}{1.3} 
    \begin{tabular}{ccccccccc}
\hline\hline
MPX           & 01    & 02    & 03    & 04    & 05    & 13     \\\hline
$\Delta\mu$ stat.  &0.0023-0.0017 & 0.0249-0.0190 &0.0702-0.0524 &0.0435-0.0333   & 0.0679-0.0514  &0.0229-0.0175  \\
$\Delta\mu$ total (0.03 sys.)  & 0.0301-0.0300  & 0.0390-0.0355  & 0.0763-0.0604 & 0.0529-0.0448 & 0.0742-0.0595 & 0.0378-0.0347  \\
\hline\hline
\end{tabular}
\label{tab:uncert}
\end{table*}

\begin{table*}[htbp]
\small
  \caption{Same as Table~\ref{tab:uncert}, but for LHC fill 3249.
          The systematic uncertainties result from luminosity fluctuations 
          not described by the fit function 
          (taking $\sigma_{\rm sys}^{\rm fct} =0.02$ and 0.03),
          shown in Fig.~\ref{fig:fill3249res}.
          LHC fill 3249.
}
 \centering
\renewcommand{\arraystretch}{1.3} 
    \begin{tabular}{ccccccccc}
\hline\hline
MPX           & 01    & 02    & 03    & 04    & 05    & 13     \\\hline
$\Delta\mu$ stat.  &0.0022-0.0019 & 0.0237-0.0204 &0.0668-0.0570 &0.0419-0.0354   & 0.0653-0.0556  &0.0220-0.0187  \\
$\Delta\mu$ total (0.02 sys.)  & 0.0201-0.0201  & 0.0310-0.0286  & 0.0607-0.0604 & 0.0464-0.0407 & 0.0682-0.0591 & 0.0297-0.0274  \\
$\Delta\mu$ total (0.03 sys.)  & 0.0301-0.0301  & 0.0383-0.0363  & 0.0732-0.0644 & 0.0515-0.0464 & 0.0652-0.0632 & 0.0372-0.0353  \\
\hline\hline
\end{tabular}
\label{tab:uncert3249}
\end{table*}

\begin{table*}[htbp]
\small
  \caption{Widths of the Gaussian fits to the pull distributions for MPX01-05 and MPX13.
          The widths are given for uncertainties resulting from the number of hits scaled
          by a factor $\sqrt{N_{\rm hit}/N_{\rm cl}}$ for each MPX device,
          given in Table~\ref{tab:stat}.
          The widths are also given for statistical and systematic 
          uncertainties added in quadrature, where the systematic uncertainties 
          result from luminosity fluctuations not described by the fit function
          (taking 0.02 and 0.03).
          Table~\ref{tab:uncert3249} indicates the sizes of the uncertainty applied.
          LHC fill 3249.
}
  
 \centering
\renewcommand{\arraystretch}{1.3} 
    \begin{tabular}{ccccccccc}
\hline\hline
MPX                                  & 01    & 02    & 03    & 04    & 05    & 13     \\\hline
Width pull stat. uncert.             &12.69  & 2.76  & 2.74  & 2.08  & 2.06  & 2.90   \\
Width pull total uncert. (0.02 sys.) & 1.24  & 2.07  & 2.57  & 1.78  & 1.93  & 1.96   \\
Width pull total uncert. (0.03 sys.) & 0.80  & 1.54  & 2.42  & 1.67  & 1.81  & 1.53   \\
\hline\hline
\end{tabular}
\label{tab:pull3249}
\end{table*}

\clearpage

\section{Conclusions}
\label{sec:conclusions}

The network of MPX devices installed in the ATLAS detector cavern 
has successfully taken data from 2008 to 2013. 
The study presented here focuses on the proton-proton collisions from May to November 2012. 
This study has demonstrated that the MPX network 
is well suited for luminosity monitoring. 
The slopes of the long-term time-stability of the luminosity measurements  both
from the hit and heavy blob (thermal neutron) counting are below 1\%. 
The uncertainties for the hit analysis are dominated by the systematic effects.
For the heavy blob (thermal neutron) analysis, the statistical uncertainties 
describe the fluctuations.
The MPX network has been used as well to study 
all the van der Meer scans performed in 2012 in detail. 
It is demonstrated that the MPX luminosity measurements can cope with a collision 
rate 1/1000 of the one  characteristic for  physics data-taking.
Although not specifically designed for luminosity measurements, the MPX network 
gives reliable supplementary 
information for the luminosity determination of LHC proton-proton collisions. 
It is demonstrated that the precision of the MPX network is sufficient to study the underlying
mechanisms of the LHC luminosity reduction.
The evaluation of the variations around the fitted time-dependence of the LHC luminosity
is found to give important information on the precision of the individual MPX devices.
This study shows that the relative uncertainty on the luminosity measurement 
is below 0.3\% for one minute intervals.

A network of TPX devices (upgraded successors of the MPX devices) has been installed as the replacement 
of the MPX network in preparation for the Run-2 LHC operation~\cite{upgrade}.

\section*{Acknowledgment}
The authors would like to thank  warmly 
the ATLAS Luminosity Group for useful discussions and  interactions.
The project is supported by the 
Ministry of Education, Youth and Sports of the Czech Republic under projects number 
MSM 68400029, LA 08032, LG 13009 and LG 13031, and the Natural Sciences and 
Engineering Research Council of Canada (NSERC).
Calibration measurements were performed at the Prague Van-de-Graaff accelerator 
funded by the Ministry of Education, Youth and Sports of the Czech Republic
under project number LM 2011030.

\bibliographystyle{IEEEtran}

\bibliography{biblio}

\begin{thebibliography}{10}
\providecommand{\url}[1]{#1}
\csname url@samestyle\endcsname
\providecommand{\newblock}{\relax}
\providecommand{\bibinfo}[2]{#2}
\providecommand{\BIBentrySTDinterwordspacing}{\spaceskip=0pt\relax}
\providecommand{\BIBentryALTinterwordstretchfactor}{4}
\providecommand{\BIBentryALTinterwordspacing}{\spaceskip=\fontdimen2\font plus
\BIBentryALTinterwordstretchfactor\fontdimen3\font minus
  \fontdimen4\font\relax}
\providecommand{\BIBforeignlanguage}[2]{{%
\expandafter\ifx\csname l@#1\endcsname\relax
\typeout{** WARNING: IEEEtran.bst: No hyphenation pattern has been}%
\typeout{** loaded for the language `#1'. Using the pattern for}%
\typeout{** the default language instead.}%
\else
\language=\csname l@#1\endcsname
\fi
#2}}
\providecommand{\BIBdecl}{\relax}
\BIBdecl

\bibitem{atlasCollaboration:2013}
{ATLAS Collaboration}, ``{The ATLAS Experiment at the CERN Large Hadron
  Collider},'' \emph{JINST}, vol.~3, p. S08003, 2008.

\bibitem{medipixCollaboration:2013}
{Medipix-2 Collaboration}, Project Webpage [Online]. Available: \\
  http://medipix.web.cern.ch/medipix, 2015.

\bibitem{mpx2ProjectProposal:2006}
M.~Campbell, C.~Leroy, S.~Pospisil, and M.~Suk, ``{Measurement of Spectral
  Characteristics and Composition of Radiation in ATLAS by MEDIPIX2-USB
  Devices},'' Project Proposal [Online]. Available:
  https://edms.cern.ch/document/815615, 2006.

\bibitem{analysisRadiaField:2013}
M.~Campbell \emph{et~al.}, ``{Analysis of Radiation Field in ATLAS Using
  2008-2011 Data from the ATLAS-MPX Network},'' {ATL-GEN-PUB-2013-001}, 2013.

\bibitem{improvedLumiDet:2013}
{ATLAS Collaboration}, ``{Improved luminosity determination in pp collisions at
  $\rm \sqrt{s} =7~TeV$ using the ATLAS detector at the LHC},'' \emph{EPJC},
  vol.~73, p. 2518, 2013.

\bibitem{cms:2013}
{CMS Collaboration}, ``{CMS Luminosity Based on Pixel Cluster Counting - Summer
  2013 Update},'' 2013, {CMS Public Analysis Summary CMS-PAS-LUM-13-001}.

\bibitem{sopczak:2013}
A.~Sopczak, ``{Luminosity Monitoring in ATLAS with MPX Detectors},'' {on behalf
  of the ATLAS and Medipix-2 Collaborations, in Proc. IPRD13, 7-10 October
  2013, Siena, Italy. {\em JINST}, vol. 9, p. C01027, 2014}.

\bibitem{vdm}
S.~van~der Meer, ``{Calibration of the effective beam height in the ISR},''
  {ISR-PO/68-31 CERN Report}, 1968.

\bibitem{lhc}
LHC, Perfromance and Statistics \\
  http://lhc-statistics.web.cern.ch/LHC-Statistics, 2015.

\bibitem{LHC94}
K.~Eggert, K.~Honkavaara, and A.~Morsch, ``{Luminosity considerations for the
  LHC},'' 1994, {CERN-AT-94-04, LHC-NOTE-263}.

\bibitem{cmsCollaboration}
{CMS Collaboration}, ``{The CMS experiment at the CERN LHC},'' \emph{JINST},
  vol.~3, p. S08004, 2008.

\bibitem{upgrade}
C.~Leroy, S.~Pospisil, M.~Suk, and Z.~Vykydal, ``{Proposal to Measure Radiation
  Field Characteristics, Luminosity and Induced Radioactivity in ATLAS with
  TIMEPIX Devices},'' Project Proposal [Online]. Available:
  http://cds.cern.ch/record/1646970, 2014.

\end{thebibliography}

\end{document}